\documentclass[11pt]{article}
\usepackage{amsmath}
\usepackage{epsfig}

\newcommand{\ord}[1]{\mathcal{O}(#1)}
\newcommand{\del}{\partial}

\newcommand{\ket}[1]{|#1\rangle}
\newcommand{\dcirc}{\mathop{{}^\circ_\circ}}
\newcommand{\ntom}
{\quad \xrightarrow[\,\,{\cal W}_n \to {\cal W}_m\,]{} \quad}

\begin{document}

\baselineskip=17pt

\begin{titlepage}
\rightline{\tt arXiv:0707.4472}
\rightline{\tt DESY 07-110}
\rightline{\tt MIT-CTP-3851}
\begin{center}

\vskip 2.0cm
{\Large \bf {Exact marginality in open string field theory:
a general framework}}\\
\vskip 1.5cm

{\large {Michael Kiermaier${}^1$ and Yuji Okawa${}^2$}}

\vskip 1.0cm

{\it {${}^1$ Center for Theoretical Physics}}\\
{\it {Massachusetts Institute of Technology}}\\
{\it {Cambridge, MA 02139, USA}}\\
mkiermai@mit.edu

\vskip 0.5cm

{\it {${}^2$ DESY Theory Group}}\\
{\it {Notkestrasse 85}}\\
{\it {22607 Hamburg, Germany}}\\
yuji.okawa@desy.de

\vskip 1.0cm

{\bf Abstract}
\end{center}

\noindent
We construct analytic solutions
of open bosonic string field theory
for any exactly marginal deformation
in any boundary conformal field theory
when properly renormalized operator products
of the marginal operator are given.
We explicitly provide such renormalized operator products
for a class of marginal deformations
which include the deformations
of flat D-branes in flat backgrounds
by constant massless modes of the gauge field
and of the scalar fields on the D-branes,
the cosine potential for a space-like coordinate,
and the hyperbolic cosine potential
for the time-like coordinate.
In our construction
we use integrated vertex operators,
which are closely related
to finite deformations
in boundary conformal field theory,
while previous analytic solutions
were based on unintegrated vertex operators.
We also introduce a modified star product
to formulate string field theory
around the deformed background.

\end{titlepage}

\newpage

\baselineskip=16pt
\tableofcontents
\baselineskip=17pt

\section{Introduction}
\setcounter{equation}{0}

String field theory\footnote{
See \cite{Taylor:2003gn, Sen:2004nf, Rastelli:2005mz, Taylor:2006ye}
for reviews.} can potentially be
a background-independent formulation of string theory.
In the current formulation of string field theory, however,
we first need to choose
one conformal field theory (CFT)
describing a consistent background of string theory.
The crucial question is then whether other string backgrounds
can be described as classical solutions of string field theory.
In particular, for each exactly marginal deformation of the CFT,
we expect to have a family of solutions
in string field theory labeled by the deformation parameter.

Recent remarkable developments in analytic methods
of open string field
theory~\cite{Schnabl:2005gv}--\cite{Bonora:2007tm}
enabled us to address this question in a concrete setting.
Analytic solutions for marginal deformations
when operator products of the marginal operator are regular
were constructed to all orders in the deformation parameter
in~\cite{Schnabl:2007az, Kiermaier:2007ba}
for open bosonic string field theory~\cite{Witten:1985cc}
and in \cite{Erler:2007rh, Okawa:2007ri, Okawa:2007it}
for open superstring field theory~\cite{Berkovits:1995ab}.
When the operator product of the marginal operator is singular,
analytic solutions were constructed to third order
in the deformation parameter in \cite{Kiermaier:2007ba}.
Recently, analytic solutions for the deformation
generated by the zero mode of the gauge field
were constructed in~\cite{Fuchs:2007yy} by a different approach
and extended to open superstring field theory
in~\cite{Fuchs:2007gw}.
While the equation of motion is satisfied
to all orders in the deformation parameter,
a closed form expression for a solution satisfying
the reality condition on the string field
has not been presented in~\cite{Fuchs:2007yy, Fuchs:2007gw}.
For earlier study of marginal deformations in string field theory,
see~\cite{Sen:2000hx}--\cite{Kishimoto:2005bs}.

In this paper, we present a procedure to construct
a solution satisfying the reality condition
in open bosonic string field theory
for any exactly marginal deformation in any boundary CFT
when properly renormalized operator products
of the marginal operator are given.
The analytic solutions
in~\cite{Schnabl:2007az, Kiermaier:2007ba}
were constructed using unintegrated vertex operators
and $b$-ghost insertions.
Our strategy is to use integrated vertex operators,
which are closely related
to finite deformations in boundary CFT.
We assume several properties
of the properly renormalized operator products
of the marginal operator.
Since the identification of a set of
assumptions which are sufficient for the construction of a solution
is one of the main points of the paper,
we will explain these assumptions in detail in the following.
We will then present our solutions.

\subsection{Assumptions}
\label{assumptions}

When there exists an exactly marginal deformation in a given
boundary CFT, we have a family of consistent boundary conditions
labeled by the deformation parameter which we denote by $\lambda$.
Consider the boundary CFT on the upper-half plane and suppose that
we change boundary conditions
on a segment of the boundary
between $a$ and $b$.
Since the new boundary condition is also conformal,
an integral of the BRST current
along a contour vanishes
if both end points of the contour lie inside the region
between $a$ and $b$.
By $C (t_f, \, t_i)$ we denote a contour in the
upper-half plane which starts from the point $t_i$ on the real axis
and ends on $t_f$ on the real axis,
and we use $C (t_f, \, t_i)$
with $t_f < t_i$ in what follows.
We have
\begin{figure}[tb]\Large
\begin{center}
\raisebox{-.6cm}{\epsfig{figure=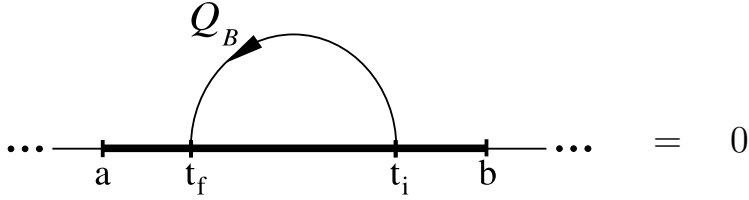, width=8cm}}
$\quad=\quad 0$
\end{center}\normalsize
\caption{Illustration of~(\ref{QexpV2}).
The bold line indicates a change of boundary conditions
on the segment between $a$ and $b$.
The integral of the BRST current in~(\ref{QexpV2})
vanishes when $a < t_f < t_i < b$.}
\label{fig1}
\end{figure}
\begin{equation}\label{QexpV2}
\int_{C (t_f, \, t_i)} \Bigl[ \, \frac{dz}{2 \pi i} \, j_B (z)
- \frac{d \bar{z}}{2 \pi i} \, \tilde{\jmath}_B (\bar{z}) \,
\Bigr] = 0
\quad \text{when} \quad a < t_f < t_i < b \,,
\end{equation}
where $j_B (z)$ and $\tilde{\jmath}_B (\bar{z})$ are
the holomorphic and antiholomorphic components
of the BRST current, respectively. See figure~\ref{fig1}.
This identity holds inside any correlation function
of the deformed CFT as long as no operators are inserted
between the contour $C (t_f, \, t_i)$ and the real axis.
When $t_f < a < b < t_i$, there are contributions from
the points $a$ and $b$ where the boundary condition changes:
\begin{equation}
\begin{split}
& \int_{C (t_f, \, t_i)} \Bigl[ \, \frac{dz}{2 \pi i} \, j_B (z)
- \frac{d \bar{z}}{2 \pi i} \, \tilde{\jmath}_B (\bar{z}) \,
\Bigr] \\
& = \int_{C (b)} \Bigl[ \, \frac{dz}{2 \pi i} \, j_B (z) - \frac{d
\bar{z}}{2 \pi i} \, \tilde{\jmath}_B (\bar{z}) \, \Bigr] + \int_{C
(a)} \Bigl[ \, \frac{dz}{2 \pi i} \, j_B (z) - \frac{d \bar{z}}{2
\pi i} \, \tilde{\jmath}_B (\bar{z}) \, \Bigr] \,,
\end{split}
\label{contour-integral-1}
\end{equation}
where we have defined the
infinitesimal contour $C(t)$ around any point $t$ by
\begin{equation}
    C(t)\,=\, \lim_{\epsilon\to0} \, C(t-\epsilon,t+\epsilon)\,.
\end{equation}
See figure~\ref{figAss1a}.
\begin{figure}[tb]
\Large
\begin{center}
\raisebox{-.6cm}{\epsfig{figure=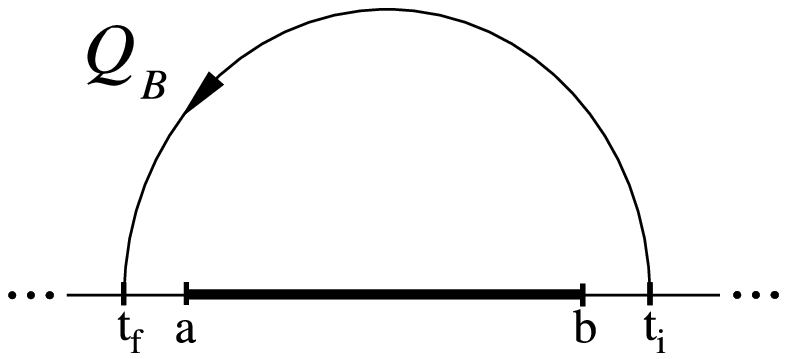, width=5cm}}
$\,=\,$
\raisebox{-.6cm}{\epsfig{figure=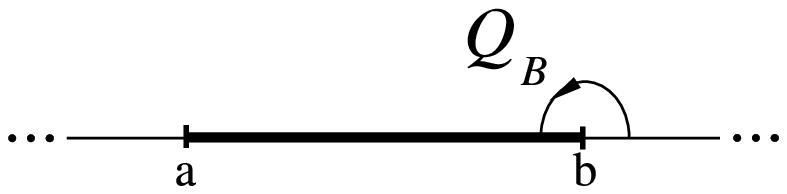, width=5cm}}
$\,+\,$
\raisebox{-.6cm}{\epsfig{figure=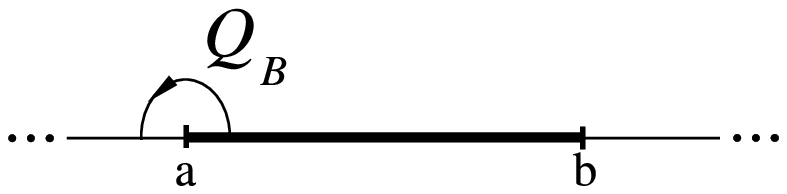, width=5cm}}
\end{center}
\normalsize
\caption{Illustration of~(\ref{contour-integral-1}).
When $t_f < a < b < t_i$,
the integral of the BRST current on the left-hand side
decomposes into a sum of two integrals
localized at the end points $a$ and $b$ of the segment.}\label{figAss1a}
\end{figure}
The nonvanishing contributions in~(\ref{contour-integral-1})
can be thought of as the BRST transformations
of the boundary-condition changing operators.
We also have
\begin{figure}[tb]
\Large
\begin{center}
\raisebox{-.8cm}{\epsfig{figure=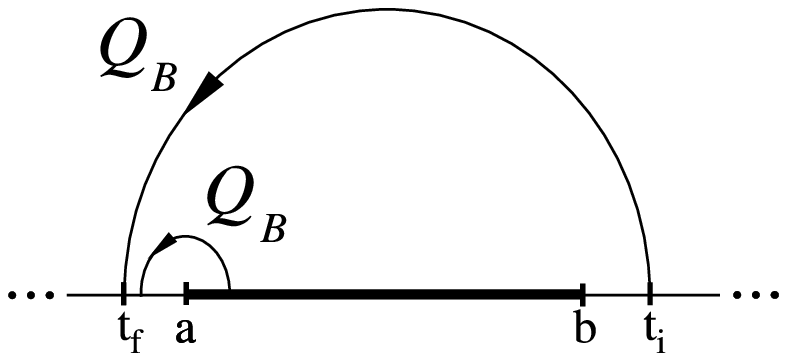, width=6cm}}
$\quad=\quad-\,$
\raisebox{-.8cm}{\epsfig{figure=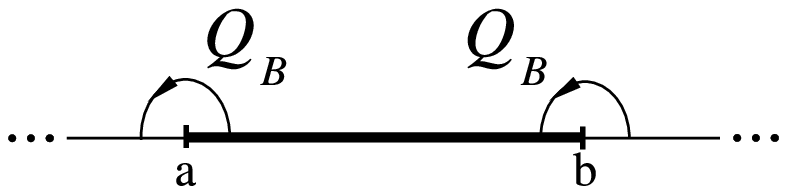, width=6cm}}
\end{center}
\normalsize
\caption{Illustration of~(\ref{contour-integral-2}).
With the presence of the BRST integral localized at $a$,
the integral along $C(t_f,t_i)$ on the left-hand side
localizes only at the other end point $b$
because of the nilpotency of the BRST transformation.
}\label{figAss2a}
\end{figure}
\begin{equation}
\begin{split}
& \int_{C (t_f, \, t_i)} \Bigl[ \, \frac{dz}{2 \pi i} \, j_B (z) -
\frac{d \bar{z}}{2 \pi i} \, \tilde{\jmath}_B (\bar{z}) \, \Bigr] \,
\int_{C (a)} \Bigl[ \, \frac{dz}{2 \pi i} \, j_B (z) - \frac{d
\bar{z}}{2 \pi i} \, \tilde{\jmath}_B (\bar{z}) \,
\Bigr] \\
& = {}- \int_{C (a)} \Bigl[ \, \frac{dz}{2 \pi i} \, j_B (z) -
\frac{d \bar{z}}{2 \pi i} \, \tilde{\jmath}_B (\bar{z}) \, \Bigr] \,
\int_{C (b)} \Bigl[ \, \frac{dz}{2 \pi i} \, j_B (z) - \frac{d
\bar{z}}{2 \pi i} \, \tilde{\jmath}_B (\bar{z}) \, \Bigr] \,,
\end{split}
\label{contour-integral-2}
\end{equation}
where again $t_f < a < b < t_i$, as shown in figure~\ref{figAss2a}.

The boundary CFT with a different boundary condition
on a segment between $a$ and $b$ discussed above
can also be described
in the boundary CFT with the original boundary condition
on the whole real axis by inserting
an exponential of the marginal operator $V(t)$
integrated over the segment between $a$ and $b$,
\begin{equation}\label{exp-before-renormalization}
\exp \biggl[ \, \lambda \int_a^b dt \, V(t) \, \biggr]
= 1 + \lambda \int_a^b dt \, V(t)
+ \frac{\lambda^2}{2!} \int_a^b dt_1 \int_a^b dt_2 \,
V(t_1) \, V(t_2) + \, \ldots \, ,
\end{equation}
into the correlation function.
When operator products of the marginal operator are singular,
we need to renormalize
the operator~(\ref{exp-before-renormalization})
properly to make it well defined,
and we denote the renormalized operator by
\begin{equation}
[ \, e^{\lambda V(a,b)} \, ]_r \,,
\end{equation}
where
\begin{equation}
V(a,b) \equiv \int_a^b dt \, V(t) \,.
\end{equation}
Then the equations (\ref{contour-integral-1})
and (\ref{contour-integral-2})
can be translated
into the following assumptions
on the operator~$[ \, e^{\lambda V(a,b)} \, ]_r$.\\

\noindent
{\it 1. The BRST transformation of
the operator $[ \, e^{\lambda V(a,b)} \, ]_r$
takes the following form:}
\begin{equation}
Q_B \cdot [ \, e^{\lambda V(a,b)} \, ]_r
= [ \, e^{\lambda V(a,b)} \, O_R (b) \, ]_r
- [ \, O_L (a) \, e^{\lambda V(a,b)} \, ]_r \,,
\tag{I}\label{1}
\end{equation}
{\it where $O_L (a)$ and $O_R (b)$ are some local operators
at $a$ and $b$, respectively.}\\

\begin{figure}[tb]
\Large
\begin{center}
\raisebox{-.6cm}{\epsfig{figure=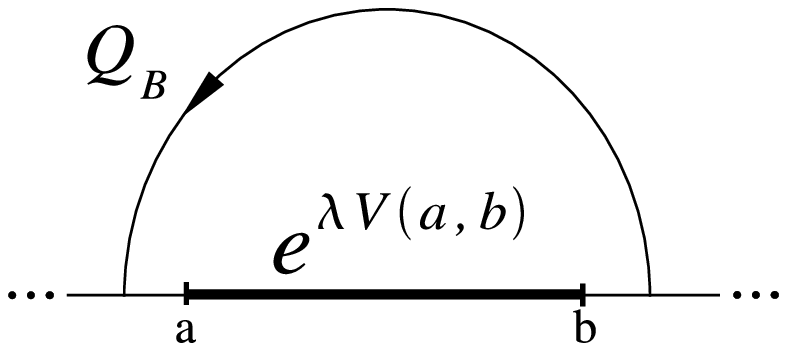, width=5cm}}
$\,=\,$
\raisebox{-.6cm}{\epsfig{figure=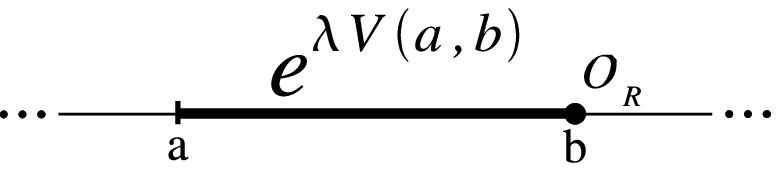, width=5cm}}
$\,-\,$
\raisebox{-.6cm}{\epsfig{figure=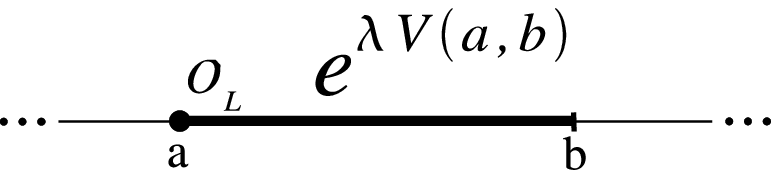, width=5cm}}\\
\end{center}
\normalsize
\caption{
Illustration of the assumption~(\ref{1}).
The BRST transformation on the operator
$[ \, e^{\lambda V(a,b)} \, ]_r$ generates
local operators $O_L(a)$ and $O_R(b)$
at the end points of the segment.
Compare this figure with figure \ref{figAss1a}.
}\label{figAss1b}
\end{figure}

\begin{figure}[tb]
\Large
\begin{center}
\raisebox{-.6cm}{\epsfig{figure=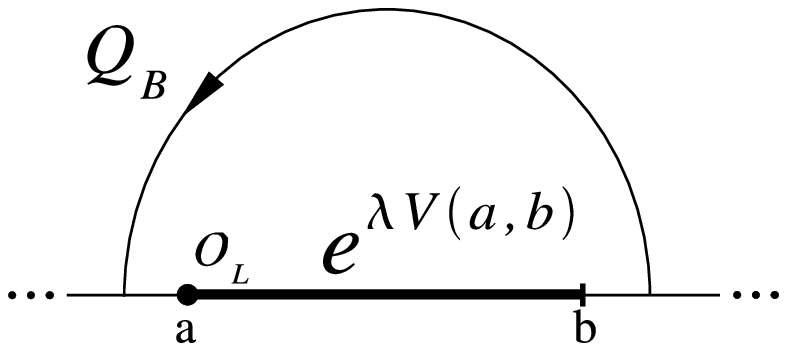, width=5cm}}
$\,=\,-$
\raisebox{-.6cm}{\epsfig{figure=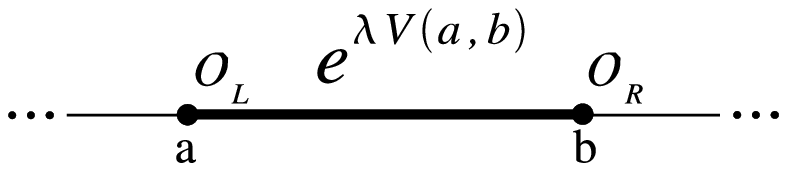, width=5cm}}
\end{center}
\normalsize
\caption{
Illustration of the assumption (\ref{2}).
The BRST transformation on the operator
$[ \, O_L(a) \, e^{\lambda V(a,b)} \, ]_r$
generates the local operator $O_R (b)$.
Compare this figure with figure \ref{figAss2a}.
}\label{figAss2b}
\end{figure}

\noindent
{\it 2. The BRST transformation of
the operator $[ \, O_L (a) \, e^{\lambda V(a,b)} \, ]_r$
is given by}
\begin{equation}
Q_B \cdot [ \, O_L (a) \, e^{\lambda V(a,b)} \, ]_r
= {}- [ \, O_L (a) \, e^{\lambda V(a,b)} \, O_R (b) \, ]_r \,.
\tag{II}\label{2}
\end{equation}\\
\noindent
These are our first two assumptions.
They are illustrated in figures~\ref{figAss1b} and~\ref{figAss2b}.


We can also introduce different boundary conditions
on different segments on the boundary by inserting
\begin{equation}
[ \, \prod_{i=1}^n \, e^{\lambda_i V (a_i, a_{i+1} )} \, ]_r \,
\label{multiple-regions}
\end{equation}
with $a_i < a_{i+1}$ for $i = 1, 2, \ldots , n$
into the correlation function.
We make the following two assumptions on this operator.\\

\noindent
{\it 3. Replacement. When $\lambda_{i+1} = \lambda_i$, the product
$e^{\lambda_i V (a_i, a_{i+1} )} \,
e^{\lambda_{i+1} V (a_{i+1}, a_{i+2} )}$
inside the operator (\ref{multiple-regions})
can be replaced by
$e^{\lambda_i V (a_i, a_{i+2} )}$}:
\begin{equation}
[ \, \ldots \, e^{\lambda_i V (a_i, a_{i+1} )} \,
e^{\lambda_i V (a_{i+1}, a_{i+2} )} \, \ldots \, ]_r
= [ \, \ldots \, e^{\lambda_i V (a_i, a_{i+2} )} \,
\ldots \, ]_r \,.
\tag{III}\label{3}
\end{equation}\\
\noindent
{\it 4. Factorization. When $\lambda_j$ vanishes,
the renormalized product~(\ref{multiple-regions})
factorizes as follows:}
\begin{equation}
[ \, \ldots \, e^{\lambda_{j-1} V (a_{j-1}, a_{j} )} \,
e^{\lambda_{j+1} V (a_{j+1}, a_{j+2} )} \, \ldots \, ]_r
=
[ \, \ldots \, e^{\lambda_{j-1} V (a_{j-1}, a_{j} )} \,]_r\,
[ \, e^{\lambda_{j+1} V (a_{j+1}, a_{j+2} )} \, \ldots \, ]_r
\,.
\tag{IV}\label{4}
\end{equation}\\
\noindent
We also assume that (\ref{3}) and (\ref{4}) hold
when $O_L (a_1)$, $O_R (a_{n+1})$, or both of them
are inserted in~(\ref{multiple-regions}).

A change of boundary conditions
on a segment between $a$ and $b$ is local
and independent of other regions of the Riemann surface
where the boundary CFT is defined.
Thus the operator $[ \, e^{\lambda V(a,b)} \, ]_r$
should be independent of the global shape
of the Riemann surface.
However, renormalization schemes
such as the standard normal ordering
can depend on the global shape of the surface
through the propagator,
and normal ordered products of nonlocal operators
generically do depend on the surface.
We consider boundary conformal field theory
defined  on a family of semi-infinite cylinders ${\cal W}_n$
obtained from the upper-half plane of $z$
by the identification $z \sim z + n + 1$
and make the following assumption.\\

\noindent
{\it 5. Locality. The operators $[ \, e^{\lambda V(a,b)} \, ]_r$
and $[ \, O_L (a) \, e^{\lambda V(a,b)} \, ]_r$
defined on ${\cal W}_n$ coincide
with those defined on ${\cal W}_m$ with $m > n$}:
\begin{equation}
\begin{split}
[ \, e^{\lambda V(a,b)} \, ]_r
~ \text{on} ~ {\cal W}_n ~
& = ~ [ \, e^{\lambda V(a,b)} \, ]_r
~ \text{on} ~ {\cal W}_m \,, \\
[ \, O_L(a) \, e^{\lambda V(a,b)} \, ]_r
~ \text{on} ~ {\cal W}_n ~
& = ~ [ \, O_L(a) \, e^{\lambda V(a,b)} \, ]_r
~ \text{on} ~ {\cal W}_m \,.
\end{split}
\tag{V}\label{5}
\end{equation}\\
\indent
Finally, $e^{\lambda V(a,b)}$
is classically invariant under the reflection
where $V(t)$ is replaced by $V(a+b-t)$,
and we assume that
$[ \, e^{\lambda V(a,b)} \, ]_r$ preserves this symmetry.\\

\noindent
{\it 6. Reflection. The operator $[ \, e^{\lambda V(a,b)} \, ]_r$
is invariant under the reflection
where $V(t)$ is replaced by $V(a+b-t)$}:
\begin{equation}
\biggl[ \, \exp \biggl( \, \lambda \int_a^b dt \, V(a+b-t) \,
\biggr) \, \biggr]_r
= \biggl[ \, \exp \biggl( \, \lambda \int_a^b dt \, V(t) \,
\biggr) \, \biggr]_r \,.
\tag{VI}\label{6}
\end{equation}

\subsection{Solutions}
\label{solution}

We believe that all of these assumptions are satisfied
for any exactly marginal deformation
in any boundary CFT
if the composite operators are properly renormalized.
When the operator
$[ \, e^{\lambda V(a,b)} \, ]_r$ expanded in $\lambda$ as
\begin{equation}
[ \, e^{\lambda V(a,b)} \, ]_r
= \sum_{n=0}^\infty \lambda^n \,
[ \, V^{(n)}(a,b)\, ]_r \,,
\end{equation}
where
\begin{equation}
[ \, V^{(n)} (a,b)\, ]_r
\equiv \frac{1}{n!} \, [ \, ( V(a,b) )^n \, ]_r
\quad\text{ for }\quad n\geq 1
\quad\text{ and }\quad [\,V^{(0)}(a,b)\,]_r\equiv1 \,,
\end{equation}
is given,
we claim that solutions
to the equation of motion can be constructed in the following way.

We first define a state $U$ by
\begin{equation}
U \equiv 1 + \sum_{n=1}^\infty \, \lambda^n \, U^{(n)} \,,
\end{equation}
where
\begin{equation}
\langle \, \phi \,,\, U^{(n)} \, \rangle
= \langle \, f \circ \phi (0) \, \,
[ \, V^{(n)} (1,n) \, ]_r \, \rangle_{{\cal W}_n} \,.
\end{equation}
Here and in what follows we denote a generic state
in the Fock space by $\phi$ and its corresponding operator
in the state-operator mapping by $\phi (0)$.
The conformal transformation $f(\xi)$ is
\begin{equation}
f(\xi) = \frac{2}{\pi} \, \arctan \xi \,,
\end{equation}
and we denote the conformal transformation of $\phi (\xi)$ under the
map $f(\xi)$ by $f \circ \phi (\xi)$.
The correlation function is evaluated
on the surface ${\cal W}_n$,
which we defined above when stating the
locality assumption~(\ref{5}).
We represent it in the region of the upper-half plane of $z$
where $-1/2 \le {\rm Re} \, z \le 1/2 + n$.

If the assumption~(\ref{1}) is satisfied,
the BRST transformation of the operator
$[ \, V^{(n)}(a,b) \, ]_r$ takes the form
\begin{equation}
Q_B \cdot [ \, V^{(n)}(a,b) \, ]_r
= \sum_{r=1}^n \, [  \, V^{(n-r)}(a,b) \, O_R^{(r)}(b)\, ]_r
{}- \sum_{l=1}^n \, [ \, O_L^{(l)}(a) \, V^{(n-l)}(a,b) \, ]_r\,,
\end{equation}
where $O_L$ and $O_R$ are expanded as follows:
\begin{equation}
O_L = \sum_{n=1}^\infty \, \lambda^n \, O_L^{(n)} \,, \qquad
O_R = \sum_{n=1}^\infty \, \lambda^n \, O_R^{(n)} \,.
\end{equation}
Thus the BRST transformation of $U$ can be split into two pieces:
\begin{equation}
Q_B U = A_R - A_L
\end{equation}
with
\begin{equation}
A_L = \sum_{n=1}^\infty \, \lambda^n \, A_L^{(n)} \,, \qquad
A_R = \sum_{n=1}^\infty \, \lambda^n \, A_R^{(n)} \,,
\end{equation}
where
\begin{equation}
\begin{split}
    \langle \, \phi \,, A_L^{(n)} \, \rangle &=
    \sum_{l=1}^n \langle \, f \circ \phi (0) \, [ \, O_L^{(l)} (1) \, V^{(n-l)} (1,n) \, ]_r \, \rangle_{{\cal W}_n},\\
    \langle \, \phi \,, A_R^{(n)} \, \rangle &=
     \sum_{r=1}^n \langle \, f \circ \phi (0) \, [  \, V^{(n-r)} (1,n) \, O_R^{(r)} (n)\, ]_r \, \rangle_{{\cal W}_n}  \,.
\end{split}
\end{equation}
We then define $\Psi_L$ and $\Psi_R$ by
\begin{equation}
\Psi_L \equiv A_L \ast U^{-1} \,, \qquad
\Psi_R \equiv U^{-1} \ast A_R \,,
\end{equation}
where $U^{-1}$ is well defined perturbatively in $\lambda$
because $U=1+\ord{\lambda}$.
We show that $\Psi_L$ and $\Psi_R$ satisfy the equation of motion,
\begin{equation}
Q_B \Psi_L + \Psi_L \ast \Psi_L = 0 \,, \qquad
Q_B \Psi_R + \Psi_R \ast \Psi_R = 0 \,,
\end{equation}
though they do not satisfy the reality condition on the string field.
They are related by the gauge transformation
generated by $U$:
\begin{equation}
\Psi_R \, = \, U^{-1}\ast \Psi_L\ast U + U^{-1}\ast Q_B U \,.
\label{Psi_L-to-Psi_R}
\end{equation}
A solution $\Psi$ satisfying the reality condition
is obtained from $\Psi_L$ or $\Psi_R$
by gauge transformations as follows:
\begin{equation}
\begin{split}
\Psi & = \frac{1}{\sqrt{U}} \ast \Psi_L \ast \sqrt{U} +
\frac{1}{\sqrt{U}} \ast Q_B \, \sqrt{U}\\
&= \sqrt{U} \ast \Psi_R \ast \frac{1}{\sqrt{U}}
+ \sqrt{U} \ast Q_B \, \frac{1}{\sqrt{U}} \\
& = \frac{1}{2} \, \biggl[ \, \frac{1}{\sqrt{U}} \ast \Psi_L \ast
\sqrt{U} + \sqrt{U} \ast \Psi_R \ast \frac{1}{\sqrt{U}} +
\frac{1}{\sqrt{U}} \ast Q_B \, \sqrt{U}
- Q_B \, \sqrt{U} \ast \frac{1}{\sqrt{U}} \, \biggr] \,,
\end{split}
\end{equation}
where $\sqrt{U}$ and $1/\sqrt{U}$ are defined perturbatively
in $\lambda$.
The three expressions are equivalent
because of the relation (\ref{Psi_L-to-Psi_R}).
This solution is the main result of the paper.
In section \ref{singular_explicit},
we explicitly construct $[ \, e^{\lambda V(a,b)} \, ]_r$
satisfying all the assumptions
and apply the general result
to obtain solutions for a class of marginal deformations
which include the deformations
of flat D-branes in flat backgrounds
by constant massless modes of the gauge field
and of the scalar fields on the D-branes,
the cosine potential for a space-like coordinate,
and the hyperbolic cosine potential
for the time-like coordinate.

The operators $O_R^{(1)}$ and $O_L^{(1)}$ are
\begin{equation}
O_R^{(1)} = O_L^{(1)} = cV
\end{equation}
for any marginal deformation. This follows only from the fact
that the marginal operator is
a primary field of dimension one.
When operator products of the marginal operator are regular,
there are no higher-order terms and thus
$O_R = O_L = \lambda \, cV$.
For any exactly marginal deformation
where the singular part of the operator product
of the marginal operator with itself is
\begin{equation}
V (t) \, V(0) \sim \frac{1}{t^2} \,,
\end{equation}
the operators $O_L^{(2)}$ and $O_R^{(2)}$ are
\begin{equation}
O_R^{(2)} = {}- O_L^{(2)} = \frac{1}{2} \, \partial c\,.
\end{equation}
For the class of marginal deformations
to be considered in section \ref{singular_explicit},
there are no higher-order terms
and the exact expressions of $O_R$ and $O_L$ are
\begin{equation}
O_R = \lambda \, cV + \frac{\lambda^2}{2} \, \partial c \,, \qquad
O_L = \lambda \, cV - \frac{\lambda^2}{2} \, \partial c \,.
\end{equation}

\subsection{The organization of the paper}
\label{organization}

In section~\ref{regular} we first revisit the problem
of constructing solutions for marginal deformations
with regular operator products.
In~\S~\ref{regular_PsiL} we construct a solution $\Psi_L$
to the string field theory equation of motion
using integrated vertex operators
without $b$-ghost insertions.
The solution $\Psi_L$, however, does not satisfy
the reality condition on the string field.
In \S~\ref{regular_real}
we construct a gauge transformation which connects
$\Psi_L$ and its conjugate solution $\Psi_R$,
and then we generate a real solution $\Psi$
using the gauge transformation.
During the construction of this gauge transformation,
we find an important identity.
It leads us to discover a class of states $U_\alpha$,
which generalize the wedge states $W_\alpha$
in a deformed background.
We study the properties of $U_\alpha$
in~\S~\ref{regular_generalizewedge}.

In the process of constructing the gauge transformation
that connects $\Psi_L$ and $\Psi_R$,
we also find another expression of the solution $\Psi_L$.
We study the new form of $\Psi_L$ in~\S~\ref{another}
and prove that it satisfies the equation of motion
using the properties of $U_\alpha$.
The new form of $\Psi_L$ can be generalized
to marginal deformations with singular operator products.
In~\S~\ref{singular_general_PsiL}
we construct $\Psi_L$ for the singular case
using the operator $[ \, e^{\lambda V(a,b)} \, ]_r$, and we prove
in~\S~\ref{singular_general_proof}
and in appendix~\ref{QBAL-appendix}
that it satisfies the equation of motion
under the assumptions stated in~\S~\ref{assumptions}.
We then generate a real solution $\Psi$
for the singular case
in~\S~\ref{singular_general_real}
by an appropriate gauge transformation
as in the regular case in~\S~\ref{regular_real}.

In section~\ref{singular_explicit}
we explicitly construct the operator
$[ \, e^{\lambda V(a,b)} \, ]_r$
satisfying the assumptions stated in~\S~\ref{assumptions}
for a class of marginal operators
with singular operator products defined in~\S~\ref{class}.
We give several examples of marginal operators
included in this class in~\S~\ref{examples}.
In~\S~\ref{explicitRenormalization}
we construct $[ \, e^{\lambda V(a,b)} \, ]_r$
for the class of marginal operators,
and we prove in~\S~\ref{BRST}
and in appendix~\ref{proof-appendix}
that the assumptions stated
in~\S~\ref{assumptions}
are satisfied.
We discuss conformal properties of the operator
$[ \, O_L(a) \, e^{\lambda V(a,b)} \, ]_r$
in~\S~\ref{conformal}.

In section~\ref{discussion_deformed}
we discuss string field theory around the deformed background
and demonstrate that it can be elegantly formulated
in terms of a new set of algebraic structures
by defining a deformed star product,
deformed inner product, and deformed BRST operator.
Section~\ref{discussion} is for discussion,
and in appendix~\ref{discussion_FKP}
we explain the relation to the previous work
by Fuchs, Kroyter and Potting in~\cite{Fuchs:2007yy}
for the special case of marginal deformations
corresponding to the constant mode of the gauge field.

\section{Marginal deformations
with regular operator products}\label{regular}
\setcounter{equation}{0}

\subsection{Solutions using integrated vertex operators}\label{regular_PsiL}

When we calculate $n$-point scattering amplitudes for open bosonic
strings on the disk, we use three {\it unintegrated} vertex
operators and $n-3$ {\it integrated} vertex operators.
The unintegrated vertex operator takes the form $cV$,
where $c$ is the $c$ ghost
and $V$ is a matter primary field of dimension one.
The unintegrated vertex operator is
invariant under the BRST transformation:
\begin{equation}
Q_B \cdot cV (t)
\equiv \int_{C (t)} \Bigl[ \, \frac{dz}{2 \pi i} \, j_B (z)
- \frac{d \bar{z}}{2 \pi i} \, \tilde{\jmath}_B (\bar{z}) \,
\Bigr] \, cV (t) = 0 \,.
\label{QcV}
\end{equation}
The integrated vertex operator is an integral of $V$
on the boundary.
The BRST transformation of $V$ is a total derivative,
\begin{equation}
Q_B \cdot V (t) = \partial_t \, [ \, cV (t) \, ] \,,
\end{equation}
and thus the integrated vertex operator
is invariant under the BRST transformation
up to nonvanishing terms from the boundaries
of the integral region:
\begin{equation}\label{QV(a,b)}
Q_B \cdot V(a,b)
= Q_B \cdot \int_a^b dt \, V(t)
= \int_a^b dt \, \partial_t \, [ \, cV(t) \, ]
= cV(b) - cV(a) \,.
\end{equation}
The vertex operator $V$ generates a marginal deformation
of the boundary CFT.
When the deformation is exactly marginal, we
expect a corresponding solution $\Psi$
to the equation of motion
of open string field theory~\cite{Witten:1985cc}:
\begin{equation}
Q_B \Psi + \Psi \ast \Psi = 0 \,.
\end{equation}
In~\cite{Schnabl:2007az, Kiermaier:2007ba},
analytic solutions for marginal deformations in open
bosonic string field theory were constructed to all orders in the
deformation parameter $\lambda$ when operator products $V(t_1) \,
V(t_2) \ldots V(t_n)$ of the marginal operator are regular.
The solution in~\cite{Schnabl:2007az, Kiermaier:2007ba}
takes the form of an expansion in $\lambda$,
\begin{equation}
    \Psi=\sum_{n=1}^\infty \lambda^n \, \Psi^{(n)} \,,
\end{equation}
and the equation of motion for $\Psi^{(n)}$ is
\begin{equation}\label{EOMexpansion}
    Q_B\Psi^{(n)}=-\sum_{i=1}^{n-1}\Psi^{(n-i)}\ast\Psi^{(i)} \,.
\end{equation}
In the solution constructed
in~\cite{Schnabl:2007az, Kiermaier:2007ba},
$\Psi^{(n)}$ is made of $n$
unintegrated vertex operators and $n-1$ $b$-ghost insertions. In
this section, we construct $\Psi^{(n)}$ using
one unintegrated and $n-1$ integrated vertex
operators when operator products of the marginal operator
are regular.

We choose the first term $\Psi^{(1)}$ of the solution to be
\begin{equation}\label{psi1cft}
\langle \, \phi, \Psi^{(1)} \, \rangle = \langle \, f \circ \phi (0)
\, c V (1) \, \rangle_{{\cal W}_1} \,.
\end{equation}
This satisfies the linearized equation of motion.
The starting point of our construction is the observation that
$\Psi_L^{(2)}$ made of one unintegrated vertex operator
and one integrated vertex operator given by
\begin{equation}
\langle \, \phi \,, \Psi_L^{(2)} \, \rangle
= \langle \, f \circ \phi (0) \, cV (1) \, V(1,2) \,
\rangle_{{\cal W}_2}
= \int_1^2 dt \, \langle
\, f \circ \phi (0) \, cV (1) \, V(t) \, \rangle_{{\cal W}_2}
\label{regPsiL^(2)}
\end{equation}
solves the equation of motion
$Q_B \Psi_L^{(2)} = {}- \Psi^{(1)} \ast \Psi^{(1)}$.
This can be shown as follows:
\begin{equation}
\begin{split}
\langle \, \phi \,, Q_B \, \Psi_L^{(2)} \, \rangle & = {}- \int_1^2
dt \, \langle \, f \circ \phi (0) \, cV (1) \, \partial_t \, [ \, cV
(t) \, ] \,
\rangle_{{\cal W}_2} \\
& = {}- \langle \, f \circ \phi (0) \, cV (1) \, cV (2) \,
\rangle_{{\cal W}_2} \\
& = {}- \langle \, \phi \,, \Psi^{(1)} \ast \Psi^{(1)} \, \rangle
\,,
\end{split}
\end{equation}
where we have used the formulas (\ref{QcV}) and (\ref{QV(a,b)}),
and
\begin{equation}
\lim_{t_2 \to t_1} cV(t_1) \, cV(t_2) = 0 \,,
\end{equation}
which follows from the condition that the operator product
$V(t_1) \, V(t_2)$ is regular in the limit $t_2 \to t_1$.

Let us next construct a solution to $\ord{\lambda^3}$.
We look for $\Psi_L^{(3)}$ which satisfies
\begin{equation}
Q_B \, \Psi_L^{(3)}
= {}- \Psi^{(1)} \ast \Psi_L^{(2)}
- \Psi_L^{(2)} \ast \Psi^{(1)} .
\end{equation}
The right-hand side is given by
\begin{equation}\label{Psi1Psi2}
\begin{split}
{}- \langle \, \phi \,, \Psi^{(1)} \ast \Psi_L^{(2)}
+ \Psi_L^{(2)} \ast \Psi^{(1)} \, \rangle
& = {}- \langle \, f \circ \phi (0) \,
cV (1) \, cV (2) \, V(2,3) \, \rangle_{{\cal W}_3} \\
& \quad ~ {}- \langle \, f \circ \phi (0) \,
cV (1) \, V(1,2) \, cV (3) \, \rangle_{{\cal W}_3} \,.
\end{split}
\end{equation}
First consider the state $\Psi^{(3)}_{L1}$ defined by
\begin{equation}
\langle \, \phi \,, \Psi^{(3)}_{L1} \, \rangle
= \langle \, f \circ \phi (0) \,
cV (1) \, V (1,2) \, V(2,3) \, \rangle_{{\cal W}_3} \,.
\end{equation}
The BRST transformation of $\Psi^{(3)}_{L1}$ is
\begin{equation}
\begin{split}
\langle \, \phi \,, Q_B \, \Psi^{(3)}_{L1} \, \rangle
& = {}- \langle \, f \circ \phi (0) \,
cV (1) \, cV (2) \, V(2,3) \, \rangle_{{\cal W}_3} \\
& \quad ~ {}- \langle \, f \circ \phi (0) \,
cV (1) \, V (1,2) \, cV(3) \, \rangle_{{\cal W}_3} \\
& \quad ~{}+ \, \langle \, f \circ \phi (0) \,
cV (1) \, V (1,2) \, cV(2) \, \rangle_{{\cal W}_3} \,.
\end{split}
\label{Q_B-Psi^(3)_1}
\end{equation}
The first two terms precisely give
$- \Psi^{(1)} \ast \Psi_L^{(2)} - \Psi_L^{(2)} \ast \Psi^{(1)}$.
To cancel the last term, consider $\Psi^{(3)}_{L2}$
defined by
\begin{equation}
\langle \, \phi \,, \Psi^{(3)}_{L2} \, \rangle
= \frac{1}{2} \,
\langle \, f \circ \phi (0) \,
cV (1) \, \left(V(1,2)\right)^2 \, \rangle_{{\cal W}_3} \,.
\end{equation}
Using the formula
\begin{equation}\label{QVabn}
Q_B \cdot \left(V(a,b)\right)^n
= n \, [ \, \left(V(a,b)\right)^{n-1} cV(b) - cV(a) \left(V(a,b)\right)^{n-1} \, ] \,,
\end{equation}
which holds for marginal operators
with regular operator products,
the BRST transformation of $\Psi^{(3)}_{L2}$
can be calculated as follows:
\begin{equation}
\langle \, \phi \,, Q_B \, \Psi^{(3)}_{L2} \, \rangle
= {}- \langle \, f \circ \phi (0) \,
cV (1) \, V (1,2) \, cV(2) \, \rangle_{{\cal W}_3} \,.
\end{equation}
This cancels the last term on the right-hand side
of (\ref{Q_B-Psi^(3)_1}).
Therefore, $\Psi_L^{(3)}$ can be constructed
by adding $\Psi^{(3)}_{L2}$ to $\Psi^{(3)}_{L1}$:
\begin{equation}
\begin{split}
\langle \, \phi \,, \Psi_L^{(3)} \, \rangle
& = \langle \, \phi \,, \Psi^{(3)}_{L1}
+ \Psi^{(3)}_{L2} \, \rangle \\
& = \langle \, f \circ \phi (0) \, cV (1) \, V (1,2) \, V(2,3) \,
\rangle_{{\cal W}_3}
+ \frac{1}{2} \, \langle \, f \circ \phi (0) \,
cV (1) \, \left(V(1,2)\right)^2 \, \rangle_{{\cal W}_3} \,.
\end{split}
\end{equation}

To generalize this solution to higher orders,
it turns out to be crucial to rewrite $\Psi_L^{(3)}$
in a different form.
Using a path-ordered expression
for $\Psi^{(3)}_{L2} \,$,
$\Psi_L^{(3)}$ can also be written as
\begin{figure}[tb]
\centerline{\hbox{\epsfig{figure=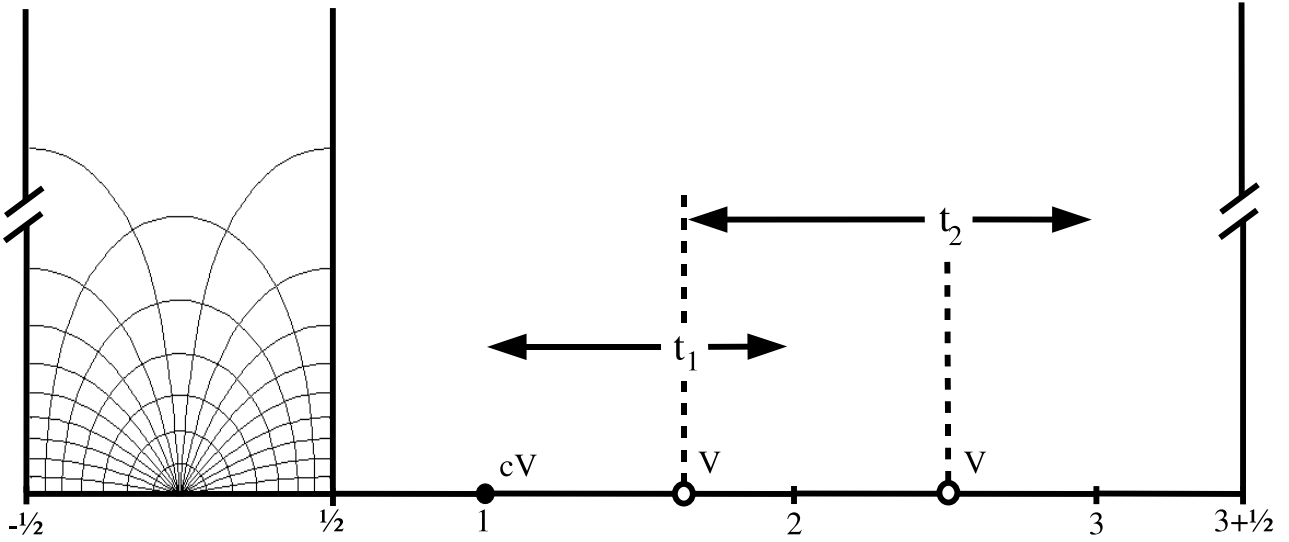, width=17.cm}}}
\caption{Illustration of $\Psi_L^{(3)}$.
The solid dot represents the $cV$ insertion,
and the circles represent the two $V$ insertions.
The left $V$ is integrated from $1$ to $2$\,,
and the right $V$ is integrated from
the position of the left $V$ to $3$\,.
}\label{figsolutionPsi3}
\end{figure}
\begin{equation}
\begin{split}
\langle \, \phi \,, \Psi_L^{(3)} \, \rangle
& = \int_1^2 dt_1 \, \int_2^3 dt_2 \, \langle \,
f \circ \phi (0) \, cV (1) \, V (t_1) \, V(t_2) \,
\rangle_{{\cal W}_3} \\
& \qquad {}+ \int_1^2 dt_1 \, \int_{t_1}^2 dt_2 \,
\langle \, f \circ \phi (0) \,
cV (1) \, V (t_1) \, V(t_2) \, \rangle_{{\cal W}_3} \\
& = \int_1^2 dt_1 \, \int_{t_1}^3 dt_2 \,
\langle \, f \circ \phi (0) \,
cV (1) \, V (t_1) \, V(t_2) \, \rangle_{{\cal W}_3} \,.
\end{split}
\end{equation}
See figure~\ref{figsolutionPsi3}.
It is instructive to see how $\Psi_L^{(3)}$ in this form
satisfies the equation of motion.
The BRST transformation of
$\Psi_L^{(3)}$ is given by
\begin{equation}\label{regQPsiL3}
\begin{split}
\langle \, \phi \,, Q_B\Psi_L^{(3)} \, \rangle = &-\int_1^2 dt_1 \,
\int_{t_1}^3 dt_2 \, \langle \, f \circ \phi (0) \, cV (1) \,
\partial_{t_1} \, [ \, cV(t_1) \, ] \, V(t_2) \, \rangle_{{\cal W}_3}\\
&  -\int_1^2 dt_1 \, \int_{t_1}^3 dt_2 \, \langle \, f \circ \phi
(0) \, cV (1) \, V(t_1) \,
\partial_{t_2} \, [ \, cV(t_2) \, ]  \, \rangle_{{\cal W}_3} \,.
\end{split}
\end{equation}
The integral region of $t_2$ depends on $t_1$.
The first line on the right-hand side of (\ref{regQPsiL3})
can be calculated as follows:
\begin{equation}\label{Q_B-Psi_L^(3)-1}
\begin{split}
&-\int_1^2 dt_1 \,
\int_{t_1}^3 dt_2 \, \langle \, f \circ \phi (0) \, cV (1) \,
\partial_{t_1} \, [ \, cV(t_1) \, ] \, V(t_2) \, \rangle_{{\cal W}_3}\\
& = -\int_1^2 dt_1 \, \partial_{t_1}
\biggl[ \, \int_{t_1}^3 dt_2 \, \langle \, f \circ \phi (0) \,
cV (1) \, cV(t_1) \, V(t_2) \, \biggr] \, \rangle_{{\cal W}_3}
- \int_1^2 dt_1 \, \langle \, f \circ \phi (0) \, cV (1) \,
cV^2(t_1) \, \rangle_{{\cal W}_3}\\
& = -\int_2^3 dt_2 \, \langle \, f \circ \phi (0) \, cV (1) \,
cV(2) \, V(t_2) \, \rangle_{{\cal W}_3}
- \int_1^2 dt_1 \, \langle \, f \circ \phi (0) \, cV (1) \,
cV^2(t_1) \, \rangle_{{\cal W}_3} \\
& = {}- \langle \, \phi \,, \Psi^{(1)} \ast \Psi_L^{(2)} \, \rangle
- \int_1^2 dt_1 \, \langle \, f \circ \phi (0) \, cV (1) \,
cV^2(t_1) \, \rangle_{{\cal W}_3} \,. \\
\end{split}
\end{equation}
The calculation of the second line
on the right-hand side of (\ref{regQPsiL3}) is straightforward:
\begin{equation}\label{Q_B-Psi_L^(3)-2}
\begin{split}
& -\int_1^2 dt_1 \, \int_{t_1}^3 dt_2 \, \langle \, f \circ \phi
(0) \, cV (1) \, V(t_1) \,
\partial_{t_2} \, [ \, cV(t_2) \, ]  \, \rangle_{{\cal W}_3} \\
& = -\int_1^2 dt_1 \, \langle \, f \circ \phi
(0) \, cV (1) \, V(t_1) \, cV(3) \, \rangle_{{\cal W}_3}
+\int_1^2 dt_1 \, \langle \, f \circ \phi
(0) \, cV (1) \, cV^2(t_1) \, \rangle_{{\cal W}_3} \\
& = {}- \langle \, \phi \,, \Psi_L^{(2)} \ast \Psi^{(1)} \, \rangle
+\int_1^2 dt_1 \, \langle \, f \circ \phi
(0) \, cV (1) \, cV^2(t_1) \, \rangle_{{\cal W}_3} \,.
\end{split}
\end{equation}
Note that the two terms with $cV^2$, which arise from collisions of
$cV$ and $V$, cancel each other.
We have thus reconfirmed that the equation of motion
at $\ord{\lambda^3}$ is satisfied.

This form of $\Psi_L^{(3)}$ can be generalized
to $\Psi_L^{(n)}$ for any $n$ as follows:
\begin{equation}
\begin{split}
    \langle \, \phi \,, \Psi_L^{(n)} \, \rangle &=
    \Bigl\langle \, f \circ \phi (0) \,
     \, c V(1)
    \int_1^{2} dt_1 \int_{t_1}^{3}
dt_2\int_{t_2}^4dt_3  \, \ldots \, \int_{t_{n-2}}^{n}
dt_{n-1}
\, V(t_1) \, V(t_2) \,V(t_3)\, \ldots \, V(t_{n-1})\Bigr\rangle_{{\cal W}_n} \\
    &=\Bigl\langle \, f \circ \phi (0) \,
     \, c V(1)
    \prod_{j=1}^{n-1} \, \int_{t_{j-1}}^{j+1}dt_j \, V(t_j)\,\Bigr\rangle_{{\cal W}_n}
    \qquad \text{ with} \quad t_0\equiv1 \,.
\end{split}
\end{equation}
See figure~\ref{solutionPsin}.
It is straightforward to show that $\Psi_L^{(n)}$
satisfies the equation of motion:
\begin{figure}[tb]
\centerline{\hbox{\epsfig{figure=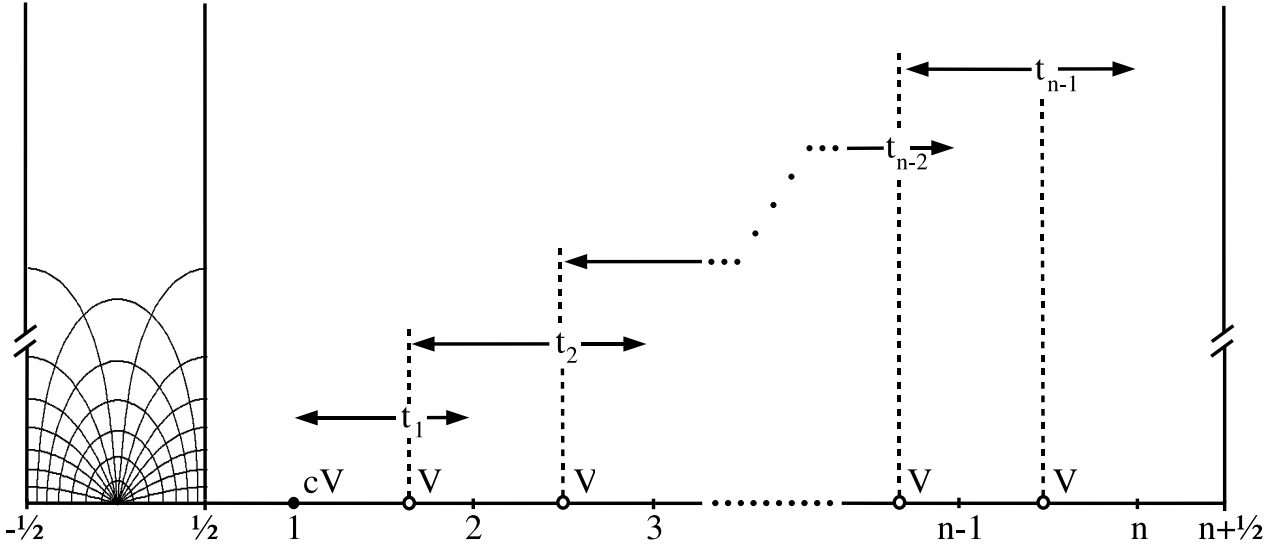, width=17.cm}}}
\caption{Illustration of $\Psi_L^{(n)}$.
The solid dot represents the $cV$ insertion,
and the circles represent the $V$ insertions.
The integration region of $t_j$
is from $t_{j-1}$ to $j+1$.
}\label{solutionPsin}
\end{figure}
\begin{equation}
\begin{split}
    \langle \, \phi & \,, Q_B\Psi_L^{(n)} \, \rangle\\
    =&-\sum_{i=1}^{n-1} \, \Bigl\langle \, f \circ \phi (0) \,
     c V(1)
    \prod_{j=1}^{i-1} \int_{t_{j-1}}^{j+1}dt_j V(t_j)
     \, \int_{t_{i-1}}^{i+1}dt_i \,
    \partial_{t_i} \, [ \, cV(t_i) \, ] \,
    \prod_{k=i+1}^{n-1} \int_{t_{k-1}}^{k+1}dt_k V(t_k) \,
    \Bigr\rangle_{{\cal W}_n} \\
    =&-\sum_{i=1}^{n-1} \, \Bigl\langle \, f \circ \phi (0) \,
     c V(1)
    \prod_{j=1}^{i-1} \int_{t_{j-1}}^{j+1}dt_j V(t_j)
     \,\, cV(i+1)  \,
     \int_{i+1}^{i+2}dt_{i+1}\ldots\int_{t_{n-2}}^{n}dt_{n-1}\,
    V(t_{i+1})\ldots V(t_k) \,
    \Bigr\rangle_{{\cal W}_n} \\
    &+\sum_{i=2}^{n-1} \, \Bigl\langle \, f \circ \phi (0) \,
     c V(1)
    \prod_{j=1}^{i-1} \int_{t_{j-1}}^{j+1}dt_j V(t_j)
     \,\, cV(t_{i-1})  \,
     \int_{t_{i-1}}^{i+2}dt_{i+1}\ldots\int_{t_{n-2}}^{n}dt_{n-1}\,
    V(t_{i+1})\ldots V(t_k) \,
    \Bigr\rangle_{{\cal W}_n} \\
    &+\sum_{i=1}^{n-2} \, \Bigl\langle \, f \circ \phi (0) \,
     c V(1)
    \prod_{j=1}^{i-1} \int_{t_{j-1}}^{j+1}dt_j V(t_j)
     \, \int_{t_{i-1}}^{i+1}dt_i \,  cV(t_i)  \,
     \partial_{t_i}\Biggl[ \, \prod_{k=i+1}^{n-1}
    \int_{t_{k-1}}^{k+1}dt_k V(t_k) \, \Biggr] \,
    \Bigr\rangle_{{\cal W}_n} \,.
\end{split}
\end{equation}
By carrying out the differentiation in the last line,
we find that the last line precisely
cancels the second line on the right-hand side.
The remaining first line on the right-hand side
is a sum of ${}- \Psi^{(i)} \ast \Psi^{(n-i)}$ over $i$.
We have thus shown
\begin{equation}\label{QregPsiLn}
    \langle \, \phi \,,\, Q_B\Psi_L^{(n)} \, \rangle
    =- \sum_{i=1}^{n-1} \, \langle \, \phi \,, \Psi^{(i)}
    \ast \Psi^{(n-i)} \, \rangle\,.
\end{equation}
It is convenient to introduce the following notation:
\begin{equation}\label{regVL}
\begin{split}
V_L^{(n)} (1,n+1) &\equiv \int_1^{2} dt_1
\int_{t_1}^{3}dt_2 \int_{t_2}^{4}dt_3
 \, \ldots \, \int_{t_{n-1}}^{n+1}
dt_{n}
\, V(t_1) \, V(t_2) \, \ldots \, V(t_{n}) \quad\text{ for }\quad
n \geq 1 \,, \\
V_L^{(0)}(1,1) &\equiv 1 \,.
\end{split}
\end{equation}
The superscript $(n)$ indicates the number of operators
and $(1,n+1)$ indicates the region
where operators are inserted, although this notation
is slightly redundant because the number of operators
and the length of the region are correlated
for $V_L^{(n)} (1,n+1)$.
The solution $\Psi_L^{(n)}$ can now be compactly written as
\begin{equation}\label{regPsiLn}
    \langle \, \phi \,, \Psi_L^{(n)} \, \rangle =
    \langle \, f \circ \phi (0) \,
     \, c V(1) \,\, V_L^{(n-1)} (1,n)\,\rangle_{{\cal W}_n} \,.
\end{equation}
The state $\Psi_L$ defined by
\begin{equation}
\Psi_L = \sum_{n=1}^\infty \lambda^n \, \Psi_L^{(n)}
\end{equation}
thus solves the equation of motion to all orders in $\lambda$.

\subsection{Solutions satisfying the reality condition}\label{regular_real}

The solution $\Psi_L$ constructed in the previous subsection
satisfies the equation of motion,
but it does not satisfy the reality condition
on the string field.
In this subsection, we construct a solution satisfying
the reality condition from $\Psi_L$.

\subsubsection{The reality condition}\label{subsubreality}

The string field $\Psi$ must have a definite parity
under the combination of the Hermitean conjugation (hc)
and the inverse BPZ conjugation ($\text{bpz}^{-1}$)
to guarantee that the string field theory action
is real~\cite{Gaberdiel:1997ia}.
We define the conjugate $A^\ddagger$ of a string field $A$ by
\begin{equation}
A^\ddagger \equiv \text{bpz}^{-1} \circ \text{hc} \, (A) \,.
\end{equation}
With this definition,
the following relations hold:
\begin{eqnarray}
(Q_B A)^\ddagger &=& {}- (-1)^A \, Q_B A^\ddagger \,,
\label{conjugation-with-Q_B} \\
(A \ast B)^\ddagger &=& B^\ddagger \ast A^\ddagger \,.
\end{eqnarray}
Here and in what follows a string field in the exponent of $(-1)$
denotes its Grassmann property:
it is $0$ mod $2$ for a Grassmann-even state
and $1$ mod $2$ for a Grassmann-odd state.
Since the string field $\Psi$ is Grassmann odd,
it must be {\it even}
under the conjugation $\Psi^\ddagger = \Psi$
in order that $Q_B \Psi$ and $\Psi \ast \Psi$ have the same parity.
We will say that a string field of ghost number one is
\emph{real} when it is even under the conjugation.

The class of states we use in constructing solutions
for marginal deformations are made of wedge states
with insertions of $cV$ and $V$. Let us consider the conjugate
of a state in this class.
The wedge state $W_\alpha$~\cite{Rastelli:2000iu}
is even under the conjugation
$W_\alpha^\ddagger = W_\alpha$ because it is constructed from
the $SL(2,R)$-invariant vacuum $\ket{0}$
satisfying $\ket{0}^\ddagger=\ket{0}$
by acting with BPZ-even Virasoro generators
$L_{-2}, \, L_{-4}, \, \ldots \,$.
The first term $\Psi^{(1)}$ in the solution
must be even $( \Psi^{(1)} )^\ddagger = \Psi^{(1)}$,
as we discussed above.
Therefore, the conjugate of $W_\alpha \ast \Psi^{(1)} \ast W_\beta$
is $W_\beta \ast \Psi^{(1)} \ast W_\alpha$.
This means that the operator $cV(t)$ on ${\cal W}_n$
is mapped to $cV(n+1-t)$ under the conjugation:
\begin{equation}
cV(t) ~ \longrightarrow ~ cV(n+1-t)
\quad \text{on} \quad {\cal W}_n \,.
\end{equation}
Its derivative $\partial_t \, [ \, cV(t) \, ]$ at $t=a$
is then mapped to ${}- \partial_t \, [ \, cV(t) \, ]$ at $t=n+1-a$.
Since $\partial_t \, [ \, cV(t) \, ]$ is the BRST transformation
of $V(t)$, this means that $Q_B \cdot V(a)$ is mapped
to ${}- Q_B \cdot V(n+1-a)$ on ${\cal W}_n$.
It then follows from (\ref{conjugation-with-Q_B}) that
$V(t)$ is mapped under the conjugation as follows:
\begin{equation}
V(t) ~ \longrightarrow ~ V(n+1-t)
\quad \text{on} \quad {\cal W}_n \,.
\end{equation}
It is straightforward to generalize the argument
to the case with multiple operator insertions.
The conjugate of the state made of the wedge state $W_n$
with $cV(t_1), V(t_2), V(t_3), \ldots, V(t_m)$
is therefore the state made of $W_n$ with
$V(n+1-t_m), V(n+1-t_{m-1}), \ldots , V(n+1-t_2), cV(n+1-t_1)$.

The state $\Psi_L^{(n)}$ with $n \geq 2$ does not satisfy
the reality condition. Indeed, the operator $V_L^{(n-1)} (1,n)$
defined in~(\ref{regVL}) is mapped as
\begin{equation}
\begin{split}
\int_1^{2} dt_1 \int_{t_1}^{3}
dt_2&\int_{t_2}^{4}dt_3  \, \ldots \, \int_{t_{n-2}}^{n}
dt_{n-1}
\, V(t_1) \, V(t_2) \,  \ldots \, V(t_{n-1})\\
 ~ \longrightarrow ~
& \int_1^2 dt_1 \int_{t_1}^3 dt_2 \int_{t_2}^4 dt_3 \,
\ldots \, \int_{t_{n-2}}^n dt_{n-1}
\, V(n+1-t_{n-1}) \, V(n+1-t_{n-2}) \, \ldots \, V(n+1-t_1) \\
& = \int_{n-1}^n dt'_1 \int_{n-2}^{t'_1}
dt'_2 \int_{n-3}^{t'_2} dt'_3 \ldots \int_1^{t'_{n-2}} dt'_{n-1} \,
V(t'_{n-1}) \, V(t'_{n-2}) \ldots V(t'_1)
\end{split}
\end{equation}
under the conjugation, where $t'_i = n+1-t_i$.
We denote the conjugate of $\Psi_L^{(n)}$ by $\Psi_R^{(n)}$.
It is given by
\begin{equation}
    \langle \, \phi \,, \Psi_R^{(n)} \, \rangle =
    \langle \, \phi \,, ( \Psi_L^{(n)} )^\ddagger \, \rangle =
    \langle \, f \circ \phi (0) \,
     \,
    \, V_R^{(n-1)} (1,n) \,\, c V(n)\,\rangle_{{\cal W}_n} \,,
\end{equation}
where we defined
\begin{equation}
\begin{split}
V_R^{(n)} (1,n+1) &\equiv \int_{n}^{n+1} dt_1
\int_{n-1}^{t_1} dt_2 \int_{n-2}^{t_2} dt_3
\ldots \int_1^{t_{n-1}} dt_{n} \,
V(t_{n}) \, V(t_{n-1}) \ldots V(t_1)
\quad\text{ for }\quad n\geq1\,,\\
V_R^{(0)} (1,1)&\equiv1\,.
\end{split}
\end{equation}
If $\Psi$ satisfies the equation of motion,
its conjugate $\Psi^\ddagger$ also satisfies the equation of motion
because
\begin{equation}
Q_B \Psi^\ddagger + \Psi^\ddagger \ast \Psi^\ddagger
= (Q_B \Psi + \Psi \ast\Psi)^\ddagger = 0 \,.
\end{equation}
Therefore, $\Psi_R$ defined by
\begin{equation}
\Psi_R = \sum_{n=1}^\infty \lambda^n \, \Psi_R^{(n)}
\end{equation}
satisfies the equation of motion.

\subsubsection{Gauge transformation}

We have found two solutions $\Psi_L$ and $\Psi_R$,
and we expect that they are related by a gauge transformation
generated by some gauge parameter $U$:
\begin{equation}\label{PsiRL}
    \Psi_R \, = \, U^{-1}\ast \Psi_L\ast U + U^{-1}\ast Q_B U
      \,.
\end{equation}
For a physical gauge transformation
which relates two string fields satisfying the reality condition,
the gauge parameter $U$ must satisfy the unitarity relation
$U^\ddagger = U^{-1}$.
As we will see later, the gauge parameter $U$
that relates $\Psi_L$ and $\Psi_R$ is even under the conjugation:
$U^\ddagger = U$.
The component fields of $\Psi_L$ and $\Psi_R$
which do not satisfy the reality condition
are thus related through the component fields of $U$
which also violate the reality condition on the gauge parameter.

Let us now construct $U$ which relates $\Psi_L$ and $\Psi_R$.
It is convenient to rewrite the equation~(\ref{PsiRL}) as follows:
\begin{equation}\label{QU}
    Q_B U\, = \, U \ast \Psi_R \, - \,\Psi_L\ast U
      \,.
\end{equation}
We can expand $U$ as
\begin{equation}
U = \sum_{n=0}^\infty \lambda^n \, U^{(n)}
\quad \text{with} \quad U^{(0)} = 1 \,,
\end{equation}
and we solve the equation perturbatively in $\lambda$.
We can choose
\begin{equation}
U^{(1)}=0
\end{equation}
because $\Psi_L^{(1)} = \Psi_R^{(1)}$ and therefore $Q_B U^{(1)}=0$.
The equation for $U^{(2)}$ is
\begin{equation}
\langle \, \phi, Q_B \, U^{(2)} \, \rangle
= \langle \, \phi, \Psi_R^{(2)} \, \rangle
- \langle \, \phi, \Psi_L^{(2)} \, \rangle
= \langle \, f \circ \phi (0) \,
[ \, V(1,2) \, cV (2) - cV (1) \, V(1,2) \, ] \,
\rangle_{{\cal W}_2} \,.
\end{equation}
This can be easily solved using the formula (\ref{QVabn}),
and a solution is
\begin{equation}
\langle \, \phi, U^{(2)} \, \rangle
= \frac{1}{2} \, \langle \, f \circ \phi (0) \,
\left(V(1,2)\right)^2 \, \rangle_{{\cal W}_2}\,.
\end{equation}
We can construct $U^{(n)}$ at higher orders recursively in this way.
However, we can infer $U^{(n)}$ from the structure of (\ref{QU}).
If we assume that $U$ can be written
without using $c$ ghosts,
the only $c$ ghost is inserted at $t=n$
in the $\ord{\lambda^n}$ term of
$\langle \, \phi, U \ast \Psi_R \, \rangle$
when represented on ${\cal W}_n$
and at $t=1$ on ${\cal W}_n$
in the $\ord{\lambda^n}$ term of
$\langle \, \phi, \Psi_L \ast U \, \rangle$.
This motivates us to make the following ansatz:
\begin{equation}
\langle \, \phi, U^{(n)} \, \rangle
\propto \langle \, f \circ \phi (0) \, V^{(n)} (1,n) \,
\rangle_{{\cal W}_n} \,,
\end{equation}
where
\begin{equation}
V^{(n)} (a,b) \equiv \frac{1}{n!}\left(V(a,b)\right)^n
\quad\text{ for }\quad
n \geq 1 \,, \qquad V^{(0)}(a,b)\equiv1 \,.
\end{equation}
We in fact show that the gauge transformation $U$ in~(\ref{PsiRL})
is given by
\begin{figure}[tb]
\Large
\begin{center}
$1+\frac{\lambda^2}{2!}$
\raisebox{-1.7cm}{\epsfig{figure=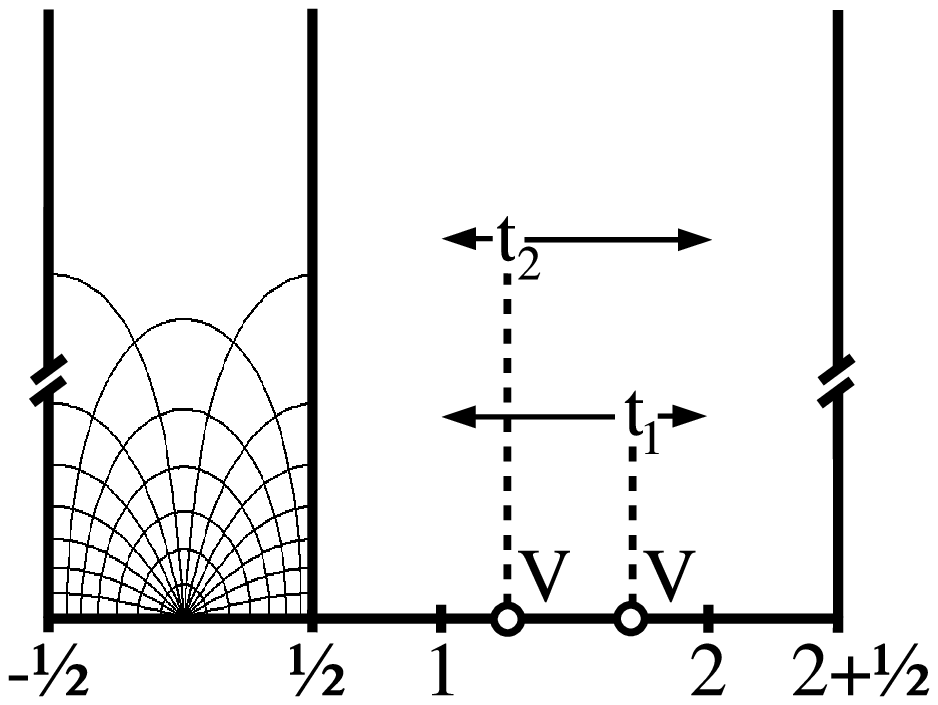, height=4.15cm}}
$\,+\,\,\frac{\lambda^3}{3!}$
\raisebox{-1.7cm}{\epsfig{figure=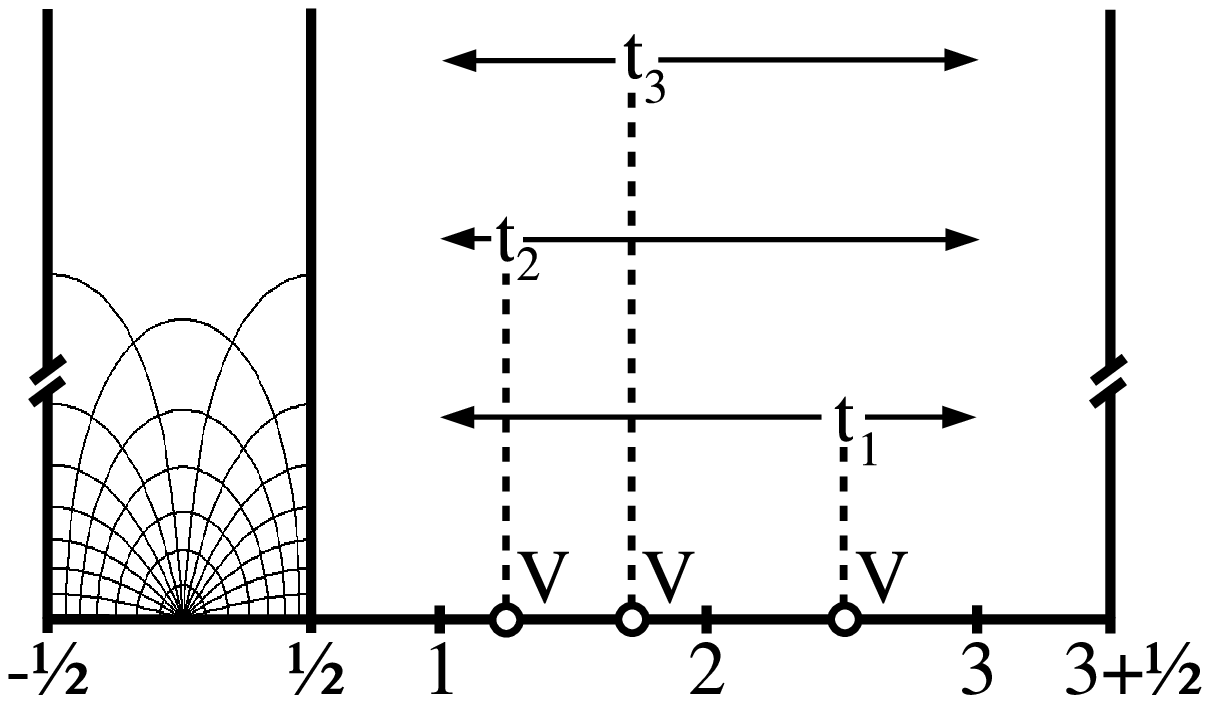, height=4.15cm}}
$+\,\,\ldots$
\end{center}
\normalsize
\caption{Illustration of the expansion
$U=1 + \lambda^2 \, U^{(2)} + \lambda^3 \, U^{(3)}
+ \ord{\lambda^4}$.}
\label{figU}
\end{figure}
\begin{equation}\label{regU}
    \langle \, \phi \,, U^{(n)} \, \rangle =
    \langle \, f \circ \phi (0) \,
     \, V^{(n)}(1,n)\,\rangle_{{\cal W}_n}\,.
\end{equation}
See figure~\ref{figU}.
The BRST transformation of $U^{(n)}$ given in~(\ref{regU}) is
\begin{equation}\label{QUlhs}
    \langle \, \phi, Q_B U^{(n)} \, \rangle
    =\bigl\langle \, f \circ \phi (0) \,
     \, \, \bigl(V^{(n-1)}(1,n)\,cV(n) \, - \, cV(1)\,V^{(n-1)}(1,n)\bigr) \,\bigr\rangle_{{\cal
     W}_n}\,,
\end{equation}
where we used~(\ref{QVabn}).
For the special case of $n=1$,
the terms on the right-hand side cancel,
which is consistent because $U^{(1)}=0$.
The $\ord{\lambda^n}$ term of
$U \ast \Psi_R \, - \,\Psi_L\ast U$
in~(\ref{QU}) is given by
\begin{equation}\label{QUrhs}
\begin{split}
    &\sum_{m=1}^{n}\langle \, f \circ \phi (0) \,
     \, \, V^{(n-m)}(1,n-m)\,V_R^{(m-1)}(n-m+1,n)\, cV(n) \,\rangle_{{\cal W}_n}\\
     &- \sum_{m=1}^{n}
     \langle \, f \circ \phi (0) \,\,\, cV(1) \,
     V_L^{(m-1)}(1,m)\,V^{(n-m)}(m+1,n)\,\rangle_{{\cal W}_n}\,.
\end{split}
\end{equation}
The proof of~(\ref{QU}) for $U$ given in~(\ref{regU})
thus reduces to showing that
\begin{equation}\label{V1toVLV0}
    \langle \, f \circ \phi (0) \,
     \, \, cV(1) \, V^{(n-1)}(1,n)\,\rangle_{{\cal W}_n}\,
     =\sum_{m=1}^{n}\langle \, f \circ \phi (0) \,
     \,\, cV(1) \,
     V_L^{(m-1)}(1,m)\,V^{(n-m)}(m+1,n)\,\rangle_{{\cal W}_n}\,
\end{equation}
and
\begin{equation}
    \langle \, f \circ \phi (0) \,
     \, \, V^{(n-1)}(1,n) \, cV(n) \,\rangle_{{\cal W}_n}\,
    =\sum_{m=1}^{n}\langle \, f \circ \phi (0) \,
     \, \, V^{(n-m)}(1,n-m)\,V_R^{(m-1)}(n-m+1,n)\, cV(n) \,
     \rangle_{{\cal W}_n}\,.
\end{equation}
Since the second equation
follows from the first one by the conjugation,
it is sufficient to show~(\ref{V1toVLV0}).
The operator $V^{(n-1)}(1,n)$ on the left-hand side can be written
in a path-ordered form as follows:
\begin{equation}
    V^{(n-1)}(1,n)=\int_1^n dt_1\int_{t_1}^n dt_2\ldots \int_{t_{n-2}}^n
    dt_{n-1} \, V(t_1)\ldots V(t_{n-1}) \,.
\end{equation}
We now decompose the integration region $1\leq t_1\leq t_2\leq
\ldots\leq t_{n-1}\leq n$ in the following way:
\begin{equation}\label{integrationdecomposition}
\begin{split}
& t_1 \ge 2 \,, \\
& t_1 \le 2 \,,\, t_2 \ge 3 \,, \\
& t_1 \le 2 \,,\, t_2 \le 3 \,,\, t_3 \ge 4 \,, \\
& \qquad \qquad \vdots \\
& t_1 \le 2 \,,\, t_2 \le 3 \,,\, \ldots \,,\, t_{m-1} \le m \,,\,
t_{m} \ge m+1 \,,\\
& \qquad \qquad \vdots  \\
& t_1 \le 2 \,,\, t_2 \le 3 \,,\, t_3 \le 4 \,,\, \ldots\ldots\ldots
\,,\, t_{n-2} \le n-1 \,,\, t_{n-1} \ge n \,,
 \\
& t_1 \le 2 \,,\, t_2 \le 3 \,,\, t_3 \le 4 \,,\, \ldots\ldots\ldots
\,,\, t_{n-2} \le n-1 \,,\, t_{n-1} \le n  \,.
\end{split}
\end{equation}
This decomposition of the integration region
precisely matches the right-hand side
of~(\ref{V1toVLV0}).
For example, the fourth line of~(\ref{integrationdecomposition})
corresponds to the integration region for the product
of the operators $V_L^{(m-1)}(1,m) \, V^{(n-m)}(m+1,n)$.
Furthermore, the fifth line vanishes because of
the vanishing integration range $n\leq t_{n-1}\leq n$.
This is consistent with the right-hand side of~(\ref{V1toVLV0})
because $V^{(1)}(n,n)=0$.
The last line
is nonvanishing and corresponds to
$V_L^{(n-1)}(1,n) \, V^{(0)}(n+1,n) = V_L^{(n-1)}(1,n)$,
where we used $V^{(0)}(a,b) \equiv 1$.
We conclude that
\begin{equation}\label{decomposeVn}
    V^{(n-1)}(1,n)\,
     =\,\sum_{m=1}^{n}\, V_L^{(m-1)}(1,m)\,V^{(n-m)}(m+1,n) \,,
\end{equation}
and we have thus shown~(\ref{V1toVLV0}). This completes the proof
that $U$ is the gauge transformation that relates $\Psi_L$ and
$\Psi_R$.

\subsubsection{Construction of a real solution}

The state $U$ takes the form
\begin{equation}
U = 1 + \sum_{n=2}^\infty \lambda^n \, U^{(n)} \,,
\end{equation}
and $U^{(n)}$ is even under the conjugation:
$(U^{(n)})^\ddagger = U^{(n)}$.
If a state $X$ is even under the conjugation,
then $\ln (1+X)$ defined by
\begin{equation}
\ln (1+X) \equiv \sum_{n=1}^\infty \frac{(-1)^{n+1}}{n} \,
\underbrace{\, X \ast X \ast \ldots \ast X \,}_{n}
\end{equation}
is also even.
If a state $Y$ is even, then $\exp \, (a \, Y)$ with real $a$ defined by
\begin{equation}
\exp \, (a \, Y) \equiv 1+\sum_{n=1}^\infty \frac{a^n}{n!} \,
\underbrace{\, Y \ast Y \ast \ldots \ast Y \,}_{n}
\end{equation}
is also even.
Therefore, $(1+X)^{-1}$, $\sqrt{1+X}$ and $1/\sqrt{1+X}$ defined by
\begin{equation}
\begin{split}
(1+X)^{-1} & \equiv
\exp \left[ \, - \ln (1+X) \, \right]
= 1 + \sum_{n=1}^\infty \, (-1)^n \,
\underbrace{\, X \ast X \ast \ldots \ast X \,}_{n} \,, \\
\sqrt{1+X} & \equiv
\exp \left[ \, \frac{1}{2} \, \ln (1+X) \, \right] \,, \qquad
\frac{1}{\sqrt{1+X}} \equiv
\exp \left[ \, - \frac{1}{2} \, \ln (1+X) \, \right]
\end{split}
\end{equation}
are all even if $X^\ddagger = X$.
We define $U^{-1}$, $\sqrt{U}$, and $1/\sqrt{U}$ in this way,
which are well defined to all orders in $\lambda$
and are even under the conjugation.

We can now construct a real solution $\Psi$ from $\Psi_L$
as follows:
\begin{equation}\label{Psireal}
\begin{split}
\Psi & \equiv \frac{1}{\sqrt{U}} \ast \Psi_L \ast \sqrt{U} +
\frac{1}{\sqrt{U}} \ast Q_B \, \sqrt{U}\\
&= \sqrt{U} \ast \Psi_R \ast \frac{1}{\sqrt{U}}
+ \sqrt{U} \ast Q_B \, \frac{1}{\sqrt{U}} \\
& = \frac{1}{2} \, \biggl[ \,
\frac{1}{\sqrt{U}} \ast \Psi_L \ast \sqrt{U}
+ \sqrt{U} \ast \Psi_R \ast \frac{1}{\sqrt{U}} +
\frac{1}{\sqrt{U}} \ast Q_B \, \sqrt{U} - Q_B \, \sqrt{U}
\ast \frac{1}{\sqrt{U}} \, \biggr] \,.
\end{split}
\end{equation}
The second expression is obtained from the first one
using $Q_B U = U \ast \Psi_R - \Psi_L \ast U$,
and $\Psi$ manifestly satisfies the reality condition
in the third expression because of the relations
$\Psi_L^\ddagger=\Psi_R$,
$(\sqrt{U} \, )^\ddagger = \sqrt{U}$,
$(1/\sqrt{U} \, )^\ddagger = 1/\sqrt{U}$, and
$(Q_B \, \sqrt{U} \, )^\ddagger
= {}- Q_B \, \sqrt{U}$.
The state $\Psi$ also satisfies the equation of motion
because it is obtained from the solution $\Psi_L$
by the gauge transformation generated by $\sqrt{U}$.

We have successfully constructed
real analytic solutions for marginal deformations
with regular operator products.
To summarize, our solution takes the form
\begin{equation}
\Psi  = \frac{1}{\sqrt{U}} \ast \Psi_L \ast \sqrt{U} +
\frac{1}{\sqrt{U}} \ast Q_B \, \sqrt{U} \,,
\end{equation}
where $\Psi_L$ and $U$ are defined by
\begin{equation}
\begin{split}
\Psi_L & = \sum_{n=1}^\infty \lambda^n \, \Psi^{(n)}_L \,, \\
  \langle \, \phi \,, \Psi^{(n)}_L\, \rangle
  & = \,\langle \, f \circ \phi (0) \,
  \, c V(1) \,\, V_L^{(n-1)} (1,n)\,\rangle_{{\cal W}_n}\, \\
 &=  \int_1^{2} dt_1 \int_{t_1}^{3}dt_2
 \, \ldots \, \int_{t_{n-2}}^{n} dt_{n-1}
 \,\langle \, f \circ \phi (0) \,\, c V(1)\,
 \, V(t_1) \, V(t_2) \, \ldots \, V(t_{n-1})
 \,\rangle_{{\cal W}_n}\,,\\
U & = 1 + \sum_{n=2}^\infty \lambda^n \, U^{(n)} \,, \\
\langle \, \phi \,, U^{(n)} \, \rangle & =
\langle \, f \circ \phi (0) \,
V^{(n)} (1,n)\,\rangle_{{\cal W}_n}\, \\
& = \frac{1}{n!}
\int_1^n dt_1 \int_1^n dt_2 \ldots \int_1^n dt_n \,
\langle \, f \circ \phi (0) \,
V(t_1) \, V(t_2) \, \ldots V(t_n) \,
\rangle_{{\cal W}_n} \,.
\end{split}
\end{equation}

\subsection{Generalization of wedge states}\label{regular_generalizewedge}

In the previous subsection, we found the identity (\ref{decomposeVn}).
It is simply a consequence of the decomposition
of the integral region (\ref{integrationdecomposition}).
The identity (\ref{decomposeVn}) can be generalized in the following way.
We define $V_{L,\alpha}^{(n)} (1,n+\alpha)$ for $\alpha \ge 0$ by
\begin{equation}\label{regVLalpha}
\begin{split}
V_{L,\alpha}^{(n)} (1,n+\alpha) &\equiv \int_1^{1+\alpha} dt_1
\int_{t_1}^{2+\alpha} dt_2 \int_{t_2}^{3+\alpha} dt_3 \, \ldots \,
\int_{t_{n-1}}^{n+\alpha} dt_{n} \, V(t_1) \, V(t_2) \, \ldots \,
V(t_{n}) \quad\text{ for }\quad n\geq1\,,\\
V_{L,\alpha}^{(0)} (1,\alpha)&\equiv1\,.
\end{split}
\end{equation}
This reduces to $V_L^{(n)} (1,n+1)$ defined in (\ref{regVL})
when $\alpha=1$.
We then find that
\begin{equation}\label{decomposeVnalphabeta}
  V^{(n)}(1,n+\alpha+\beta)\,=\,\sum_{m=0}^n
  V^{(m)}_{L,\alpha}(1,m+\alpha) \,
  V^{(n-m)}(m+\alpha+1,n+\alpha+\beta)
\end{equation}
for any non-negative real numbers $\alpha$ and $\beta$.
This identity reduces to (\ref{decomposeVn}) when $\alpha=1$, $\beta=0$.
This generalized identity
can be shown, as before, by decomposing
the path-ordered integration region
$1\leq t_1\leq t_2\leq\ldots\leq t_n\leq n+\alpha+\beta$
of $V^{(n)}(1,n+\alpha+\beta)$
in the following way:
\begin{equation}
\begin{split}
& t_1 \ge 1+\alpha \,, \\
& t_1 \le 1+\alpha \,,\, t_2 \ge 2+\alpha \,, \\
& t_1 \le 1+\alpha \,,\, t_2 \le 2+\alpha \,,\, t_3 \ge 3+\alpha \,, \\
& \qquad \qquad \vdots \\
& t_1 \le 1+\alpha \,,\, t_2 \le 2+\alpha \,,\, \ldots \,,\, t_{m}
\le m+\alpha \,,\,
t_{m+1} \ge m+1+\alpha \,,\\
& \qquad \qquad \vdots  \\
& t_1 \le 1+\alpha \,,\, t_2 \le 2+\alpha \,,\, t_3 \le 3+\alpha
\,,\, \ldots\ldots\ldots \,,\, t_{n-1} \le n-1+\alpha \,,\, t_{n}
\ge n+\alpha \,,
 \\
& t_1 \le 1+\alpha \,,\, t_2 \le 2+\alpha \,,\, t_3 \le 3+\alpha
\,,\, \ldots\ldots\ldots \,,\, t_{n-1} \le n-1+\alpha \,,\, t_{n}
\le n+\alpha \,.
\end{split}
\end{equation}
This identity can be promoted to a relation of string fields.
We define $U_\alpha$ and $U_{L,\alpha}$ with $\alpha \ge 0$ by
\begin{equation}\label{U_alpha}
\begin{split}
    U_\alpha \equiv
    \sum_{n=0}^\infty \, \lambda^n \, U_\alpha^{(n)} \,, \qquad
    U_{L,\alpha} \equiv
    \sum_{n=0}^\infty \, \lambda^n \, U_{L,\alpha}^{(n)} \,,
\end{split}
\end{equation}
where
\begin{equation}
\begin{split}
    \langle \, \phi \,, U_\alpha^{(n)} \, \rangle & =
    \langle \, f \circ \phi (0) \, \, \,
    V^{(n)} (1,n+\alpha) \, \rangle_{{\cal W}_{n+\alpha}}
    \quad \text{for} \quad n + \alpha > 0 \,, \quad U_0^{(0)} = 1 \,, \\
    \langle \, \phi \,, U_{L,\alpha}^{(n)} \, \rangle & =
    \langle \, f \circ \phi (0) \, \, \,
    V_{L,\alpha}^{(n)} (1,n+\alpha) \,
    \rangle_{{\cal W}_{n+\alpha}}
    \quad \text{for} \quad n+\alpha > 0 \,, \quad U_{L,0}^{(0)} = 1 \,.
\end{split}
\end{equation}
The gauge parameter $U$ in the previous subsection is thus
\begin{equation}
U = U_0\,,
\end{equation}
and the solution $\Psi_L$ in (\ref{regPsiLn}) is $U_{L,1}$
with an extra insertion of $\lambda \, cV(1)$.
It then follows from (\ref{decomposeVnalphabeta}) that
\begin{equation}\label{Ualphabeta}
    U_{\alpha+\beta}=U_{L,\alpha}\ast U_\beta\,.
\end{equation}
When $\beta=0$, we have
\begin{equation}
U_\alpha = U_{L,\alpha} \ast U \,,
\end{equation}
where we have used $U_0 = U$.
As we discussed in the previous subsection,
the inverse of $U$ is well defined to all orders in $\lambda$.
We thus find that
\begin{equation}
U_{L,\alpha} = U_\alpha \ast U^{-1} \,.
\end{equation}
It follows from this and (\ref{Ualphabeta}) that
\begin{equation}\label{genWedgeAlgebra}
U_{\alpha+\beta} = U_\alpha \ast U^{-1} \ast U_\beta \,.
\end{equation}
The state $U_\alpha$ is $W_\alpha + \ord{\lambda}$
for $\alpha > 0$,
where $W_\alpha$ is the well-known wedge state
defined by
\begin{equation}
\langle \, \phi \,, W_\alpha \, \rangle
= \langle \, f \circ \phi (0) \,
 \rangle_{{\cal W}_\alpha} \,.
\end{equation}
The relation (\ref{genWedgeAlgebra})
for positive $\alpha$ and $\beta$
thus reduces to the famous relation
$W_{\alpha+\beta} = W_\alpha \ast W_\beta$
when $\lambda=0$,
and the state $U_\alpha$ can be thought of as
a generalization of the wedge state $W_\alpha$.
When $\alpha$ is a positive integer,
$U_\alpha$ can be written in terms of $U_1$ and $U^{-1}$:
\begin{equation}
\begin{split}
& U_2 = U_1 \ast U^{-1} \ast U_1 \,, \\
& U_3 = U_1 \ast U^{-1} \ast
U_1 \ast U^{-1} \ast U_1 \,, \\
& U_4 = U_1 \ast U^{-1} \ast U_1 \ast U^{-1} \ast
U_1 \ast U^{-1} \ast U_1 \,, \\
& \qquad \qquad \vdots \\
\end{split}
\end{equation}
This structure indicates a modification of the star product
for finite $\lambda$ defined by
\begin{equation}\label{modified-star}
A \star B \equiv A \ast U^{-1} \ast B \,,
\end{equation}
and the relation (\ref{genWedgeAlgebra}) can be written as
\begin{equation}
U_{\alpha+\beta} = U_\alpha \star U_\beta \,.
\end{equation}
On a technical level,
the relation (\ref{genWedgeAlgebra}) will play an important role
in the next section for the construction of solutions
associated with general marginal deformations.
On a more conceptual level,
we will see in section~\ref{discussion_deformed}
that the modified star product~(\ref{modified-star})
naturally appears in the string field theory action
expanded around a deformed background.

\section{Marginal deformations
with singular operator products}
\setcounter{equation}{0}
\label{singular_general}

\subsection{Another form of the solution
with regular operator products}
\label{another}

In the process of constructing a real solution from $\Psi_L$
in the previous section, we proved that
\begin{equation}\label{QUagain}
Q_B \, U = U \ast \Psi_R - \Psi_L \ast U\,.
\end{equation}
As we have seen in~(\ref{QUlhs}),
the BRST transformation of $U$
can be decomposed into two pieces:
\begin{equation}\label{QUARAL}
Q_B \, U = A_R - A_L \,,
\end{equation}
where $A_L$ and $A_R$ are given by
\begin{figure}[tb]
\Large
\begin{center}
$\lambda$
\hskip-.1cm\raisebox{-1.45cm}{\epsfig{figure=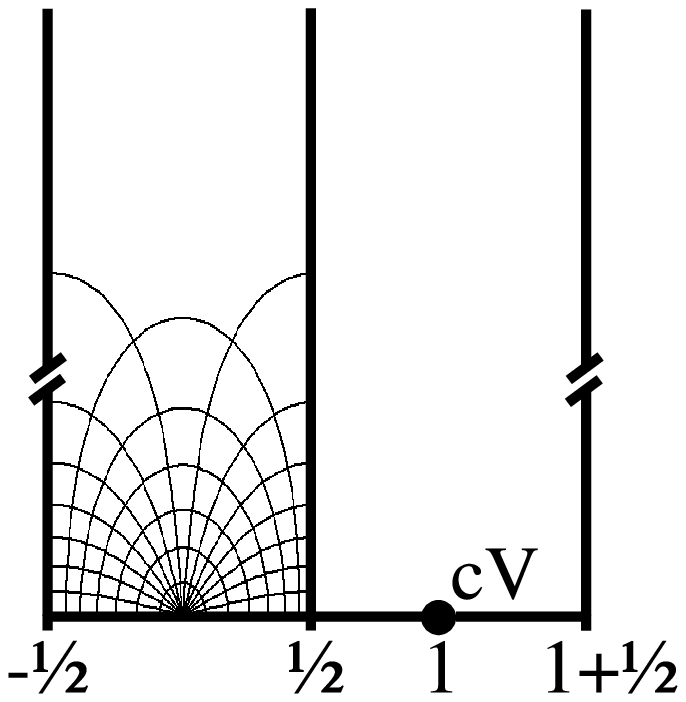, height=3.4cm}} \hskip-.2cm
$+\,\,\lambda^2$
\hskip-.1cm \raisebox{-1.45cm}{\epsfig{figure=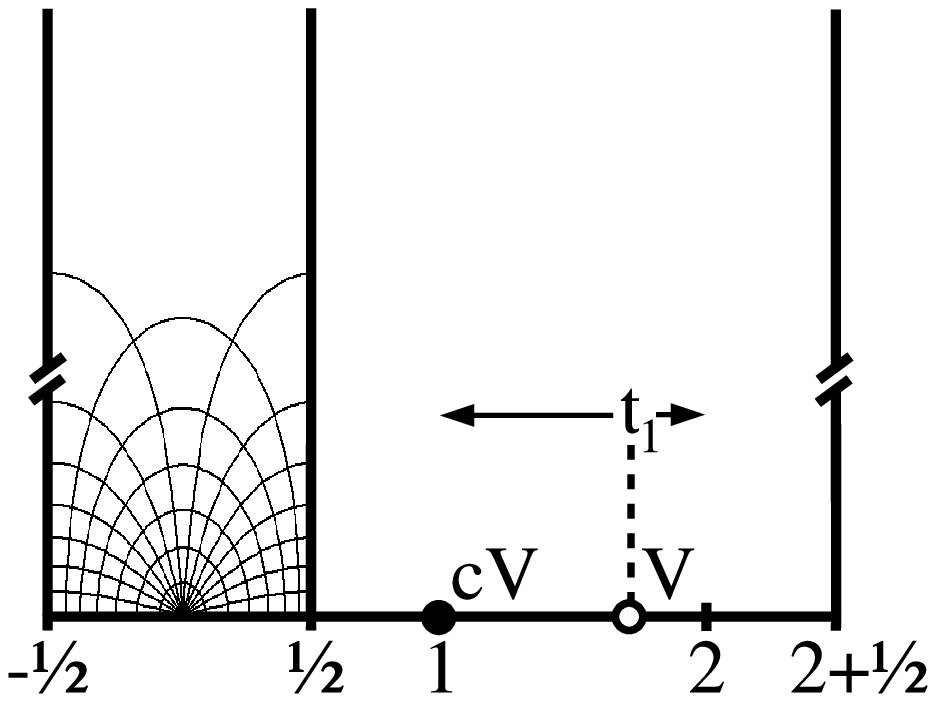, height=3.4cm}} \hskip-.2cm
$+\,\,\frac{\lambda^3}{2}$
\hskip-.1cm \raisebox{-1.45cm}{\epsfig{figure=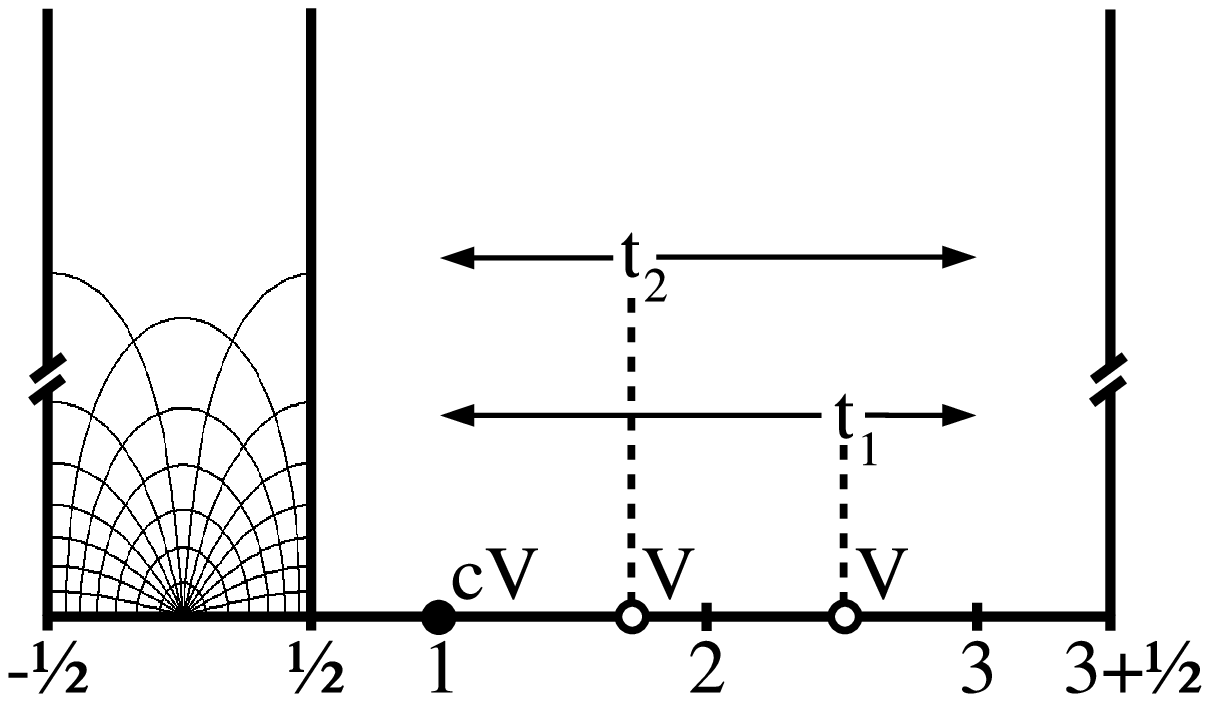, height=3.4cm}} \hskip-.2cm
$+\,\ldots$
\end{center}
\normalsize
\caption{Illustration of the expansion
$A_L= \lambda \, A_L^{(1)} + \lambda^2 \, A_L^{(2)}
+ \lambda^3 \, A_L^{(3)} + \ord{\lambda^4}$.}\label{figALreg}
\end{figure}
\begin{equation}\label{ALARregular}
\begin{split}
    \langle \, \phi \,, A_L \, \rangle &=
    \sum_{n=1}^\infty \lambda^n \,
    \langle \, f \circ \phi (0) \, cV (1) \, V^{(n-1)} (1,n) \,
    \rangle_{{\cal W}_n} \,, \\
    \langle \, \phi \,, A_R \, \rangle &=
    \sum_{n=1}^\infty \lambda^n \,
    \langle \, f \circ \phi (0) \, V^{(n-1)} (1,n) \, cV (n) \,
    \rangle_{{\cal W}_n} \,.
\end{split}
\end{equation}
See figure~\ref{figALreg}.
At $\ord{\lambda^n}$ with $n \ge 2$,
$A_L$ and $A_R$ account for
the term with $cV(1)$ and the term with $cV(n)$
in $Q_BU^{(n)}$, respectively.
At $\ord{\lambda}$, $Q_B \, U$ vanishes
because $U^{(1)} = 0$, but we have chosen
$A_L$ and $A_R$ at $\ord{\lambda}$
to be $\lambda \, \Psi^{(1)}$ for later convenience.

In the proof of~(\ref{QUagain}),
we have actually shown that
\begin{equation}\label{ALPsiLU}
A_L = \Psi_L \ast U \,, \qquad
A_R = U \ast \Psi_R \,.
\end{equation}
As we discussed in the previous section,
the inverse of $U$ is well defined to all orders in $\lambda$.
We thus obtain new expressions for $\Psi_L$ and $\Psi_R$:
\begin{equation}\label{defPsiLPsiR}
\Psi_L = A_L \ast U^{-1} \,, \qquad
\Psi_R = U^{-1} \ast A_R \,.
\end{equation}
We have shown that $\Psi_L$
with $\Psi_L^{(n)}$ in the form of~(\ref{regPsiLn})
satisfies the equation of motion. Let us now see how $\Psi_L$
in the new form satisfies the equation of motion.
The BRST transformation of $\Psi_L$ can be calculated as follows:
\begin{equation}\label{rewriteQPsiL}
\begin{split}
    Q_B\Psi_L \, &= \, Q_B\,(A_L \ast U^{-1})\\
    &=(Q_B A_L)\ast U^{-1}+A_L\ast U^{-1}\ast(Q_B U)\ast U^{-1}\\
    &=(Q_B A_L)\ast U^{-1}+A_L\ast U^{-1}\ast(A_R-A_L)\ast U^{-1}\\
    &=(Q_B A_L +A_L\ast U^{-1}\ast A_R)\ast U^{-1}-A_L\ast U^{-1}\ast A_L\ast U^{-1}\\
    &=(Q_B A_L +A_L\ast U^{-1}\ast A_R)\ast U^{-1}-\Psi_L \ast \Psi_L \,.
\end{split}
\end{equation}
Therefore, the equation of motion is satisfied if
\begin{equation}\label{QAL}
    {}- Q_B A_L = A_L\ast U^{-1}\ast A_R \,.
\end{equation}
The left-hand side of the equation can be calculated as follows:
\begin{equation}\label{Q_B-A_L}
    {}- \langle \, \phi \,, Q_B A_L \, \rangle =
    \sum_{n=2}^\infty \lambda^n \,
    \langle \, f \circ \phi (0) \,
    cV (1) \, V^{(n-2)} (1,n) \, cV (n) \, \rangle_{{\cal W}_n}\,.
\end{equation}
Let us next consider the structure
of the state $A_L \ast U^{-1} \ast A_R$
on the right-hand side of~(\ref{QAL}).
The $\ord{\lambda^n}$ terms of $A_L$ and $A_R$
are made of the wedge state $W_n$ with operator insertions.
The inverse $U^{-1}$ can be written as a linear combination
of string products made of $\lambda^n \, U^{(n)}$,
and their $\ord{\lambda^n}$ terms
are again made of the wedge state $W_n$ with operator insertions.
It thus follows that all of the $\ord{\lambda^n}$ terms
of $A_L \ast U^{-1} \ast A_R$ are made of $W_n$
with operator insertions.
This is consistent with the structure of~(\ref{Q_B-A_L}).
Furthermore, the insertions of $\lambda \, cV$
on the surface ${\cal W}_n$ are always
$\lambda \, cV (1)$ and $\lambda \, cV (n)$,
which is again
consistent with the structure of~(\ref{Q_B-A_L}).
Finally, let us consider the structure
of integrated vertex operators.
The state ${}-Q_B A_L$ takes the form of the state $U_2$
defined in~(\ref{U_alpha})
with insertions of $\lambda \, cV$.
Similarly, $A_L$ and $A_R$ take the form of $U_1$
with an insertion of $\lambda \, cV$.
The equation (\ref{QAL}) thus
follows from~(\ref{genWedgeAlgebra}) with $\alpha=\beta=1$:
\begin{equation}
U_2 = U_1 \ast U^{-1} \ast U_1 \,.
\end{equation}
We conclude that $\Psi_L$ of the form given in~(\ref{defPsiLPsiR})
satisfies the equation of motion.

\subsection{Generalization to the case
with singular operator products}\label{singular_general_PsiL}

The form $\Psi_L=A_L\ast U^{-1}$ for the solution can be generalized
to the case where operator products of the marginal operator
are singular.
As we discussed in the introduction,
let us denote the properly renormalized operator
implementing the change of the boundary condition
between the points $a$ and $b$
by $[ \, e^{\lambda \, V(a,b)} \, ]_r$,
which is given in the form of an expansion in $\lambda$:
\begin{equation}
[ \, e^{\lambda \, V(a,b)} \, ]_r
= \sum_{n=0}^\infty \, \frac{\lambda^n}{n!} \,
[ \, \left(V(a,b)\right)^n \, ]_r \,,
= \sum_{n=0}^\infty \lambda^n \,
[ \, V^{(n)}(a,b) \, ]_r \,.
\end{equation}
We define $U$ in the general case by
\begin{equation}
U \equiv \sum_{n=0}^\infty \, \lambda^n \, U^{(n)} \,,
\end{equation}
where
\begin{equation}
\langle \, \phi \,, U^{(n)} \, \rangle
= \langle \, f \circ \phi (0) \, \,
[ \, V^{(n)} (1,n) \, ]_r \, \rangle_{{\cal W}_n} \,.
\end{equation}

As we discussed in the introduction, we assume that
the BRST transformation of $[ \, e^{\lambda \, V(a,b)} \, ]_r$
for any exactly marginal deformation
takes the form
\begin{equation}\label{singQexpVab}
Q_B \cdot [ \, e^{\lambda \, V(a,b)} \, ]_r
= [ \, e^{\lambda \, V(a,b)} \, O_R (b) \, ]_r
- [ \, O_L (a) \, e^{\lambda \, V(a,b)} \, ]_r \,,
\end{equation}
where $O_L$ and $O_R$ are $\lambda$-dependent, Grassmann-odd
local operators.
The operators $O_L$ and $O_R$ are closely related
and mapped to each other
under the conjugation discussed in~\S~\ref{subsubreality}
when the reflection assumption~(\ref{6}) is satisfied.
We will discuss the relation between $O_L$ and $O_R$
in more detail in~\S~\ref{singular_general_real},
but it is relevant only when generating
a real solution from $\Psi_L$
and we do not need to assume any relation
between $O_L$ and $O_R$ in the construction
of the solution $\Psi_L$.
In the case of marginal deformations with regular operator products,
we see from~(\ref{QVabn}) that
\begin{equation}
Q_B \cdot e^{\lambda \, V(a,b)}
= \lambda \, \bigl[ \, e^{\lambda \, V(a,b)} \, cV (b)
- cV (a) \, e^{\lambda \, V(a,b)} \, \Bigr]
\end{equation}
and identify
\begin{equation}\label{regOLOR}
    O^{regular}_L=O^{regular}_R=\lambda \, cV \,.
\end{equation}
In the case of marginal deformations
with singular operator products,
there can be corrections to $O_L$ and $O_R$,
which are determined from the BRST transformation
of $[ \, V^{(n)} (a,b) \, ]_r$
in the form
\begin{equation}
Q_B \cdot [ \, V^{(n)}(a,b) \, ]_r
= \sum_{r=1}^n \, [  \, V^{(n-r)}(a,b) \, O_R^{(r)}(b)\, ]_r
{}- \sum_{l=1}^n \, [ \, O_L^{(l)}(a) \, V^{(n-l)}(a,b) \, ]_r\,,
\end{equation}
where $O_L$ and $O_R$ are expanded as follows:
\begin{equation}
O_L = \sum_{n=1}^\infty \, \lambda^n \, O_L^{(n)} \,, \qquad
O_R = \sum_{n=1}^\infty \, \lambda^n \, O_R^{(n)} \,.
\end{equation}
The operators $O_L^{(1)}$ and $O_R^{(1)}$ are determined from
the BRST transformation of $[ \, V^{(1)}(a,b) \, ]_r$.
Since $[ \, V^{(1)}(a,b) \, ]_r$ does not require
any renormalization, we find
\begin{equation}
Q_B \cdot [ \, V^{(1)}(a,b) \, ]_r
= Q_B \cdot V(a,b)
= cV(b) - cV(a)
\end{equation}
for any dimension-one primary field $V$.
Thus the operators $O_L^{(1)}$ and $O_R^{(1)}$ are determined to be
\begin{equation}
O_L^{(1)} = O_R^{(1)} = cV
\end{equation}
for any marginal deformation.
Similarly, the operators $O_L^{(n)}$ and $O_R^{(n)}$ with $n \ge 2$
are determined from the BRST transformation of
$[ \, V^{(n)}(a,b) \, ]_r$ with $n \ge 2$\,,
but we do not need any specific information on these operator
in the construction of solutions.
The BRST transformation of $U$ is then given by
\begin{equation}\label{QUsing}
Q_B \, U = A_R - A_L \,,
\end{equation}
where
\begin{equation}
A_L \equiv \sum_{n=1}^\infty \, \lambda^n \, A_L^{(n)} \,, \qquad
A_R \equiv \sum_{n=1}^\infty \, \lambda^n \, A_R^{(n)} \,,
\end{equation}
with
\begin{figure}[tb]
\Large
\begin{center}
$\frac{1}{2}$
\hskip-.2cm
\raisebox{-1.3cm}{\epsfig{figure=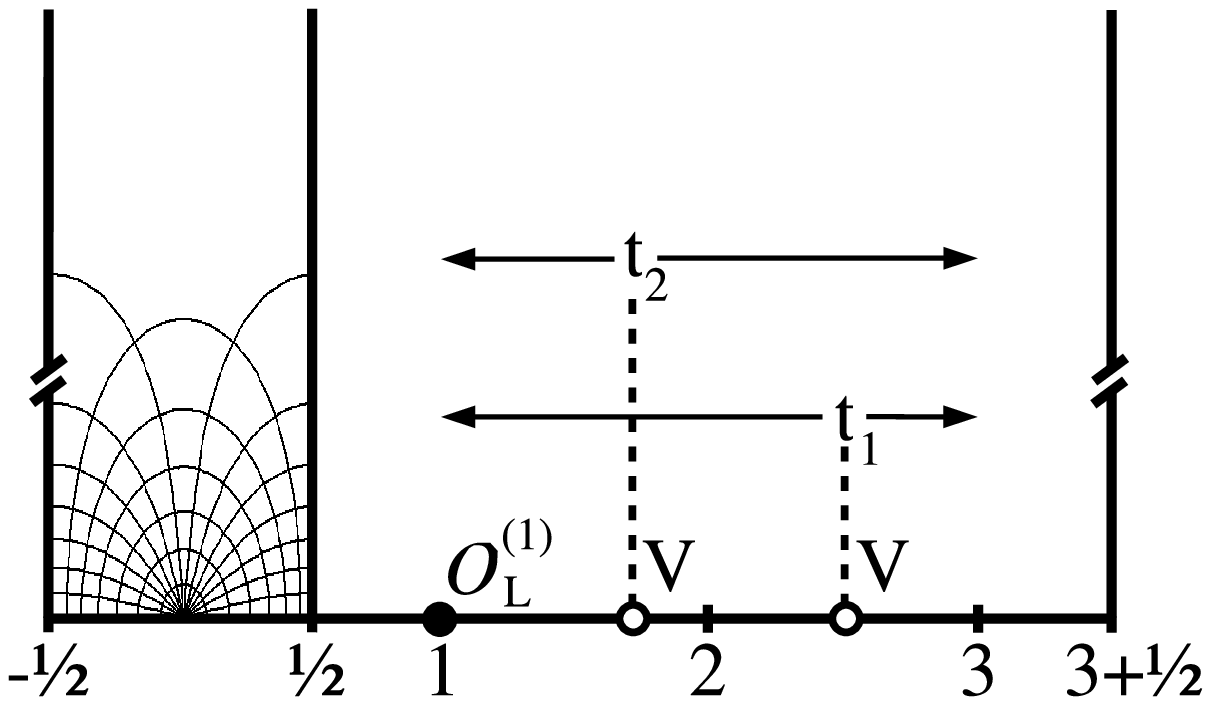, height=3.2cm}}
\hskip-.2cm $+$
\raisebox{-1.3cm}{\epsfig{figure=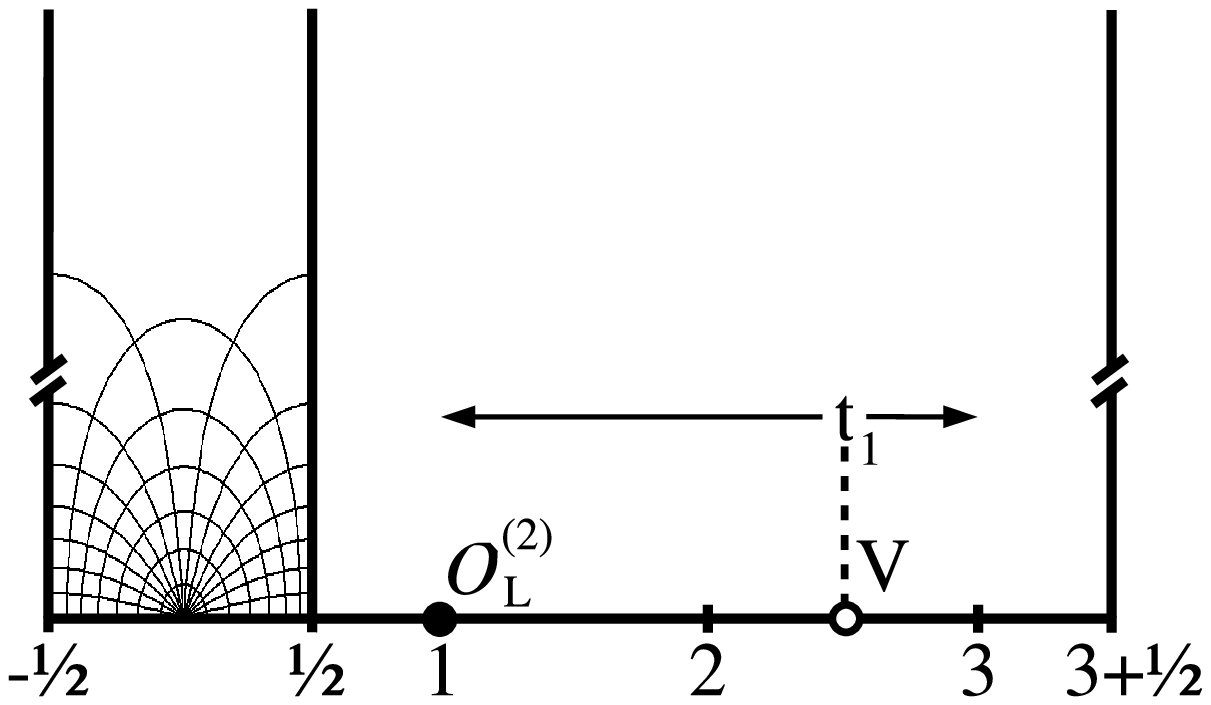,height=3.2cm}}
\hskip-.2cm $+$
\raisebox{-1.3cm}{\epsfig{figure=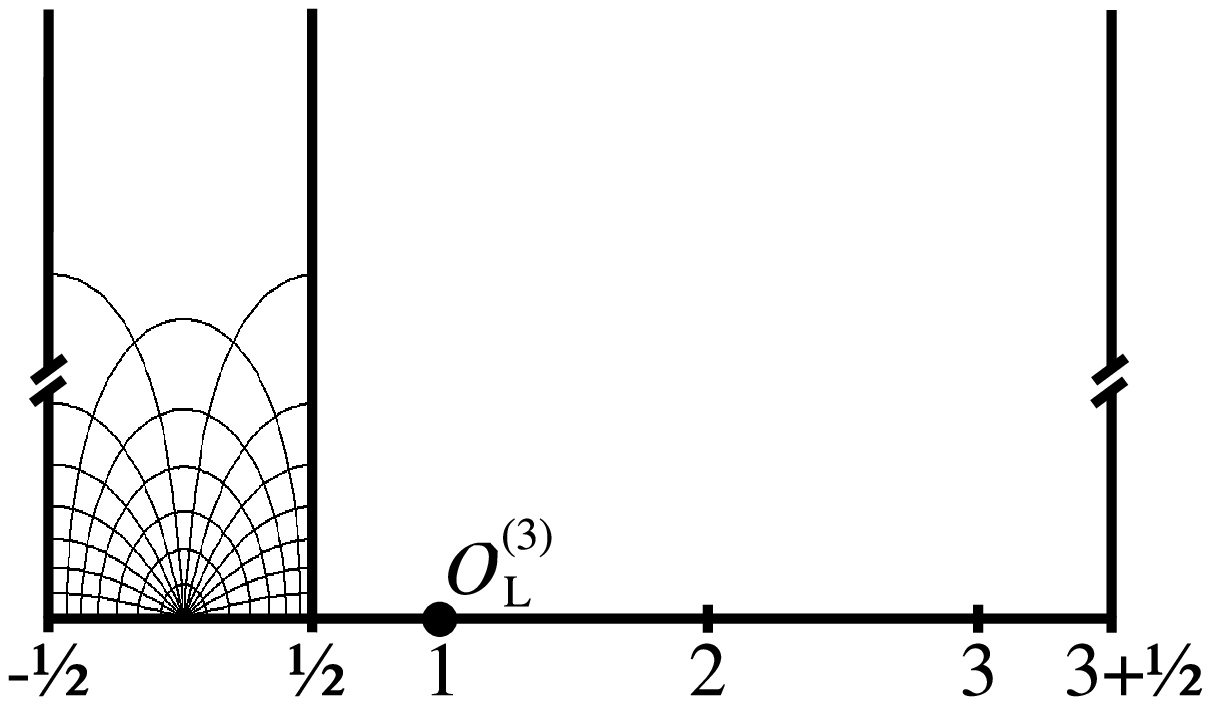,height=3.2cm}}
\end{center}
\normalsize
\caption{Illustration of $A_L^{(3)}$.}\label{figAL3}
\end{figure}
\begin{equation}
\begin{split}
    \langle \, \phi \,, A_L^{(n)} \, \rangle &=
    \sum_{l=1}^n \langle \, f \circ \phi (0) \, [ \, O_L^{(l)} (1) \, V^{(n-l)} (1,n) \, ]_r \, \rangle_{{\cal W}_n},\\
    \langle \, \phi \,, A_R^{(n)} \, \rangle &=
     \sum_{r=1}^n \langle \, f \circ \phi (0) \, [  \, V^{(n-r)} (1,n) \, O_R^{(r)} (n)\, ]_r \, \rangle_{{\cal W}_n}  \,.
\end{split}
\end{equation}
See figure~\ref{figAL3}.
We have defined $A_L^{(1)}$ and $A_R^{(1)}$
to be $\Psi^{(1)}$ as in the regular case.

We now \emph{define} $\Psi_L$ by
\begin{equation}
\Psi_L \equiv A_L \ast U^{-1} \,,
\end{equation}
and we conclude from the calculation~(\ref{rewriteQPsiL}),
where we only used the relation $Q_B U = A_R - A_L$,
that $\Psi_L$ satisfies the equation of motion if
\begin{equation}\label{condQAL}
    {}- Q_B A_L = A_L\ast U^{-1}\ast A_R \,.
\end{equation}
So far we have only used the assumption~(\ref{1})
on the BRST transformation
of $[\,e^{\lambda V(a,b)}\,]_r$.
We show in the next subsection that the equation~(\ref{condQAL})
holds when the
assumptions~(\ref{2})--(\ref{5})
stated in the introduction are satisfied.

\subsection{Proof that the equation of motion is satisfied}
\label{singular_general_proof}

Let us first examine the left-hand side of~(\ref{condQAL}).
{}From the assumption~(\ref{2})
on the BRST transformation of
$[ \, O_L (a) \, e^{\lambda \, V (a,b)} \, ]_r$,
it is given by
\begin{equation}
\begin{split}
{}- \langle \, \phi \,, Q_B A_L^{(n)} \, \rangle &=
 \sum_{l+r\leq n} \langle \, f \circ \phi (0) \,
[ \, O_L^{(l)} (1) \, V^{(n-l-r)} (1,n) \, O_R^{(r)} (n) \, ]_r \,
\rangle_{{\cal W}_n}\,.
\end{split}
\end{equation}
If we define $U_\alpha$ for $\alpha\geq0$ in the singular case by
\begin{equation}
U_\alpha \equiv\sum_{n=0}^\infty \, \lambda^n \, U_\alpha^{(n)}
\end{equation}
with
\begin{equation}
\langle \, \phi \,, U_\alpha^{(n)} \, \rangle =
\langle \, f \circ \phi (0) \, \, [ \, V^{(n)} (1,n+\alpha) \, ]_r \,
\rangle_{{\cal W}_{n+\alpha}}
\quad \text{for} \quad n + \alpha > 0 \,, \qquad
U_0^{(0)} \equiv 1 \,,
\end{equation}
then ${}- Q_B A_L$ can be constructed from $U_{l+r}$
by inserting $\lambda^l \, O_L^{(l)}$ and $\lambda^r \, O_R^{(r)}$
and by summing over $l$ and $r$. We schematically write the state
in the following way:
\begin{equation}
{}- Q_B A_L \sim \sum_{l, \, r} \, \Bigl( \, U_{l+r}
~ \text{with} ~ \lambda^l \, O_L^{(l)}
~ \text{and} ~ \lambda^r \, O_R^{(r)} \, \Bigr) \,.
\end{equation}
The state $A_L$ on the right-hand side of~(\ref{condQAL})
can be constructed from $U_l$ by inserting $\lambda^l \, O_L^{(l)}$
and by summing over $l$.
Similarly, the state $A_R$
can be constructed from $U_r$ by inserting $\lambda^r \, O_R^{(r)}$
and by summing over $r$.
Therefore, the state $A_L \ast U^{-1} \ast A_R$
can be schematically expressed as follows:
\begin{equation}
\begin{split}
A_L \ast U^{-1} \ast A_R
& \sim \sum_l \, \Bigl( \, U_l
~ \text{with} ~ \lambda^l \, O_L^{(l)} \, \Bigr) \ast U^{-1} \ast
\sum_r \, \Bigl( \, U_r
~ \text{with} ~ \lambda^r \, O_R^{(r)} \, \Bigr) \\
& \sim \sum_{l, \, r} \, \Bigl( \, U_l \ast U^{-1} \ast U_r
~ \text{with} ~ \lambda^l \, O_L^{(l)}
~ \text{and} ~ \lambda^r \, O_R^{(r)} \, \Bigr) \,.
\end{split}
\end{equation}
The equation
${}- Q_B A_L = A_L\ast U^{-1}\ast A_R$ thus
follows if the relation
\begin{eqnarray}\label{Ulr}
    U_{l+r} \, = \, U_l\ast U^{-1} \ast U_r
\end{eqnarray}
with additional operator insertions
of $O_L^{(l)}$ and $O_R^{(r)}$ holds
for the singular case.

Motivated by this observation,
we first show that
the relation $U_{l+r} \, = \, U_l\ast U^{-1} \ast U_r$
holds for the singular case
if the assumptions of replacement~(\ref{3}), factorization~(\ref{4}),
and locality~(\ref{5}) are satisfied.
It is then straightforward to generalize the proof
by taking into account the insertions
of $O_L^{(l)}$ and $O_R^{(r)}$
and show the equation~(\ref{condQAL}).
Instead of presenting a lengthy formal proof,
we demonstrate how these equations hold
using concrete examples
and then explain how the proof generalizes.

Let us consider the equation
$U_2=U_1\ast U^{-1}\ast U_1$ at $\ord{\lambda^2}$.
Since $U^{-1} = 1 - \lambda^2 \, U^{(2)}+\ord{\lambda^3}$,
it can be written as follows:
\begin{equation}
U_2^{(2)} = U_1^{(0)} \ast U_1^{(2)} + U_1^{(1)} \ast U_1^{(1)} +
U_1^{(2)} \ast U_1^{(0)} - U_1^{(0)} \ast U^{(2)} \ast U_1^{(0)}
\,.
\end{equation}
All the terms are made of the wedge state $W_4$
with operator insertions.
In the regular case, the equation was a consequence
of the following relation of the operator insertions
on~${\cal W}_4$:
\begin{equation}\label{regV14^2}
\left(V(1,4)\right)^2 = \left(V(2,4)\right)^2 + 2 \, V(1,2) \, V(3,4)
+ \left(V(1,3)\right)^2 - \left(V(2,3)\right)^2 \,.
\end{equation}
In the singular case, we need to show
\begin{equation}\label{singV14^2}
[ \, \left(V(1,4)\right)^2 \, ]_r = [ \, \left(V(2,4)\right)^2 \, ]_r
+ 2 \, [ \, V(1,2) \, ]_r \, [ \, V(3,4) \, ]_r
+ [ \, \left(V(1,3)\right)^2 \, ]_r
- [ \, \left(V(2,3)\right)^2 \, ]_r \,.
\end{equation}
Note that we have implicitly used the locality assumption~(\ref{5}).
The operators $[ \, \left(V(2,4)\right)^2 \, ]_r$
and $[ \, \left(V(1,3)\right)^2 \, ]_r$
on the right-hand side were originally defined on~${\cal W}_3$
and $[ \, \left(V(2,3)\right)^2 \, ]_r$
was defined on~${\cal W}_2$.
They are now inserted on~${\cal W}_4$
in the same forms because of the assumption~(\ref{5}).
We next use the factorization assumption~(\ref{4}) of the following form:
\begin{equation}
[ \, e^{\lambda_1 V(1,2)} \, e^{\lambda_2 V(3,4)} \, ]_r
= [ \, e^{\lambda_1 V(1,2)} \, ]_r \,
[ \, e^{\lambda_2 V(3,4)} \, ]_r \,.
\end{equation}
The relation at $\ord{\lambda_1 \, \lambda_2}$ is
\begin{equation}
[ \, V(1,2) \, V(3,4) \, ]_r
= [ \, V(1,2) \, ]_r \, [ \, V(3,4) \, ]_r \,.
\end{equation}
Thus the right-hand side of~(\ref{singV14^2}) can be written as
\begin{equation}
\begin{split}
& [ \, \left(V(2,4)\right)^2 \, ]_r
+ 2 \, [ \, V(1,2) \, ]_r \, [ \, V(3,4) \, ]_r
+ [ \, \left(V(1,3)\right)^2 \, ]_r
- [ \, \left(V(2,3)\right)^2 \, ]_r \\
& = [ \, \left(V(2,4)\right)^2 \, ]_r
+ 2 \, [ \, V(1,2) \, V(3,4) \, ]_r
+ [ \, \left(V(1,3)\right)^2 \, ]_r
- [ \, \left(V(2,3)\right)^2 \, ]_r \,.
\end{split}
\end{equation}
We then use the assumption~(\ref{3}) of replacement in the final step.
It follows from the assumption~(\ref{3}) that
\begin{equation}
[ \, e^{\lambda V(a,c)} \, ]_r
= [ \, e^{\lambda V(a,b)} \, e^{\lambda V(b,c)} \, ]_r
\end{equation}
for $a < b < c$. At $\ord{\lambda^2}$, we obtain
the following formula:
\begin{equation}\label{III-lambda^2}
[ \, \left( V(a,c) \right)^2 \, ]_r
= [ \, \left( V(a,b) \right)^2 \, ]_r
+ 2 \, [ \, V(a,b) \, V(b,c) \, ]_r
+ [ \, \left( V(b,c) \right)^2 \, ]_r \,.
\end{equation}
We thus find
\begin{equation}\label{replacement-examples}
\begin{split}
[ \, \left(V(2,4)\right)^2 \, ]_r
& = [ \, ( \, V(2,3) + V(3,4) \, )^2 \, ]_r \\
& = [ \, \left(V(2,3)\right)^2 \, ]_r + 2 \, [ \, V(2,3) \, V(3,4) \, ]_r
+ [ \, \left(V(3,4)\right)^2 \, ]_r \,, \\
[ \, \left(V(1,3)\right)^2 \, ]_r
& = [ \, ( \, V(1,2) + V(2,3) \, )^2 \, ]_r \\
& = [ \, \left(V(1,2)\right)^2 \, ]_r + 2 \, [ \, V(1,2) \, V(2,3) \, ]_r
+ [ \, \left(V(2,3)\right)^2 \, ]_r \,. \\
\end{split}
\end{equation}
For the operator $[ \, \left(V(1,4)\right)^2 \, ]_r$
on the left-hand side of~(\ref{singV14^2}),
we use the formula~(\ref{III-lambda^2}) recursively and obtain
\begin{equation}
\begin{split}
[ \, \left(V(1,4)\right)^2 \, ]_r
& = [ \, ( \, V(1,2) + V(2,3) + V(3,4) \, )^2 \, ]_r \\
& = [ \, \left(V(1,2)\right)^2 \, ]_r + [ \, \left(V(2,3)\right)^2 \, ]_r
+ [ \, \left(V(3,4)\right)^2 \, ]_r \\
& \quad ~ + 2 \, [ \, V(1,2) \, V(2,3) \, ]_r + 2 \, [ \, V(2,3) \,
V(3,4) \, ]_r
+ 2 \, [ \, V(1,2) \, V(3,4) \, ]_r \,. \\
\end{split}
\end{equation}
We can explicitly confirm
that the equation~(\ref{singV14^2}) is satisfied.
However, the coefficients in the basis
\begin{equation}
\begin{split}
& \bigl\{ \, [ \, \left(V(1,2)\right)^2 \, ]_r \,,\,
[ \, \left(V(2,3)\right)^2 \, ]_r \,,\,
[ \, \left(V(3,4)\right)^2 \, ]_r \,, \\
& \quad [ \, V(1,2) \, V(2,3) \, ]_r \,,\,
[ \, V(2,3) \, V(3,4) \, ]_r \,,\,
[ \, V(1,2) \, V(3,4) \, ]_r \, \bigr\}
\end{split}
\end{equation}
are guaranteed to match on both sides of~(\ref{singV14^2})
because they are the same as those
in the regular case
where the corresponding identity~(\ref{regV14^2}) has been shown.

This proof can be generalized
to $U_{l+r} = U_l \ast U^{-1} \ast U_r$ at $\ord{\lambda^n}$
for any positive integers $l$, $r$, and $n$.
The state $U_{l+r}^{(n)}$ can be expressed
in terms of $[ \, V^{(n)}(1,l+r+n) \, ]_r$ on ${\cal W}_{l+r+n}$.
Because of the assumption~(\ref{5}),
the terms of $U_l \ast U^{-1} \ast U_r$ at $\ord{\lambda^n}$
can also be expressed in terms of products of the form
\begin{equation}
\prod_j \, [ \, V^{(k_j)}(a_j,b_j) \, ]_r
\end{equation}
on ${\cal W}_{l+r+n}$,
where positive integers $k_j$, $a_j$, and $b_j$ satisfy
$1 \le a_j < b_j \le l+r+n$, $b_j < a_{j+1}$ and $\sum_j k_j = n$.
Using the factorization assumption~(\ref{4}),
the products can be written in the form
\begin{equation}
[ \, \prod_j \, V^{(k_j)}(a_j,b_j) \, ]_r
\end{equation}
on ${\cal W}_{l+r+n}$.
Finally, we use the replacement assumption~(\ref{3})
to expand both sides of the equation
$U_{l+r} = U_l \ast U^{-1} \ast U_r$
in the basis
\begin{equation}
\Bigl\{ \, [ \, \prod_{i=1}^{l+r+n-1} \,
V^{(\ell_i)}(i,i+1) \, ]_r \, \Bigr\} \,,
\end{equation}
where $\ell_i$'s are non-negative integers
with $\sum_{i=1}^{l+r+n-1} \ell_i = n$.
The coefficients in the basis are guaranteed to match
on both sides of $U_{l+r} = U_l \ast U^{-1} \ast U_r$
because the equation holds in the regular case.
This completes the proof of $U_{l+r} = U_l \ast U^{-1} \ast U_r$
in the singular case to all orders in $\lambda$.

The proof of ${}- Q_B A_L = A_L\ast U^{-1}\ast A_R$
is essentially parallel
using the assumptions~(\ref{3}) and~(\ref{4})
of replacement and factorization
with additional insertions of $O_L$ and $O_R$.
We provide the details of the proof in appendix~\ref{QBAL-appendix}.
We thus conclude that $\Psi_L$ given by
\begin{equation}
\Psi_L \,=\, A_L\ast U^{-1}
\end{equation}
solves the equation of motion
for any exactly marginal deformations
satisfying the
assumptions~(\ref{1})--(\ref{5}).

\subsection{Construction of a real solution}\label{singular_general_real}

It is straightforward to construct a real solution
$\Psi$ from $\Psi_L$ as we did in~\S~\ref{regular_real}
for marginal deformations with regular operator products.
The state $U$ satisfies $U^\ddagger = U$
under the assumption~(\ref{6}) of reflection.
It then follows from~(\ref{conjugation-with-Q_B})
that $(Q_B U)^\ddagger = {}- Q_B U$
and thus $(A_R - A_L)^\ddagger = {} A_L - A_R$. From this
we conclude that the local operators $O_L$ and $O_R$ are mapped
under the conjugation discussed in~\S~\ref{subsubreality} as follows:
\begin{equation}\label{conjOLOR}
O_L (t) ~ \longrightarrow ~ O_R (n+1-t) \,, \quad
O_R (t) ~ \longrightarrow ~ O_L (n+1-t)
\quad \text{on} \quad {\cal W}_n \,.
\end{equation}
We thus find
\begin{equation}
A_L^\ddagger = A_R \,, \qquad A_R^\ddagger = A_L \,.
\end{equation}
In the case of marginal deformations with regular operator products,
$O_L$ and $O_R$ are both $\lambda \, cV$
and are indeed mapped as~(\ref{conjOLOR}).

We define $\Psi_R$ by
\begin{equation}
\Psi_R \equiv U^{-1} \ast A_R\,.
\end{equation}
As in the regular case,
the state $\Psi_R$ is the conjugate of $\Psi_L$:
\begin{equation}
\Psi_R = \Psi_L^\ddagger\,.
\end{equation}
It satisfies the equation of motion
and obeys the relation $Q_B U = U \ast \Psi_R - \Psi_L \ast U$.
We conclude that $\Psi$ given by
\begin{equation}
\begin{split}
\Psi & = \frac{1}{\sqrt{U}} \ast \Psi_L \ast \sqrt{U} +
\frac{1}{\sqrt{U}} \ast Q_B \, \sqrt{U}\\
&= \sqrt{U} \ast \Psi_R \ast \frac{1}{\sqrt{U}}
+ \sqrt{U} \ast Q_B \, \frac{1}{\sqrt{U}} \\
& = \frac{1}{2} \, \biggl[ \, \frac{1}{\sqrt{U}} \ast \Psi_L \ast
\sqrt{U} + \sqrt{U} \ast \Psi_R \ast \frac{1}{\sqrt{U}} +
\frac{1}{\sqrt{U}} \ast Q_B \, \sqrt{U} - Q_B \, \sqrt{U}
\ast \frac{1}{\sqrt{U}} \, \biggr]
\end{split}
\end{equation}
is real and satisfies the equation of motion.
The solution $\Psi$ can also be expressed
in terms of $A_L$ and $A_R$ in the following way,
which might be more convenient
for an explicit expansion in $\lambda$:
\begin{equation}
\begin{split}
\Psi & = \frac{1}{\sqrt{U}} \ast A_L \ast \frac{1}{\sqrt{U}} +
\frac{1}{\sqrt{U}} \ast Q_B \, \sqrt{U}\\
&= \frac{1}{\sqrt{U}} \ast A_R \ast \frac{1}{\sqrt{U}}
+ \sqrt{U} \ast Q_B \, \frac{1}{\sqrt{U}} \\
& = \frac{1}{2} \, \biggl[ \, \frac{1}{\sqrt{U}} \ast (A_L+A_R) \ast
\frac{1}{\sqrt{U}} + \frac{1}{\sqrt{U}} \ast Q_B \, \sqrt{U}
- Q_B \, \sqrt{U} \ast \frac{1}{\sqrt{U}} \, \biggr] \,.
\end{split}
\end{equation}

\section{Explicit construction}\label{singular_explicit}
\setcounter{equation}{0}

We have separated the construction of solutions
for marginal deformations in open string field theory
into two steps.
In the previous section,
we have presented the general construction of solutions
in open string field theory from
the operator $[ \, e^{\lambda V(a,b)} \, ]_r$.
The second step is then to construct
such properly renormalized operators
satisfying the assumptions stated in the introduction
for concrete examples of exactly marginal deformations.
This is a problem in the boundary CFT
and independent of string field theory.
In this section,
we carry out the second step
for a class of marginal deformations
with singular operator products
by constructing $[ \, e^{\lambda V(a,b)} \, ]_r$ explicitly.

\subsection{A class of marginal deformations
with singular operator products}\label{class}

The dependence of the two-point function
$\langle \, V(t_1) \, V(t_2) \, \rangle$ on $t_1$ and $t_2$
for a dimension-one primary field $V$
is completely fixed by conformal symmetry.
When the singular part of the operator product expansion (OPE)
of $V$ with itself is given by
\begin{equation}\label{doublepoleOPE}
V (t) \, V(0) \sim \frac{1}{t^2} \,,
\end{equation}
the operator product $V(t_1) \, V(t_2)$ can be made finite
in the limit $t_1 \to t_2$ by subtracting
$\langle \, V(t_1) \, V(t_2) \, \rangle$ from it.\footnote{
When the double-pole term $1/t^2$ in the OPE $V(t) \, V(0)$
is nonvanishing, we normalize $V(t)$ such that the coefficient
of the double-pole term is unity.
If the state $\Psi^{(1)}$ using $V$ with this normalization
is odd instead of even under the conjugation
discussed in~\S~\ref{subsubreality},
we set $\lambda = i \, \tilde{\lambda}$
and take $\tilde{\lambda}$ to be real
when constructing the real solution $\Psi$
in~\S~\ref{singular_general_real}.}
We define $\dcirc V(t_1) \, V(t_2) \dcirc$ for $t_1 \ne t_2$ by
\begin{equation}
\dcirc V(t_1) \, V(t_2) \dcirc
\equiv V(t_1) \, V(t_2) - G(t_1,t_2) \,,
\end{equation}
where
\begin{equation}
G(t_1,t_2) \equiv \langle \, V(t_1) \, V(t_2) \, \rangle \,.
\end{equation}
Note that
the correlation function $\langle \, V(t_1) \, V(t_2) \, \rangle$
depends on the Riemann surface where the boundary CFT is defined,
and thus the definition of $\dcirc V(t_1) \, V(t_2) \dcirc$ also
depends on the Riemann surface.

The OPE of $V$ with itself, however, can have other singular terms.
For example, the singular part of the OPE can be
\begin{equation}
V(t) \, V(0) \sim \frac{1}{t^2}
+ \frac{1}{t} \, \widetilde{V} (0)
\label{single-pole-term}
\end{equation}
with some dimension-one primary field $\widetilde{V}$,
which can be proportional to $V$ itself.
The operator $\dcirc V(t_1) \, V(t_2) \dcirc$
is not finite if the single-pole term with $\widetilde{V}$
is nonvanishing.
We will discuss the case with the OPE~(\ref{single-pole-term})
in more detail in~\S~\ref{BRST}.

The operator $\dcirc V(t_1) \, V(t_2) \dcirc$ coincides
with the ordinary normal-ordered product
$: V(t_1) \, V(t_2) :$ and is thus manifestly finite 
for $V(t) = i \, \partial_t X^\mu (t) / \sqrt{2 \alpha'}$,
where $X^\mu$ is a space-like coordinate
along the D-brane.
However, it is in general different from
$: V(t_1) \, V(t_2) :$ when $V$ is a composite operator.
For example, when $V(t)$ is given by
\begin{equation}
V (t) = \sqrt{2}:\cos\biggl(\frac{X^\mu (t)}{\sqrt{\alpha'}}\biggr): \,,
\end{equation}
we can write $\dcirc V(t_1) \, V(t_2) \dcirc$ as
\begin{equation}
\dcirc V(t_1) \, V(t_2) \dcirc
= \, G (t_1,t_2)^{-1}
: \cos \biggl(\frac{X^\mu (t_1) + X^\mu (t_2)}{\sqrt{\alpha'}}\biggr) :
{}+ G(t_1,t_2) \, \biggl[ \,
: \cos \biggl(\frac{X^\mu (t_1) - X^\mu (t_2)}{\sqrt{\alpha'}}\biggr) : - 1 \,
\biggr] \,,
\end{equation}
which is not the same as the normal-ordered product:
\begin{equation}
\dcirc V(t_1) \, V(t_2) \dcirc
\ne {}: V(t_1) \, V(t_2) :
{}= 2 :\cos\biggl(\frac{X^\mu (t_1)}{\sqrt{\alpha'}}\biggr) \,
\cos\biggl(\frac{X^\mu (t_2)}{\sqrt{\alpha'}}\biggr): \,.
\end{equation}

We similarly define
$\dcirc V(t_1) \, V(t_2) \, \ldots \, V(t_{n}) \dcirc$
for arbitrary $n$ with $t_i \ne t_j$ recursively as follows:
\begin{equation}\label{genNOrec}
\begin{split}
 \dcirc V(t_1) \dcirc &\equiv V(t_1) \,, \\
 \dcirc V(t_1) \, V(t_2) \, \ldots \, V(t_n) \dcirc
 &\equiv
\dcirc V(t_1) \, V(t_2) \, \ldots \, V(t_{n-1}) \dcirc \, V(t_n) \\
& \quad ~ {}- \sum_{i=1}^{n-1} \, G(t_i,t_n)\,\dcirc V(t_1) \,
V(t_2) \, \ldots \, V(t_{i-1}) \, V(t_{i+1}) \, \ldots V(t_{n-1})
\dcirc \,
\end{split}
\end{equation}
for $n > 1$ and $t_i\neq t_j$.
This can be formally written in the following form:
\begin{equation}\label{genNO}
\dcirc\prod_i  V(t_i)\dcirc =
\exp\,\biggl(-\frac{1}{2}\int dt_1dt_2\, G(t_1,t_2) \,
\frac{\delta}{\delta V(t_1)}\frac{\delta}{\delta V(t_2)}\,\biggr)
\prod_i \, V(t_i)\quad\text{ for }\quad t_i\neq t_j \,.
\end{equation}
For $V(t) = i \, \partial_t X^\mu (t) / \sqrt{2 \alpha'}$,
the operator product
$\dcirc V(t_1) \, V(t_2) \, \ldots \, V(t_{n}) \dcirc$
again coincides with
\linebreak[4]
$:V(t_1) \, V(t_2) \, \ldots \, V(t_{n}) :$
and is regular.
In general, however,
$\dcirc V(t_1) \, V(t_2) \, \ldots \, V(t_{n}) \dcirc$
with $n \ge 3$ can be singular,
even if it is finite in the limit $t_i \to t_j$
for any pair of $i$ and $j$,
when more than two operators
simultaneously collide.
In this section, we consider a class of marginal operators $V$
which satisfy the following {\it finiteness condition}.\\

\noindent\hskip1cm
{\bf The finiteness condition.} {\it The limit}
\begin{equation}\label{finiteness}
\lim_{t \to t'} \dcirc V(t) \, V(t')^n \dcirc
\end{equation}
\hskip1cm{\it is finite for any positive integer $n$.}\\

\noindent
We explicitly construct $[ \, e^{\lambda \, V(a,b)} \, ]_r$
satisfying the assumptions stated in the introduction
for this class of marginal operators.

\subsection{Examples}\label{examples}

Let us give some examples of such marginal deformations
for D-branes in flat spacetime with Neumann or Dirichlet
boundary conditions.
As we have already mentioned,
the finiteness condition~(\ref{finiteness}) is satisfied for
\begin{equation}
V(t) = \frac{i}{\sqrt{2 \alpha'}} \, \partial_t X^\mu (t) \,,
\end{equation}
where $X^\mu$ is a space-like direction along the D-brane.
The direction $X^\mu$ can be noncompact
or can be compactified on a circle with any radius.
Similarly, the operator
\begin{equation}
V(t) = \frac{1}{\sqrt{2 \alpha'}} \, \partial_t X^0 (t)
\end{equation}
for the time-like direction also satisfies
the finiteness condition.\footnote{
We have to set $\lambda = i \, \tilde{\lambda}$
and take $\tilde{\lambda}$ to be real
for this operator
when constructing the real solution $\Psi$.}
Both of these deformations correspond to turning on
a constant mode of the gauge field on the D-brane.

The finiteness condition is also satisfied for
\begin{equation}
V(t) = \frac{1}{\sqrt{2 \alpha'}} \, \partial_\perp X^\alpha (t) \,,
\end{equation}
where $X^\alpha$ is a direction transverse to the D-brane
and $\partial_\perp$ is the derivative normal to the boundary.
The direction $X^\alpha$ can be noncompact
or can be compactified on a circle with any radius.
This deformation corresponds to displacement
of the position of the D-brane in the direction $X^\alpha$.
The condition~(\ref{finiteness})
is satisfied because the operator
$\dcirc V(t_1) \, V(t_2) \, \ldots \, V(t_{n}) \dcirc$
again coincides with
$: V(t_1) \, V(t_2) \, \ldots \, V(t_{n}) :$
and is regular.

A more nontrivial example of $V$
satisfying~(\ref{finiteness}) is
\begin{equation}\label{Yboson}
V (t) = \sqrt{2}:\cos\biggl(\frac{X^\mu (t)}{\sqrt{\alpha'}}\biggr): \,,
\end{equation}
where $X^\mu$ is again a space-like direction along the D-brane.
The direction $X^\mu$ can be noncompact
or can be compactified on a circle whose radius is
a multiple of the self-dual radius
to be consistent with the periodicity of the cosine potential.
This deformation is known to be exactly
marginal~\cite{Callan:1994ub, Polchinski:1994my,
Recknagel:1998ih, Sen:1999mh}
and interpolates Neumann and Dirichlet boundary conditions.
If we start from a D25-brane and deform the background
by this operator, we obtain a periodic array of D24-branes
at some value of the deformation parameter.
When we compactify the $X^\mu$ direction
on a circle with the self-dual radius,
the free boson for the $X^\mu$ direction can be described
by a different free boson $Y^\mu$
because of the $SU(2) \times SU(2)$ symmetry,
and the marginal operator $V(t)$ can be written
in terms of $Y^\mu$ as follows:
\begin{equation}
V (t) = \sqrt{2}:\cos\biggl(\frac{X^\mu (t)}{\sqrt{\alpha'}}\biggr):
{}= \frac{i}{\sqrt{2 \alpha'}} \, \partial_t Y^\mu (t) \,.
\end{equation}
See, for example, \S~3.1 of \cite{Sen:2004nf}.
Finiteness of
$\dcirc V(t_1) \, V(t_2) \, \ldots \, V(t_{n}) \dcirc$
at the self-dual radius is then
a consequence of Wick's theorem in the description
in terms of $Y^\mu$.
On the other hand, the finiteness is highly nontrivial
in the original description in terms of $X^\mu$.
The operator algebra of boundary operators
necessary for the calculation of
$\dcirc V(t_1) \, V(t_2) \, \ldots \, V(t_{n}) \dcirc$,
however, does not depend on the compactification radius.
Thus $\dcirc V(t_1) \, V(t_2) \, \ldots \, V(t_{n}) \dcirc$
is finite for any radius which is a multiple of
the self-dual radius and for the noncompact case as well.

The operator algebra of boundary operators
necessary for the calculation of the operator product
$\dcirc V(t_1) \, V(t_2) \, \ldots \, V(t_{n}) \dcirc$
is the same if we replace $X^\mu$ by $i X^0$.
Therefore, the marginal operator
\begin{equation}
V (t) = \sqrt{2}:\cosh\biggl(\frac{X^0(t)}{\sqrt{\alpha'}}\biggr):
\end{equation}
also satisfies the finiteness condition.
This deformation has been discussed in detail
in the context of the rolling tachyon~\cite{Sen:2002nu}.

All the operators mentioned in this subsection are known
to be exactly marginal.
In the remainder of this section,
we construct solutions in terms of
$\dcirc V(t_1) \, V(t_2) \, \ldots \, V(t_{n}) \dcirc$,
and the construction depends
on the explicit form of $V$ only
through these operator products.
Thus all the marginal deformations discussed
in this subsection are covered by our construction.

\subsection{Renormalizing operators}\label{explicitRenormalization}

For the class of marginal operators
satisfying the finiteness condition~(\ref{finiteness})
in \S~\ref{class},
we can construct finite operators $\dcirc ( V(a,b) )^n \dcirc$
for any $n$ using the point-splitting regularization.
For $n=2$, we construct $\dcirc ( V(a,b) )^2 \dcirc$ as follows:
\begin{equation}
\begin{split}
\dcirc ( V(a,b) )^2 \dcirc
= & \quad\lim_{\epsilon \to 0} \,
\int_a^{b-\epsilon} dt_1\int_{t_1+\epsilon}^{b} dt_2\,
 \Bigl(V(t_1)V(t_2)-G(t_1,t_2)\Bigr)\\
&+\lim_{\epsilon \to 0} \,\int_{a+\epsilon}^b
dt_1\int_a^{t_1-\epsilon} dt_2\,
\Bigl(V(t_1)V(t_2)-G(t_1,t_2)\Bigr)\,.
\end{split}
\end{equation}
The first line and the second line on the right-hand side
are actually identical.
We could have written $\dcirc ( V(a,b) )^2 \dcirc$
using only one of them, but we used both of them so that
the integral region reduces to the product of
$a \le t_1 \le b$ and $a \le t_2 \le b$
without any ordering constraint in the limit $\epsilon \to 0$.
The construction can be generalized to any $n$ as follows:
\begin{equation}\label{dcirc-V^n-dcirc}
\dcirc ( V(a,b) )^n \dcirc
= \lim_{\epsilon \to 0} \,
\int_{\Gamma^{(n)}_\epsilon} dt_1 dt_2 \ldots dt_n \sum_{0 \le k \le
n/2} \frac{(-1)^k \, n!}{2^k \, k! \, (n-2k)!} \, \prod_{i=1}^k \,
G(t_i,t_{i+k}) \, \prod_{j=2k+1}^n V(t_j) \,,
\end{equation}
where the integral region $\Gamma^{(n)}_\epsilon$ is
\begin{equation}
\Gamma^{(n)}_\epsilon: \quad
a \le t_i \le b \quad \text{for} \quad i= 1, 2, \ldots , n
\quad \text{with} \quad | \, t_i - t_j \, | \ge \epsilon
\quad \text{for} \quad i \ne j \,.
\end{equation}
The finiteness condition~(\ref{finiteness})
guarantees that the limit $\epsilon\to 0$ is well defined
and finite for any $n$.
We then define $\dcirc e^{\lambda \, V(a,b)} \dcirc$
by its expansion in $\lambda$:
\begin{equation}
\dcirc e^{\lambda \, V(a,b)} \dcirc \equiv \sum_{n=0}^\infty
\frac{\lambda^n}{n!} \dcirc ( V(a,b) )^n \dcirc \,.
\label{dcirc-exp-dcirc}
\end{equation}

The definition of $\dcirc e^{\lambda \, V(a,b)} \dcirc$ depends
on the Riemann surface where the boundary CFT is defined
through the propagator $G(t_1,t_2)$.
When we calculate star products of string fields
involving the operators in the expansion (\ref{dcirc-exp-dcirc}),
the operators defined on ${\cal W}_n$
are embedded in a surface ${\cal W}_m$ with $m \geq n$,
and the operators in the expansion (\ref{dcirc-exp-dcirc})
are not invariant.
Thus we {\it cannot} simply set
$[ \, e^{\lambda V(a,b)} \, ]_r
\equiv \dcirc e^{\lambda V(a,b)} \dcirc$
because the locality assumption (\ref{5})
on $[ \, e^{\lambda V(a,b)} \, ]_r$ is not satisfied.

Let us study the issue more explicitly
in a simpler example.
The operator $\dcirc V(a) \, V(a,b) \dcirc$ is given by
\begin{equation}
\dcirc V(a) \, V(a,b) \dcirc
= \lim_{\epsilon \to 0} \,
\int_{a+\epsilon}^b dt \, \dcirc V(a) \, V(t) \dcirc
= \lim_{\epsilon \to 0} \,
\int_{a+\epsilon}^b dt \, \Bigl[ \, V(a) \, V(t)
- G(a,t) \, \Bigr] \,.
\end{equation}
We denote the propagator $G(t_1,t_2)$ on ${\cal W}_n$
by $G_n (t_1,t_2)$. Its explicit expression is
\begin{equation}\label{propagator}
G_n (t_1,t_2)
\equiv \langle \, V(t_1) \, V(t_2) \, \rangle_{{\cal W}_n}
= \frac{\pi^2}{(n+1)^2\sin^2
\Bigl( \, \dfrac{t_2-t_1}{n+1}\,\pi \Bigr)} \,.
\end{equation}
The operator $\dcirc V(a) \, V(a,b) \dcirc$
defined on ${\cal W}_n$ is thus
\begin{equation}
\dcirc V(a) \, V(a,b) \dcirc
= \lim_{\epsilon \to 0} \,
\int_{a+\epsilon}^b dt \, \Bigl[ \, V(a) \, V(t)
- \frac{\pi^2}{(n+1)^2\sin^2
\bigl( \frac{t-a}{n+1}\,\pi \bigr)} \, \Bigr]
\quad \text{on} \quad {\cal W}_n \,.
\end{equation}
When this operator is embedded in ${\cal W}_m$,
it should be written using the propagator on ${\cal W}_m$
as follows:
\begin{equation}
\begin{split}
& \lim_{\epsilon \to 0} \,
\int_{a+\epsilon}^b dt \, \Bigl[ \, V(a) \, V(t)
- \frac{\pi^2}{(n+1)^2\sin^2
\bigl( \frac{t-a}{n+1}\,\pi \bigr)} \, \Bigr] \\
& = \lim_{\epsilon \to 0} \,
\int_{a+\epsilon}^b dt \, \Bigl[ \, V(a) \, V(t)
- \frac{\pi^2}{(m+1)^2\sin^2
\bigl( \frac{t-a}{m+1}\,\pi \bigr)} \, \Bigr]
- \int_a^b dt \, \delta G(a,t) \,,
\end{split}
\end{equation}
where
\begin{equation}
\begin{split}
\delta G(t_1,t_2) \,& \equiv\, G_n(t_1,t_2)-G_m(t_1,t_2)\\
&=\frac{\pi^2}{(n+1)^2\sin^2\Bigl( \, \dfrac{t_2-t_1}{n+1}\,\pi\Bigr)}
\,-\,
\frac{\pi^2}{(m+1)^2\sin^2\Bigl( \, \dfrac{t_2-t_1}{m+1}\,\pi\Bigr)}
\\
& =\, \frac{(m-n)(2+m+n)\pi^2}{3(m+1)^2(n+1)^2}
+\ord{(t_2-t_1)^2} \,,
\end{split}
\end{equation}
and $\delta G(t_1,t_2)$ is finite in the limit $t_2 \to t_1$.
The operator $\dcirc V(a) \, V(a,b) \dcirc$
defined on ${\cal W}_n$ is thus rewritten
when embedded in ${\cal W}_m$ as
\begin{equation}
\dcirc V(a) \, V(a,b) \dcirc
\ntom
\dcirc V(a) \, V(a,b) \dcirc - \int_a^b dt \, \delta G(a,t) \,.
\label{extra-term}
\end{equation}
The notation
\begin{equation}
A \ntom B
\end{equation}
implies that $A = B$,
but $A$ is written in terms of the propagator on ${\cal W}_n$
and $B$ is written in terms of the propagator on ${\cal W}_m$.
The assumption of locality (\ref{5}) can be stated using this notation as
\begin{equation}
[ \, e^{\lambda V(a,b)} \, ]_r
\ntom [ \, e^{\lambda V(a,b)} \, ]_r \,, \qquad
[ \, O_L(a) \, e^{\lambda V(a,b)} \, ]_r
\ntom [ \, O_L(a) \, e^{\lambda V(a,b)} \, ]_r \,.
\end{equation}
As can be expected from the fact that
$O_L^{(1)} = O_R^{(1)} = cV$ in general,
we will need to define the operator
$[ \, V(a) \, e^{\lambda V(a,b)} \, ]_r$
satisfying
\begin{equation}\label{rewrite5} 
[ \, V(a) \, e^{\lambda V(a,b)} \, ]_r
\ntom [ \, V(a) \, e^{\lambda V(a,b)} \, ]_r \,.
\end{equation}
The operator $\dcirc V(a) \, V(a,b) \dcirc$
does not satisfy
\begin{equation}\label{localityVaVab}
[ \, V(a) \, V(a,b) \, ]_r
\ntom [ \, V(a) \, V(a,b) \, ]_r \,,
\end{equation}
and thus violates~(\ref{rewrite5}) at $\ord{\lambda}$.
In order to cancel the extra term in (\ref{extra-term}),
we add back a finite part
of the propagator contraction which we subtracted.
We define the renormalized contraction
$\langle \, V(a) \, V(a,b) \, \rangle_r$ by
\begin{equation}\label{contrVVab}
\langle V(a)V(a,b)\rangle_r\,\equiv\,
\lim_{\epsilon \to 0} \, \biggl[
\, \int_{a+\epsilon}^b dt \, G(a,t) -\frac{1}{\epsilon} \,
\biggr] \,.
\end{equation}
Its explicit expression on ${\cal W}_n$ is
\begin{equation}
\langle \, V(a) \, V(a,b) \, \rangle_r\,
=-\frac{\pi}{1+n}\cot\biggl(\frac{\pi (b-a)}{1+n}\biggr)
\quad \text{on} \quad {\cal W}_n \,,
\end{equation}
and it is rewritten when embedded in ${\cal W}_m$ as
\begin{equation}
\langle \, V(a) \, V(a,b) \, \rangle_r
\ntom
\langle \, V(a) \, V(a,b) \, \rangle_r
+ \int_a^b dt \, \delta G(a,t) \,.
\label{-extra-term}
\end{equation}
This allows us to define our first renormalized operator
$[ \,V(a)\,V(a,b)\, ]_r$ by
\begin{equation}\label{renVVab}
    \bigl[\,V(a)\,V(a,b)\,\bigr]_r \,\equiv\, \dcirc V(a)\,V(a,b)\dcirc \,+\, \langle \, V(a) \, V(a,b) \, \rangle_r\,.
\end{equation}
Since the extra term in~(\ref{extra-term})
is canceled by the extra terms in~(\ref{-extra-term}),
the operator $[ \,V(a)\,V(a,b)\, ]_r$ is invariant
under the embedding from ${\cal W}_n$ to ${\cal W}_m$
and thus satisfies~(\ref{localityVaVab}).
In fact, we can write $[ \,V(a)\,V(a,b)\, ]_r$
in the following form which does not depend on the propagator:
\begin{equation}\label{alternativeVaVabr}
\bigl[\,V(a)\,V(a,b)\,\bigr]_r
= \lim_{\epsilon \to 0} \, \biggl[ \,
\int_{a+\epsilon}^b dt\,V(a)\,V(t)-\frac{1}{\epsilon} \, \biggr] \,.
\end{equation}
Similarly, we can define the renormalized contraction
and the renormalized operator for $V(a,b) \, V(b)$ by
\begin{equation}\label{renVabV}
\begin{split}
\langle \, V(a,b) \, V(b) \, \rangle_r
\,&\equiv\, \lim_{\epsilon \to 0} \,
\biggl[ \, \int_{a}^{b-\epsilon}\, G(t,b) -\frac{1}{\epsilon} \,
\biggr] \,, \\
\bigl[\,V(a,b)\,V(b)\,\bigr]_r \, &\equiv\,
\dcirc V(a,b)\,V(b)\dcirc \,+\,
\langle \, V(a,b) \, V(b) \, \rangle_r\,
= \lim_{\epsilon \to 0} \, \biggl[ \,
\int_{a}^{b-\epsilon} dt\,V(t)\,V(b)
-\frac{1}{\epsilon} \, \biggr] \,.
\end{split}
\end{equation}
The renormalized contraction
$\langle \, V(a,b) \, V(b) \, \rangle_r$
on ${\cal W}_n$ is
\begin{equation}
\langle \, V(a,b) \, V(b) \, \rangle_r\,
=-\frac{\pi}{1+n}\cot\biggl(\frac{\pi (b-a)}{1+n}\biggr)
\quad \text{on} \quad {\cal W}_n \,.
\end{equation}

We use the same strategy to define $[ \, ( V(a,b) )^2 \, ]_r$.
We define $\langle \, V(a,b)^2 \, \rangle_r$ by
\begin{equation}\label{contrVab2}
\langle \, V(a,b)^2 \, \rangle_r
\,\equiv\, 2 \, \lim_{\epsilon \to 0} \,
\biggl[ \, \int_a^{b-\epsilon} dt_1
\int_{t_1+\epsilon}^{b} dt_2\,
G(t_1,t_2)-\frac{b-a-\epsilon}{\epsilon}-\ln\epsilon \,
\biggr] \,.
\end{equation}
Its expression on ${\cal W}_n$ is
\begin{equation}\label{<V(a,b)^2>_r-on-W_n}
\langle \, V(a,b)^2 \, \rangle_r
=\,\ln\biggl(
\frac{\pi^2}{(n+1)^2\sin^2\bigl(\frac{b-a}{n+1}\,\pi\bigr)}\biggr)\\
=\,\ln G_n (a,b)
\quad \text{on} \quad {\cal W}_n \,.
\end{equation}
We then define $[ \, ( V(a,b) )^2 \, ]_r$ by
\begin{equation}\label{renVab2}
\bigl[\, ( V(a,b) )^2\,\bigr]_r
\,\equiv\, \dcirc ( V(a,b) )^2\dcirc
\,+\, \langle \, V(a,b)^2 \, \rangle_r\,.
\end{equation}
Since $\dcirc ( V(a,b) )^2\dcirc$
and $\langle \, V(a,b)^2 \, \rangle_r$ defined on ${\cal W}_n$
are rewritten when embedded in ${\cal W}_m$ as
\begin{equation}
\begin{split}
\dcirc ( V(a,b) )^2\dcirc & \ntom
\dcirc ( V(a,b) )^2\dcirc {}- \Delta \,, \\
\langle \, V(a,b)^2 \, \rangle_r & \ntom
\langle \, V(a,b)^2 \, \rangle_r {}+ \Delta \,,
\end{split}
\end{equation}
where
\begin{equation}
\Delta \equiv
\int_a^bdt_1\int_a^bdt_2\,\delta G(t_1,t_2) \,,
\end{equation}
the operator $[ \, ( V(a,b) )^2 \, ]_r$ is invariant
under the embedding from ${\cal W}_n$ to ${\cal W}_m$.

The operator $[ \, e^{\lambda V(a,b)} \,]_r$ can also be defined
using $\langle \, V(a,b)^2 \, \rangle_r$ as follows:
\begin{equation}\label{renexpV}
[ \, e^{\lambda V(a,b)} \,]_r
\,\equiv\,e^{\frac{1}{2}\lambda^2
\langle \, V(a,b)^2 \, \rangle_r}\,
\dcirc e^{\lambda V(a,b)} \dcirc \,.
\end{equation}
By replacing $G_n$ in (\ref{dcirc-V^n-dcirc}) on ${\cal W}_n$
with $G_m + \delta G$, we find
\begin{equation}
\dcirc ( V(a,b) )^k \dcirc \ntom
\sum_{0 \le \ell \le k/2}
\frac{(-1)^\ell \, k!}{2^\ell \, (k-2 \ell)! \, \ell !} \,
\Delta^\ell \, \dcirc ( V(a,b) )^{k -2 \ell} \dcirc \,.
\end{equation}
It then follows from
\begin{equation}
\begin{split}
\dcirc e^{\lambda V(a,b)} \dcirc & \ntom
e^{-\frac{1}{2} \, \lambda^2 \Delta} \,
\dcirc e^{\lambda V(a,b)} \dcirc \,, \\
e^{\frac{1}{2}\lambda^2 \langle \, V(a,b)^2 \, \rangle_r}
& \ntom
e^{\frac{1}{2} \, \lambda^2 \Delta} \,\,
e^{\frac{1}{2}\lambda^2 \langle \, V(a,b)^2 \, \rangle_r}
\end{split}
\end{equation}
that the operator $[ \, e^{\lambda V(a,b)} \,]_r$ transforms as 
\begin{equation}\label{rewrite5again}
[ \, e^{\lambda V(a,b)} \,]_r
\ntom
[ \, e^{\lambda V(a,b)} \,]_r
\end{equation}
under the embedding
and thus satisfies the locality assumption (\ref{5}).
It is obvious from the definition (\ref{dcirc-V^n-dcirc})
that $[\,e^{\lambda V(a,b)}\,]_r$
is invariant when $V(t)$ is replaced by $V(a+b-t)$
and thus satisfies the reflection assumption (\ref{6}) as well.

Let us next define the operators
$[\,V(a)\,e^{\lambda V(a,b)}\,]_r$
and $[\,e^{\lambda V(a,b)}\,V(b)\,]_r$.
Using the renormalized contractions
$\langle \, V(a,b)^2 \, \rangle_r$,
$\langle \, V(a) \, V(a,b) \, \rangle_r$,
and $\langle \, V(a,b) \, V(b) \, \rangle_r$,
they are defined as follows:
\begin{equation}\label{renVexpV}
\begin{split}
[\,V(a)\,e^{\lambda V(a,b)}\,]_r\,
&\equiv\,e^{\frac{1}{2}\lambda^2\langle
V(a,b)^2\rangle_r}\, \dcirc
\, \Bigl( \, V(a)+\lambda \, \langle \, V(a) \, V(a,b) \,
\rangle_r\Bigr) \,
e^{\lambda V(a,b)}\dcirc\,,\\
[\,e^{\lambda V(a,b)}\,V(b)\,]_r\,
&\equiv\,e^{\frac{1}{2}\lambda^2\langle
V(a,b)^2\rangle_r}\, \dcirc e^{\lambda V(a,b)}
\Bigl( \, V(b)+\lambda \, \langle \, V(a,b) \, V(b) \,
\rangle_r\Bigr)
\dcirc\,.
\end{split}
\end{equation}
Let us prove that $[\,V(a)\,e^{\lambda V(a,b)}\,]_r$
satisfies the condition~(\ref{rewrite5}).
It follows from the definition of
$\dcirc V(t_1) \, V(t_2) \, \ldots V(t_n) \dcirc$ that
\begin{equation}
\dcirc V(a) \, e^{\lambda V(a,b)} \dcirc
= \lim_{\epsilon \to 0} \, \biggl[ \,
V(a-\epsilon) \dcirc e^{\lambda V(a,b)} \dcirc
{}- \lambda \, \int_a^b dt \, G(a-\epsilon,t) \,
\dcirc e^{\lambda V(a,b)} \dcirc \, \biggr] \,.
\end{equation}
We thus find
\begin{equation}\label{localityVaexpVab1}
\begin{split}
e^{\frac{1}{2}\lambda^2\langle V(a,b)^2\rangle_r}\dcirc V(a) \,
e^{\lambda V(a,b)} \dcirc
\hskip-.1cm
\ntom
\hskip-.1cm
e^{\frac{1}{2}\lambda^2\langle V(a,b)^2\rangle_r}
\dcirc V(a) \, e^{\lambda V(a,b)} \dcirc
{}- \lambda \, \int_a^b dt \, \delta G(a,t) \,
[\, e^{\lambda V(a,b)} \,]_r
\end{split}
\end{equation}
for the first term in the definition~(\ref{renVexpV})
of $[ \, V(a) \, e^{\lambda V(a,b)} \, ]_r$.
Similarly, the second term transforms as
\begin{equation}\label{localityVaexpVab2}
\begin{split}
\lambda \, \langle \, V(a) \, V(a,b) \, \rangle_r\,
[\,e^{\lambda V(a,b)}]_r
\ntom
\lambda \, \langle \, V(a) \, V(a,b) \, \rangle_r \,
[\,e^{\lambda V(a,b)}]_r
{}+ \lambda \, \int_a^b dt \, \delta G(a,t) \,
[\, e^{\lambda V(a,b)} \,]_r\,,
\end{split}
\end{equation}
where we used~(\ref{-extra-term}).
Combining~(\ref{localityVaexpVab1})
and~(\ref{localityVaexpVab2}), we have thus shown
that $[ \, V(a) \, e^{\lambda V(a,b)} \, ]_r$
satisfies~(\ref{rewrite5}).

To summarize, we have defined
$[ \, e^{\lambda V(a,b)} \, ]_r$
satisfying the assumptions of locality~(\ref{5})
and reflection~(\ref{6})
and $[ \, V(a) \, e^{\lambda V(a,b)} \, ]_r$
satisfying~(\ref{rewrite5})
for the class of marginal operators
satisfying the finiteness condition stated in \S~\ref{class}.

\subsection{The BRST transformation}\label{BRST}

Let us next calculate the BRST transformation
of $[ \, e^{\lambda V(a,b)} \, ]_r$ defined in (\ref{renexpV})
to verify that the assumption (\ref{1}) on the BRST transformation is satisfied
and determine $O_L$ and $O_R$.
The calculation at $\ord{\lambda}$
is the same as (\ref{QV(a,b)}) in the regular case
and gives $O_L^{(1)} = O_R^{(1)} = cV$.
The calculation at $\ord{\lambda^2}$
involves the OPE
of the marginal operator with itself.
We in fact expect that the assumption (\ref{1})
is {\it not} satisfied when the marginal
deformation is not exactly marginal.
It is known that the deformation associated with $V$
is not exactly marginal
if the single-pole term in~(\ref{single-pole-term})
is nonvanishing. See, for example, \cite{Recknagel:1998ih}.
In the construction of analytic solutions
in~\cite{Kiermaier:2007ba},
there was indeed an obstruction
to solve the equation of motion at $\ord{\lambda^2}$
when the single-pole term in~(\ref{single-pole-term})
is nonvanishing.
It is therefore instructive to briefly consider
the case of the more general OPE~(\ref{single-pole-term}),
\begin{equation}
V(t) \, V(0) \sim \frac{1}{t^2}
+ \frac{1}{t} \, \widetilde{V} (0) \,,
\label{single-pole-term-again}
\end{equation}
and to see how the assumption (\ref{1}) is violated
when the single-pole term with $\widetilde{V}$ is nonvanishing.
We regularize $V^{(2)}(a,b)$ as follows:
\begin{equation}
\int_a^{b-\epsilon} d t_1 \int_{t_1+\epsilon}^b d t_2 \,
V(t_1) \, V(t_2) \,.
\end{equation}
The calculation of its BRST transformation
is similar to the calculation of $Q_B \Psi_L^{(3)}$
presented in~(\ref{Q_B-Psi_L^(3)-1}) and~(\ref{Q_B-Psi_L^(3)-2}):
\begin{equation}
\begin{split}
Q_B \cdot \biggl[ \,
& \int_a^{b-\epsilon} d t_1 \int_{t_1+\epsilon}^b d t_2 \,
V(t_1) \, V(t_2) \, \biggr]
=\int_a^{b-\epsilon} d t_1 \int_{t_1+\epsilon}^b d t_2 \,
\Bigl[ \, \del_{t_1} [ \, cV(t_1) \, ] \, V(t_2)
+ V(t_1) \,\del_{t_2} [ \, cV(t_2) \, ] \, \Bigr] \\
= &\int_{a}^{b-\epsilon} d t_1 \, V(t_1)\,cV(b)
-\int_{a+\epsilon}^b d t_2 \,cV(a) \, V(t_2)
+\int_a^{b-\epsilon} d t_1 \, V(t_1) \, V(t_1+\epsilon) \,
\bigl[ \, c(t_1)-c(t_1+\epsilon) \, \bigr] \,. \\
\end{split}
\end{equation}
The last term on the right-hand side no longer vanishes
in the limit $\epsilon\to0$
when the OPE of $V$ with itself is singular
and can be calculated as follows:
\begin{equation}\label{collision}
\begin{split}
& \int_a^{b-\epsilon} dt \, V(t) \, V(t+\epsilon) \,
\Bigl[ \, c(t) - c(t+\epsilon) \, \Bigr] \\
& = \int_a^{b-\epsilon} dt \,
\Big( \, \frac{1}{\epsilon^2}
- \frac{1}{\epsilon} \, \widetilde{V} (t)
+ \ord{\epsilon^0} \, \Bigr)
\Bigl[ \, - \epsilon \, \partial c(t)
- \frac{\epsilon^2 }{2} \, \partial^2 c(t) + \ord{\epsilon^3} \,
\Bigr] \\
& = \int_a^{b-\epsilon} dt \, \Bigl[ \,
\partial c \widetilde{V} (t)
-\frac{1}{\epsilon} \, \partial c(t)
-\frac{1}{2} \, \partial^2 c(t) \, \Bigr] + \ord{\epsilon} \\
& = \int_a^{b-\epsilon} dt \, \partial c \widetilde{V} (t)
-\frac{1}{\epsilon} \, c(b-\epsilon)
+\frac{1}{\epsilon} \, c(a)
-\frac{1}{2} \, \partial c(b-\epsilon)
+\frac{1}{2} \, \partial c(a) + \ord{\epsilon} \\
& = \int_a^b dt \, \partial c \widetilde{V} (t)
-\frac{1}{\epsilon} \, c(b) +\frac{1}{\epsilon} \, c(a)
+\frac{1}{2} \, \partial c(b) +\frac{1}{2} \, \partial c(a)
+ \ord{\epsilon} \,.
\end{split}
\end{equation}
We thus obtain
\begin{equation}
\begin{split}
& Q_B\cdot\biggl[\int_a^{b-\epsilon} d t_1
\int_{t_1+\epsilon}^b d t_2 \, V(t_1) \, V(t_2)\biggr] \\
& = \Biggl[\int_{a}^{b-\epsilon} d t_1 \, V(t_1)\,V(b)
-\frac{1}{\epsilon}\Biggr] c(b)
-c(a)\Biggl[\int_{a+\epsilon}^b d t_2 \,V(a) \, V(t_2)
-\frac{1}{\epsilon}\Biggr] \\
& \quad ~ {}+\frac{\del c(b)}{2}+\frac{\del c(a)}{2}
+ \int_a^b d t_1 \partial c \widetilde{V}(t_1)+\ord \epsilon \,.
\end{split}
\label{Q_B-V^(2)}
\end{equation}
This does not take the form of the $\ord{\lambda^2}$ term of
$[ \, e^{\lambda V(a,b)} \, O_R (b) \, ]_r
- [ \, O_L (a) \, e^{\lambda V(a,b)} \, ]_r$
because of the term with $\del c\widetilde{V}$,
which is finite in the limit $\epsilon \to 0$.
The divergences in (\ref{Q_B-V^(2)}) arise
only when $V(t)$ approaches the end points of the integral region,
and any counterterms to take care of those localized divergences
will not cancel the finite integral of $\del c\widetilde{V}$
over the whole integral region.
Therefore, the assumption (\ref{1})  on the BRST transformation is not satisfied
when the single-pole term in (\ref{single-pole-term-again})
is nonvanishing.
This is consistent because the deformation
is not exactly marginal in this case, as we mentioned before.
When the single-pole term in (\ref{single-pole-term-again}) vanishes,
the result (\ref{Q_B-V^(2)}) in the limit $\epsilon \to 0$
is finite and given by
\begin{equation}
\lim_{\epsilon \to 0} \, \biggl[ \,
Q_B\cdot\Bigl[\int_a^{b-\epsilon} d t_1 \int_{t_1+\epsilon}^b d t_2 \,
V(t_1) \, V(t_2)\Bigr] \, \biggr]
= [ \, V(a,b) \, cV(b) \, ]_r
- [ \, cV(a) \, V(a,b) \, ]_r
+\frac{\del c(b)}{2}+\frac{\del c(a)}{2} \,.
\end{equation}
Note that $[ \, V(a,b) \, cV(b) \, ]_r$ and
$[ \, cV(a) \, V(a,b) \, ]_r$
given in~(\ref{alternativeVaVabr})
and~(\ref{renVabV})
emerged naturally.
We conclude that
\begin{equation}
O_R^{(1)}=O_L^{(1)}= cV\,,\qquad
O_R^{(2)}={}-O_L^{(2)}= \frac{1}{2} \, \del c
\end{equation}
for any exactly marginal deformation
with the singular OPE given by~(\ref{doublepoleOPE}).

Let us now calculate the BRST transformation of
$[\,e^{\lambda V(a,b)}\,]_r$
for the class of marginal operators
satisfying the finiteness condition~(\ref{finiteness}) 
in~\S~\ref{class}:
\begin{equation}
Q_B\cdot[\,e^{\lambda V(a,b)}\,]_r\,
= e^{\frac{1}{2}\lambda^2\langle
V(a,b)^2\rangle_r}\,\sum_{n=1}^\infty \frac{\lambda^n}{n!} \,
Q_B\cdot \dcirc \left(V(a,b)\right)^n \dcirc
\end{equation}
We use the expression~(\ref{dcirc-V^n-dcirc})
of $\dcirc \left(V(a,b)\right)^n \dcirc$
and calculate its BRST transformation as follows:
\begin{equation}\label{renQVabn}
\begin{split}
& Q_B\cdot \dcirc \left(V(a,b)\right)^n \dcirc \\
&=\,\sum_{0 \le k\le n/2} \,
\frac{(-1)^k \, n!}{2^k \, k! \, (n-2k)!} \, \lim_{\epsilon \to 0}
\int_{\Gamma^{(n)}_\epsilon} d^nt\prod_{i=1}^k \,
G(t_i,t_{i+k}) \,\, Q_B\cdot \prod_{j=2k+1}^n V(t_j)\\
&=\,n\sum_{0 \le k< n/2} \, \frac{(-1)^k \, (n-1)!}{2^k \, k! \,
(n-2k-1)!} \, \lim_{\epsilon \to 0}
\int_{\Gamma^{(n)}_\epsilon} d^nt\prod_{i=1}^k \,
G(t_i,t_{i+k}) \, \prod_{j=2k+1}^{n-1} V(t_j)\,
\del_{t_n}\bigl[cV(t_n)\bigr]\\
&=\,n \, \lim_{\epsilon \to 0}
\int_{\Gamma^{(n)}_\epsilon} d^nt\,\dcirc V(t_1)\ldots
V(t_{n-1})\dcirc\,\, \del_{t_n}\bigl[cV(t_n)\bigr] \,.
\end{split}
\end{equation}
Using~(\ref{genNOrec}), this can be written in the following way:
\begin{equation}\label{pullin1}
\begin{split}
Q_B\cdot \dcirc \left(V(a,b)\right)^n \dcirc &=\,
\lim_{\epsilon \to 0} \int_{\Gamma^{(n)}_\epsilon}
d^nt \, \biggl[ \, n\dcirc V(t_1)\ldots V(t_{n-1})\,
\del_{t_n}\bigl[cV(t_n)\bigr]\dcirc \\
&\qquad\qquad\qquad\qquad+n(n-1)\dcirc V(t_1)\ldots
V(t_{n-2})\dcirc \,
\del_{t_n}\bigl[G(t_{n-1},t_n)\,c(t_n)\bigr] \, \biggr] \,.
\end{split}
\end{equation}
The first term of the integrand on the right-hand side is finite
so that we can take the limit $\epsilon \to 0$
and carry out the integral over $t_n$.
The only divergence in the second term of the integrand
arises when $|t_n-t_{n-1}|\to0$.
The integral region therefore factorizes into
that of $t_1,\, t_2,\, \ldots t_{n-2}$
without the restriction $| t_i - t_j | \ge \epsilon$
and $\Gamma^{(2)}_\epsilon$ for $t_{n-1}$ and $t_n$.
We thus obtain
\begin{equation}\label{pullin2}
\begin{split}
Q_B\cdot \dcirc \left(V(a,b)\right)^n \dcirc &=\,n\int_a^b dt_n\,\dcirc \left(V(a,b)\right)^{n-1}\,\,\del_{t_n}\bigl[cV(t_n)\bigr]\dcirc \\
&\qquad\qquad\qquad+n(n-1)\dcirc \left(V(a,b)\right)^{n-2}\dcirc
\, \lim_{\epsilon \to 0}
\int_{\Gamma^{(2)}_\epsilon}dt_{n-1}dt_n\,\del_{t_n}\bigl[G(t_{n-1},t_n)\,c(t_n)\bigr]\\
&=\,n\dcirc \left(V(a,b)\right)^{n-1}\bigl[cV(b)-cV(a)\bigr]\dcirc \\
&\qquad\qquad\qquad+n(n-1)\dcirc \left(V(a,b)\right)^{n-2}\dcirc
\, \lim_{\epsilon \to 0}
\int_{\Gamma^{(2)}_\epsilon}dt_{n-1}dt_n\,\del_{t_n}\bigl[G(t_{n-1},t_n)\,c(t_n)\bigr] \,.
\end{split}
\end{equation}
The integral can be evaluated as follows:
\begin{equation}\label{partialGc}
\begin{split}
& \lim_{\epsilon \to 0} \int_{\Gamma^{(2)}_\epsilon} dt_1 dt_2 \,
\partial_{t_2} \bigl[ \, G(t_1,t_2) \, c(t_2) \, \bigr] \\
& = \lim_{\epsilon \to 0} \, \biggl[ \,
\int_a^{b-\epsilon} dt_1 \int_{t_1+\epsilon}^b dt_2 \,
\partial_{t_2} \bigl[ \, G(t_1,t_2) \, c(t_2) \, \bigr]
+ \int_{a+\epsilon}^b dt_1 \int_a^{t_1-\epsilon} dt_2 \,
\partial_{t_2} \bigl[ \, G(t_1,t_2) \, c(t_2) \, \bigr] \,
\biggr] \\
& = \lim_{\epsilon \to 0} \, \biggl[ \,
\int_a^{b-\epsilon} dt_1 \,
\Bigl[ \, G(t_1,b) \, c(b)
- G(t_1,t_1+\epsilon) \, c(t_1+\epsilon) \, \Bigr] \\
& \qquad \qquad
{}+ \int_{a+\epsilon}^b dt_1 \,
\Bigl[ \, G(t_1-\epsilon,t_1) \, c(t_1-\epsilon)
- G(a,t_1) \, c(a) \, \Bigr] \, \biggr] \\
& = \lim_{\epsilon \to 0} \, \biggl[ \,
\int_a^{b-\epsilon} dt \, G(t,b) \, c(b)
- \int_{a+\epsilon}^b dt \, G(a,t) \, c(a)
+ \int_a^{b-\epsilon} dt \, G(t,t+\epsilon) \,
\Bigl[ \, c(t) - c(t+\epsilon) \, \Bigr] \, \biggr] \,.
\end{split}
\end{equation}
The calculation of the last term is essentially the same
as that of~(\ref{collision})
without the term involving $\widetilde{V}$:
\begin{equation}
\begin{split}
& \int_a^{b-\epsilon} dt \, G(t,t+\epsilon) \,
\Bigl[ \, c(t) - c(t+\epsilon) \, \Bigr]
= \int_a^{b-\epsilon} dt \,
\Big( \, \frac{1}{\epsilon^2} + \ord{\epsilon^0} \, \Bigr)
\Bigl[ \, - \epsilon \, \partial c(t)
- \frac{\epsilon^2 }{2} \, \partial^2 c(t) + \ord{\epsilon^3} \,
\Bigr] \\
& = -\frac{1}{\epsilon} \, c(b) +\frac{1}{\epsilon} \, c(a)
+\frac{1}{2} \, \partial c(b) +\frac{1}{2} \, \partial c(a)
+ \ord{\epsilon} \,.
\end{split}
\end{equation}
We thus find
\begin{equation}
\begin{split}
& \lim_{\epsilon \to 0} \int_{\Gamma^{(2)}_\epsilon} dt_1 dt_2 \,
\partial_{t_2} \bigl[ \, G(t_1,t_2) \, c(t_2) \, \bigr] \\
& = \lim_{\epsilon \to 0} \, \biggl[ \,
\int_a^{b-\epsilon} dt \, G(t,b) -\frac{1}{\epsilon} \,
\biggr] \, c(b)
- \lim_{\epsilon \to 0} \, \biggl[ \,
\int_{a+\epsilon}^b dt \, G(a,t) -\frac{1}{\epsilon} \,
\biggr] \, c(a)
+\frac{1}{2} \, \partial c(b) +\frac{1}{2} \, \partial c(a) \\
& = \langle \, V(a,b) \, V(b) \, \rangle_r \, c(b)
-\langle \, V(a) \, V(a,b) \, \rangle_r \,c(a)
+\frac{1}{2} \, \partial c(b) +\frac{1}{2} \, \partial c(a) \,,
\end{split}
\end{equation}
where we have used (\ref{contrVVab}) and (\ref{renVabV}).
Combining this and (\ref{pullin2}),
the result can be written as follows:
\begin{equation}\label{Q_B-exp}
\begin{split}
Q_B \cdot \dcirc e^{\lambda V(a,b)} \dcirc
& = \lambda \, \dcirc e^{\lambda V(a,b)} \, cV(b) \dcirc
{}- \lambda \, \dcirc cV(a) \, e^{\lambda V(a,b)} \dcirc \\
& \quad ~ + \lambda^2 \, \langle \, V(a,b) \, V(b) \, \rangle_r \,
\dcirc e^{\lambda V(a,b)} \, c(b) \dcirc
{}- \lambda^2 \, \langle \, V(a) \, V(a,b) \, \rangle_r \,
\dcirc c(a) \, e^{\lambda V(a,b)} \dcirc \\
& \quad ~ + \frac{\lambda^2}{2} \,
\dcirc e^{\lambda V(a,b)} \, \partial c(b) \dcirc
{}+ \frac{\lambda^2}{2} \,
\dcirc \partial c(a) \, e^{\lambda V(a,b)} \dcirc \,.
\end{split}
\end{equation}
Note that the structures
\begin{equation}
\dcirc \Bigl(V(a)
+\lambda \, \langle \, V(a) \, V(a,b) \, \rangle_r\Bigr) \,
e^{\lambda V(a,b)} \dcirc \,, \qquad
\dcirc e^{\lambda V(a,b)} \, \Bigl(V(b)
+\lambda \, \langle \, V(a,b) \, V(b) \, \rangle_r\Bigr) \dcirc
\end{equation}
of $[\,V(a)\,e^{\lambda V(a,b)}\,]_r$
and $[\,e^{\lambda V(a,b)}\,V(b)\,]_r$
defined in~(\ref{renVexpV}) emerged naturally.
Therefore, the BRST transformation of
$[ \, e^{\lambda V(a,b)} \, ]_r$
can be written using the definitions~(\ref{renVexpV}) as follows:
\begin{equation}\label{QexpV}
Q_B\cdot[\,e^{\lambda V(a,b)}\,]_r\,
=\,\Bigl[\, e^{\lambda V(a,b)}\Bigl(\lambda
cV(b)+\frac{\lambda^2}{2}\,\del c(b)\Bigr)\,\Bigr]_r
-\Bigl[\,\Bigl(\lambda cV(a)-\frac{\lambda^2}{2}\,\del c(a)\Bigr)
e^{\lambda V(a,b)}\,\Bigr]_r\,.
\end{equation}
We have thus verified the assumption~(\ref{1})  on the BRST transformation
and determined the operators $O_L$ and $O_R$ to be
\begin{equation}\label{explicitOLOR}
\begin{split}
O_R=\lambda \, cV+\frac{\lambda^2}{2}\,\del c\,,\qquad
O_L
=\lambda \, cV-\frac{\lambda^2}{2}\,\del c \,,
\end{split}
\end{equation}
or equivalently
\begin{equation}
\begin{split}
O_R^{(1)} = O_L^{(1)} = cV \,, \qquad
O_R^{(2)} = -O_L^{(2)} = \frac{1}{2}\,\del c \,, \qquad
O_R^{(n)} = O_L^{(n)} = 0 \quad \text{for} \quad n \ge 3 \,.
\end{split}
\end{equation}
With these expressions for $O_R$ and $O_L$
and the explicit forms
of $[\,e^{\lambda V(a,b)} \,]_r$
and $[\, V (a) \, e^{\lambda V(a,b)} \,]_r$
given in~(\ref{renexpV}) and~(\ref{renVexpV}),
$\Psi_L$ and $\Psi$ can be explicitly constructed
for the class of marginal deformations
satisfying the finiteness condition~(\ref{finiteness})
in~\S~\ref{class}.
If all the assumptions~(\ref{1})--(\ref{6}) stated in the introduction
are satisfied, $\Psi_L$ and $\Psi$ are guaranteed to solve
the equation of motion.
The locality assumption~(\ref{5})
for the operator $[ \, O_L(a) \, e^{\lambda V(a,b)} \,]_r$
is satisfied because of~(\ref{rewrite5}),~(\ref{rewrite5again}),
and~(\ref{explicitOLOR}).
We have thus verified the assumptions 
(\ref{1}), (\ref{5}), and (\ref{6}).
We prove the remaining assumptions of replacement (\ref{3}) and factorization (\ref{4})
in appendix~\ref{explicit_proof}
and the assumption (\ref{2})  on the BRST transformation
in appendix~\ref{appQOLexpV}.

\subsection{Conformal properties of
$[ \, O_L(a) \, e^{\lambda \, V(a,b)} \, ]_r$}
\label{conformal}

The operator $O_L(a)$ always appears
in the combination
$[\,O_L(a) \, e^{\lambda V(a,b)}\,\ldots\,]_r$
with some $b$.
Similarly, the operator $O_R(b)$ always appears
in the combination
$[\,\ldots \, e^{\lambda V(a,b)} \, O_R(b)\,]_r$
with some $a$.
Correspondingly, the operators $O_L^{(l)}(a)$
and $O_R^{(r)}(b)$ always appear in the form
\begin{equation}
\bigl[ \, \sum_{l=1}^n \, O_L^{(l)}(a) \,
V^{(n-l)}(a,b) \ldots \, \bigr]_r \,, \qquad
\bigl[ \, \ldots \sum_{r=1}^n \, V^{(n-r)}(a,b) \,
O_R^{(r)}(b) \, \bigr]_r \,,
\end{equation}
or
\begin{equation}
\bigl[ \, \sum_{l+r \le n} \, O_L^{(l)}(a) \,
V^{(n-l-r)}(a,b) \, O_R^{(r)}(b) \, \bigr]_r \,.
\end{equation}
We do not need to require the existence of $O_L(a)$
and $O_R(b)$ as independent operators,
and we only need to define
$[ \, O_L(a) \, e^{\lambda V(a,b)} \ldots \, ]_r$
and $[ \, \ldots e^{\lambda V(a,b)} \, O_R(b) \, ]_r$
expanded in $\lambda$.
In fact, operators in these forms are expected
to transform covariantly under conformal transformations.
Let us consider conformal transformations
of the operator $[ \, O_L(a) \, e^{\lambda \, V(a,b)} \, ]_r$
we determined in~\S~\ref{BRST}
to the first nontrivial order in $\lambda$.

When we change boundary conditions
on a segment between $a$ and $b$ of the real axis,
the two end points $a$ and $b$ behave
as primary fields
under conformal transformations,
and they are often described in terms of boundary-condition
changing operators.
We thus expect that the operator $[ \, e^{\lambda V(a,b)} \, ]_r$
is mapped by a conformal transformation $g(z)$
to $g'(a)^{h(\lambda)} \, g'(b)^{h(\lambda)} \,
[ \, e^{\lambda V(g(a), \, g(b))} \, ]_r$,
where $h(\lambda)$ can be interpreted
as the dimension of the boundary-condition changing operator.
For simplicity, we assume that the segment between $a$ and $b$
is mapped by $g(z)$ to a segment on the real axis
so that the operator
$[ \, e^{\lambda V(g(a), \, g(b))} \, ]_r$
is well defined without any generalization.
Since the BRST transformation maps a primary field
to another primary field of the same dimension,
we also expect that the operator
$[ \, O_L(a) \, e^{\lambda V(a,b)} \, ]_r$
transforms covariantly and is mapped by $g(z)$
as
\begin{equation}\label{gmapsOLexpVab}
g\circ [ \, O_L(a) \, e^{\lambda V(a,b)} \, ]_r =
g'(a)^{h(\lambda)} \, g'(b)^{h(\lambda)} \,
[ \, O_L (g(a)) \, e^{\lambda V(g(a), \, g(b))} \, ]_r\,.
\end{equation}
To linear order in $\lambda$,
the conformal transformation is
\begin{equation}
g \circ \bigl[ \, \lambda \, cV(a) + \ord{\lambda^2} \, \bigr]
= \lambda \, cV(g(a)) + \ord{\lambda^2}
\end{equation}
and is consistent with
(\ref{gmapsOLexpVab}) for $h(\lambda) = \ord{\lambda}$.
At~$\ord{\lambda^2}$, we have
\begin{equation}\label{O_L-V}
[ \, O^{(1)}_L(a) \, V(a,b) \, ]_r+[ \, O^{(2)}_L(a)\, ]_r
= [ \, cV(a) \, V(a,b) \, ]_r
- \frac{1}{2} \, \partial c(a) \,.
\end{equation}
The operator $\partial c$ is not
a primary field
and thus the second term of (\ref{O_L-V})
does not transform covariantly
under conformal transformations.
In fact, the first term does not transform covariantly either
but the sum $[ \, O^{(1)}_L(a) \, V(a,b) \, ]_r+[ \, O^{(2)}_L(a)\, ]_r$
{\it does} transform covariantly.
The operator $\bigl[\,V(a)\,V(a,b)\,\bigr]_r$
is mapped by $g(z)$ as follows:
\begin{equation}
\begin{split}
& g \circ \bigl[ \, V(a) \, V(a,b) \, \bigr]_r
= \lim_{\epsilon \to 0} \, \biggl[ \,
\int_{a+\epsilon}^{b} dt \,
g'(a) \, V\bigl(g(a)\bigr) \, g'(t) \, V\bigl(g(t)\bigr)
-\frac{1}{\epsilon} \, \biggr] \\
& = \lim_{\epsilon \to 0} \, \biggl[ \,
g'(a) \int_{g(a+\epsilon)}^{g(b)} d\tilde{t}\,\,
V\bigl(g(a)\bigr)\,V\bigl(\tilde{t}\bigr)
-\frac{1}{\epsilon} \, \biggr] \\
& = g'(a) \, \lim_{\epsilon \to 0} \, \biggl[ \,
\int_{g(a+\epsilon)}^{g(b)} d\tilde{t}\,\,
V\bigl(g(a)\bigr)\,V\bigl(\tilde{t}\bigr)
-\frac{1}{g(a+\epsilon)-g(a)} \, \biggr]
+ \lim_{\epsilon \to 0} \, \biggl[ \,
\frac{g'(a)}{g(a+\epsilon)-g(a)}
-\frac{1}{\epsilon} \, \biggr] \\
& = g'(a) \, \bigl[ \, V\bigl(g(a)\bigr)\,\,
V\bigl(g(a), \, g(b)\bigr)\,\bigr]_r
-\frac{g''(a)}{2 \, g'(a)} \,,
\end{split}
\end{equation}
where $\tilde{t} = g(t)$.
If we compare this with
\begin{equation}
g \circ \partial c(a)
= \frac{d}{da} \,
\biggl[ \, \frac{c\bigl(g(a)\bigr)}{g'(a)} \, \biggr]
= \partial c\bigl(g(a)\bigr)
- \frac{g''(a)}{g'(a)^2} \, c\bigl(g(a)\bigr) \,,
\end{equation}
we find
\begin{equation}
\begin{split}
& g \circ [ \, cV(a) \, V(a,b) \, ]_r
- g \circ \frac{\partial c(a)}{2} \\
& = \bigl[ \, cV\bigl(g(a)\bigr)\,\, V\bigl(g(a),\,g(b)\bigr)\,\bigr]_r
-\frac{g''(a)}{2 \, g'(a)^2} \, c\bigl(g(a)\bigr)
-\frac{\del c\bigl(g(a)\bigr)}{2}
+ \frac{g''(a)}{2 \, g'(a)^2} \, c\bigl(g(a)\bigr)\\
& = \bigl[ \, cV\bigl(g(a)\bigr)\,\, V\bigl(g(a),\,g(b)\bigr)\,\bigr]_r
-\frac{\del c\bigl(g(a)\bigr)}{2} \,.
\end{split}
\end{equation}
This is consistent with~(\ref{gmapsOLexpVab})
at $\ord{\lambda^2}$
with $h(\lambda) = \ord{\lambda^2}$.
Note that the coefficient of the second term in (\ref{O_L-V})
had to be $-1/2$ for the noncovariant term to be canceled.
Each of these two operators $[ \, O^{(1)}_L(a) \, V(a,b) \, ]_r$
and $[ \, O^{(2)}_L(a)\, ]_r$ defined on ${\cal W}_n$ is invariant
when embedded in ${\cal W}_m$.
Thus any linear combination of the two
is invariant under the embedding from ${\cal W}_n$ to ${\cal W}_m$,
but only the combination
$[ \, O^{(1)}_L(a) \, V(a,b) \, ]_r + [ \, O^{(2)}_L(a)\, ]_r$
transforms covariantly under conformal transformations.
Although the covariance
of $[ \, e^{\lambda V(a,b)} \, ]_r$
and $[ \, O_L(a) \, e^{\lambda V(a,b)} \, ]_r$
under conformal transformations
is not required for the solution
to satisfy the equation of motion,
this calculation provides a nontrivial consistency check
of our result for the operator $O_L$.

\section{String field theory around the deformed background}
\label{discussion_deformed}
\setcounter{equation}{0}

\subsection{Action}

Now that we have constructed solutions
for general marginal deformations,
let us expand the string field theory action
around the solutions.
The string field theory action is given by
\begin{equation}
S[\Psi]
= {}- \frac{1}{g^2} \, \biggl[ \,
\frac{1}{2} \, \langle \, \Psi \,,\, Q_B \Psi \, \rangle
+\frac{1}{3} \, \langle \, \Psi \,,\, \Psi \ast \Psi \, \rangle \,
\biggr] \,,
\end{equation}
where $g$ is the open string coupling constant.
In the case of a D25-brane in flat spacetime,
$g$ is related to the D25-brane tension $T_{25}$
as $T_{25} = 1 / (2 \pi^2 g^2)$\,.
We shift the string field $\Psi$ as
\begin{equation}
\Psi=\Psi_\lambda+\delta\Psi \,,
\end{equation}
where the solution $\Psi_\lambda$ is
\begin{equation}\label{Psi_lambda}
\begin{split}
\Psi_\lambda
&= \frac{1}{2} \, \biggl[ \, \frac{1}{\sqrt{U}} \ast \Psi_L
\ast \sqrt{U} + \sqrt{U} \ast \Psi_R \ast \frac{1}{\sqrt{U}}
+ \frac{1}{\sqrt{U}} \ast Q_B \sqrt{U}
- Q_B \sqrt{U} \ast \frac{1}{\sqrt{U}} \, \biggr] \\
&= \frac{1}{2} \, \biggl[ \, \frac{1}{\sqrt{U}} \ast (A_L+A_R) \ast
\frac{1}{\sqrt{U}} +\frac{1}{\sqrt{U}} \ast Q_B \sqrt{U}
- Q_B \sqrt{U} \ast \frac{1}{\sqrt{U}} \, \biggr]\,.
\end{split}
\end{equation}
We then expand the action and obtain
\begin{equation}\label{deformedaction1}
\begin{split}
S[\Psi] & = S[\Psi_\lambda]+S[\delta\Psi]
-\frac{1}{g^2} \, \langle \, \delta\Psi \,,\,
\Psi_\lambda \ast \delta\Psi \, \rangle\\
& = S[\Psi_\lambda]+S[\delta\Psi]
-\frac{1}{2g^2} \, \Bigl[ \,
\langle \, \delta\Psi \, , \,
\frac{1}{\sqrt{U}} \ast (A_L+A_R)\ast\frac{1}{\sqrt{U}}
\ast\delta\Psi\,\rangle\\
& \quad ~ \hskip 4cm+\langle \, \delta\Psi \, , \,
\frac{1}{\sqrt{U}} \ast Q_B \sqrt{U} \ast \delta \Psi \, \rangle
{}- \langle \, \delta\Psi \, , \,
Q_B \sqrt{U} \ast \frac{1}{\sqrt{U}}\ast\delta\Psi \, \rangle \,
\Bigr] \,.
\end{split}
\end{equation}
The term linear in $\delta\Psi$ vanishes
because $\Psi_\lambda$ satisfies the equation of motion.
The term $S[\Psi_\lambda]$ only shifts the action
by an overall constant.
In fact, it should vanish for solutions
corresponding to exactly marginal deformations.
The structure of the action suggests
the following field redefinition:
\begin{equation}
\Phi\equiv\sqrt{U}\ast\delta\Psi\ast\sqrt{U}\qquad\Longrightarrow\qquad
\delta\Psi=\frac{1}{\sqrt{U}} \ast \Phi\ast \frac{1}{\sqrt{U}} \,.
\end{equation}
The term $S[\delta \Psi]$ can be expressed
in terms of the new variable $\Phi$ as follows:
\begin{equation}\label{SdeltaPsi}
\begin{split}
S[\delta\Psi] &=
S\biggl[ \, \frac{1}{\sqrt{U}} \ast \Phi\ast \frac{1}{\sqrt{U}} \,
\biggr] \\
& = {}-\frac{1}{2g^2}\Bigl\langle \,
\frac{1}{\sqrt{U}}\ast \Phi\ast\frac{1}{\sqrt{U}} \,,
\,Q_B\cdot\biggl[ \, \frac{1}{\sqrt{U}} \ast \Phi\ast
\frac{1}{\sqrt{U}} \, \biggr] \, \Bigr\rangle
{}-\frac{1}{3g^2}\bigl\langle \, \Phi \, , \,U^{-1}
\ast\Phi\ast U^{-1}\ast\Phi\ast U^{-1} \, \bigr\rangle\\
& = {}- \frac{1}{2g^2}\Bigl\langle \Phi \,,\,
U^{-1}\ast Q_B \Phi\ast U^{-1} \, \Bigr\rangle
-\frac{1}{3g^2}\bigl\langle \, \Phi \, , \,U^{-1}
\ast\Phi\ast U^{-1}\ast\Phi\ast U^{-1} \, \bigr\rangle\\
& \quad ~ {}-\frac{1}{2g^2}\Bigl\langle \, \frac{1}{\sqrt{U}}
\ast \Phi\ast\frac{1}{\sqrt{U}} \,,\,
Q_B\frac{1}{\sqrt{U}} \ast \Phi\ast
\frac{1}{\sqrt{U}} \, \Bigr\rangle
+\frac{1}{2g^2}\Bigl\langle \, \frac{1}{\sqrt{U}}\ast
\Phi\ast\frac{1}{\sqrt{U}} \,,\,\frac{1}{\sqrt{U}} \ast
\Phi\ast Q_B\frac{1}{\sqrt{U}} \, \Bigr\rangle \,.
\end{split}
\end{equation}
Using the identity
\begin{equation}
Q_B \frac{1}{\sqrt{U}}
= {}- \frac{1}{\sqrt{U}} \ast Q_B\sqrt{U}
\ast \frac{1}{\sqrt{U}} \,,
\end{equation}
it is easy to see that the last line of~(\ref{SdeltaPsi})
precisely cancels the last two terms
on the right-hand side of~(\ref{deformedaction1}).
The action around the deformed background in terms of $\Phi$
is thus given by
\begin{equation}\label{deformedaction2}
\begin{split}
S[\Psi]=S[\Psi_\lambda]&
{}-\frac{1}{2g^2}\Bigl[ \, \bigl\langle \, \Phi \,,\,
U^{-1}\ast Q_B \Phi\ast U^{-1} \, \bigr\rangle
+\bigl\langle \, \Phi  \, , \, U^{-1}\ast (A_L+A_R)\ast
U^{-1}\ast \Phi\ast U^{-1} \, \bigr\rangle \, \Bigr]\\
&{}-\frac{1}{3g^2} \, \bigl\langle \, \Phi \, , \,U^{-1}
\ast\Phi\ast U^{-1}\ast\Phi\ast U^{-1} \, \bigr\rangle\,.
\end{split}
\end{equation}
Let us now introduce the following deformed algebraic structures:
\begin{equation}\label{algstruc}
\begin{split}
A \star B & \equiv A \ast U^{-1} \ast B \,, \\
{\cal Q} A & \equiv
Q_B A + A_L \star A -(-1)^A \, A \star A_R
= Q_B A + \Psi_L \ast A -(-1)^A \, A \ast \Psi_R \,, \\
\langle\langle \, A, B \, \rangle\rangle & \equiv
\langle \, A, U^{-1} \ast B \ast U^{-1} \, \rangle \,.
\end{split}
\end{equation}
As $U=1+\ord{\lambda^2}$, $A_L = \ord{\lambda}$,
and $A_R = \ord{\lambda}$,
these structures reduce to the original star product~$*$,
BRST operator~$Q_B$, and inner product~$\langle\,\,,\,\rangle$
when $\lambda\to 0$.
The shifted action ${\cal S}[\Phi] \equiv S[\Psi]-S[\Psi_\lambda]$
in terms of the new variable $\Phi$ can be written as follows:
\begin{equation}
{\cal S}[\Phi] = {}-\frac{1}{g^2} \, \biggl[ \,
\frac{1}{2} \, \langle\langle \, \Phi \,,\,
{\cal Q}\Phi \, \rangle\rangle
+\frac{1}{3} \, \langle\langle \, \Phi \, , \,
\Phi\star\Phi \, \rangle\rangle \, \biggr] \,,
\end{equation}
where we have used
\begin{equation}
\begin{split}
& \langle \, \Phi  \, , \, U^{-1}\ast (A_L+A_R) \ast U^{-1}\ast
\Phi\ast U^{-1} \, \rangle\\
&=\langle \, \Phi  \, , \, U^{-1}\ast A_L\ast U^{-1}\ast
\Phi\ast U^{-1} \, \rangle
\,+\,\langle \, \Phi  \, , \, U^{-1}\ast
\Phi\ast U^{-1}\ast A_R\ast U^{-1} \, \rangle\,.
\end{split}
\end{equation}
Thus string field theory
around the deformed background can be described
by the star product $\star\,$, the operator ${\cal Q}$,
and the inner product
$\langle\langle\,\,,\,\rangle\rangle$.
Note that $\sqrt{U}$ and $1/\sqrt{U}$ completely disappeared
and the action is written in terms of $U^{-1}$, $A_L$, and $A_R$.

\subsection{Properties of algebraic structures around the deformed background}\label{appdeformed}

Let us verify that the new algebraic structures obey
the following relations
necessary for a consistent formulation of string field theory:
\begin{eqnarray}
{\cal Q}^2 A &=& 0 \,,
\label{Q2A}\\
{\cal Q}\, ( A \star B )
&=& ( {\cal Q} A ) \star B + (-1)^A \, A \star ( {\cal Q} B ) \,,
\label{QA*B}\\
\langle\langle \, A, B \, \rangle\rangle
& =& (-1)^{AB} \, \langle\langle \, B, A \, \rangle\rangle \,,
\label{ABBA}\\
\langle\langle \, {\cal Q} A, B \, \rangle\rangle
& =& -(-1)^{A} \langle\langle \, A, {\cal Q} B \,\rangle\rangle \,,
\label{QAB}\\
\langle\langle \, A, B \star C \, \rangle\rangle
& =& \langle\langle \, A \star B , C \, \rangle\rangle \,.
\label{AB*C}
\end{eqnarray}
Furthermore, we show that
the generalized wedge states $U_\alpha$ satisfy
\begin{equation}\label{QUalpha}
{\cal Q} \, U_\alpha =0 \,.
\end{equation}

Let us begin with~(\ref{Q2A}).
It follows from the definition of ${\cal Q}$ that
\begin{equation}
\begin{split}
{\cal Q}^2 A & =
{\cal Q} \, \bigl[ \, Q_B A
+ \Psi_L \ast A -(-1)^A \, A \ast \Psi_R \, \bigr]\\
&= Q_B^2 A+Q_B \Psi_L\ast A-\Psi_L\ast Q_B A-(-1)^A \, Q_B A \ast \Psi_R- A \ast Q_B\Psi_R\\
& \quad ~ +\Psi_L\ast\Bigl(Q_B A + \Psi_L \ast A -(-1)^A \, A \ast \Psi_R\Bigr)
+(-1)^A\,\Bigl(Q_B A + \Psi_L \ast A -(-1)^A \, A \ast \Psi_R\Bigr)\ast \Psi_R\,.
\end{split}
\end{equation}
Using $Q_B^2=0$ and the equation of motion for $\Psi_L$ and $\Psi_R$,
all the terms cancel and we find
${\cal Q}^2 A=0$.

Similarly, we can prove~(\ref{QA*B}) as follows:
\begin{equation}
\begin{split}
{\cal Q}_B \, ( A \star B )
& = Q_B A\ast U^{-1}\ast B+(-1)^A A\ast Q_BU^{-1}\ast B
+ (-1)^A A\ast U^{-1}\ast Q_BB \\
& \quad ~ + \Psi_L \ast A \ast U^{-1}\ast B
-(-1)^{A}(-1)^{B} \, A\ast U^{-1}\ast B \ast \Psi_R\\
& = {\cal Q}  A \star B +(-1)^A A \star{\cal Q} B  \\
& \quad ~ +(-1)^A A\ast Q_BU^{-1}\ast B
+(-1)^A A\ast \Psi_R \ast U^{-1}\ast B -(-1)^{A} \, A\ast U^{-1}\ast \Psi_L\ast B \,.
\end{split}
\end{equation}
The terms in the last line cancel because of the identity
\begin{equation}\label{QU-1}
Q_BU^{-1}={}-U^{-1}\ast Q_BU\ast U^{-1}
=U^{-1}\ast(A_L-A_R)\ast U^{-1}=U^{-1}\ast\Psi_L-\Psi_R\ast U^{-1}\,.
\end{equation}
This completes the proof
of~(\ref{QA*B}).

It is easy to verify~(\ref{ABBA}) using the properties
of the inner product $\langle\,\,,\,\rangle$:
\begin{equation}
\begin{split}
\langle\langle \, A, B \, \rangle\rangle
&=\langle \, A, U^{-1}\ast B\ast U^{-1} \, \rangle\\
&=\langle \, A\ast U^{-1}, B\ast U^{-1} \, \rangle\\
&=(-1)^{AB}\langle \, B\ast U^{-1}, A\ast U^{-1} \, \rangle\\
&=(-1)^{AB}\langle \, B, U^{-1}\ast A\ast U^{-1} \, \rangle\\
&=(-1)^{AB}\langle\langle \, B, A \, \rangle\rangle\,.
\end{split}
\end{equation}
To show~(\ref{QAB}), we use the corresponding identity of $Q_B$ and
the properties of $\langle\,\,,\,\rangle$. We find
\begin{equation}
\begin{split}
\langle\langle \, {\cal Q}A, B \, \rangle\rangle
& =\langle \, Q_B A + \Psi_L \ast A -(-1)^A \, A \ast \Psi_R\,,\,
U^{-1}\ast B\ast U^{-1} \, \rangle\\
&= {}-(-1)^A\langle \,  A \,,\, Q_B U^{-1}\ast B\ast U^{-1}
+ U^{-1}\ast Q_B B\ast U^{-1}
+(-1)^BU^{-1}\ast B\ast Q_B U^{-1} \, \rangle\\
& \quad ~ + (-1)^{A}(-1)^{B}
\langle \,  A\,,\, U^{-1}\ast B\ast U^{-1}\ast \Psi_L \, \rangle
-(-1)^A \langle \, A\,,\,
\Psi_R\ast U^{-1}\ast B\ast U^{-1} \, \rangle\,.
\end{split}
\end{equation}
Using the identity~(\ref{QU-1}), we obtain
\begin{equation}
\begin{split}
\langle\langle \, {\cal Q}A, B \, \rangle\rangle
=&-(-1)^A \, \langle \, A\,,\, U^{-1} \ast
\bigl( \, Q_B B + \Psi_L \ast B -(-1)^B \, B \ast \Psi_R \, \bigr)
\ast U^{-1} \, \rangle\\
=&-(-1)^A \, \langle\langle \, A, {\cal Q}B \, \rangle\rangle\,.
\end{split}
\end{equation}
Finally, the relation~(\ref{AB*C}) follows from the definitions
of the deformed structures
and the property of the inner product $\langle \, , \, \rangle$:
\begin{equation}
\begin{split}
\langle\langle \, A, B\star C \, \rangle\rangle
& =\langle \, A\,,\,
U^{-1}\ast B\ast U^{-1}\ast C \ast U^{-1}\, \rangle \\
& =\langle \, A \ast U^{-1}\ast B\,,\,
U^{-1}\ast C \ast U^{-1}\, \rangle
=\langle\langle \, A\star B,C \, \rangle\rangle\,.
\end{split}
\end{equation}
We have thus shown that the deformed algebraic structures satisfy
all the algebraic relations required for a consistent
formulation of string field theory.

Let us now show the equation~(\ref{QUalpha}), namely, that
the generalized wedge states $U_\alpha$ are
annihilated by ${\cal Q}$.
We define the generalizations $A_{L,\alpha}$ and $A_{R,\alpha}$
of $A_L$ and $A_R$, respectively, by
\begin{equation}
A_{L,\alpha} \equiv
\sum_{n=1}^\infty \, \lambda^n \, A_{L,\alpha}^{(n)} \,, \qquad
A_{R,\alpha} \equiv
\sum_{n=1}^\infty \, \lambda^n \, A_{R,\alpha}^{(n)} \,
\end{equation}
for $\alpha\geq0$, where
\begin{equation}
\begin{split}
    \langle \, \phi \,, A_{L,\alpha}^{(n)} \, \rangle &=
    \sum_{l=1}^n \langle \, f \circ \phi (0) \, [ \, O_L^{(l)} (1) \, V^{(n-l)} (1,n+\alpha) \, ]_r \, \rangle_{{\cal W}_{n+\alpha}},\\
    \langle \, \phi \,, A_{R,\alpha}^{(n)} \, \rangle &=
     \sum_{r=1}^n \langle \, f \circ \phi (0) \, [  \, V^{(n-r)} (1,n+\alpha) \, O_R^{(r)} (n+\alpha)\, ]_r \, \rangle_{{\cal W}_{n+\alpha}}  \,.
\end{split}
\end{equation}
Note that $A_L=A_{L,0}$ and $A_R=A_{R,0}$.
The states $A_{L,\alpha}$ and $A_{R,\alpha}$ satisfy
the following relations:
\begin{equation}
    Q_B U_\alpha = A_{R,\alpha}-A_{L,\alpha}\,,\qquad
    A_{L,\alpha+\beta}=A_{L,\alpha}\ast U^{-1}\ast U_\beta\,,\qquad
    A_{R,\alpha+\beta}=U_\alpha\ast U^{-1}\ast A_{R,\beta}\,,
\end{equation}
which are generalizations
of $Q_B U=A_R-A_L$ and
$U_{\alpha+\beta} = U_\alpha\ast U^{-1}\ast U_\beta$.
The first relation immediately follows from
the assumption~(\ref{1}).
The second and third relations can be shown
using the assumptions~(\ref{3})--(\ref{5})
as in the proofs of
$U_{\alpha+\beta} = U_\alpha\ast U^{-1}\ast U_\beta$
and ${}- Q_B A_L = A_L \ast U^{-1} \ast A_R$
in~\S~\ref{singular_general_proof}
and~appendix~\ref{QBAL-appendix}.
Using these relations,
it is easy to show that ${\cal Q} \, U_\alpha$ vanishes:
\begin{equation}
\begin{split}
{\cal Q} \, U_\alpha &=A_{R,\alpha}-A_{L,\alpha} + \Psi_L \ast U_\alpha - \, U_\alpha \ast \Psi_R  \\
&=U_\alpha\ast U^{-1}\ast A_{R}-A_L\ast U^{-1}\ast U_\alpha   + A_L\ast U^{-1} \ast U_\alpha - \, U_\alpha \ast U^{-1}\ast A_R \\
&=0\,.
\end{split}
\end{equation}

The state $U_1$ is expected to play the role of the $SL(2,R)$-invariant
vacuum in the deformed theory,
and $U=U_0$ is the identity state
of the deformed star algebra.
In fact,
\begin{equation}
 U \star A = U \ast U^{-1} \ast A = A \,, \qquad
 A \star U = A \ast U^{-1} \ast U = A \,.
\end{equation}

\section{Discussion}\label{discussion}
\setcounter{equation}{0}

The main result of the paper is
the construction of analytic solutions
of open bosonic string field theory
for general marginal deformations.
We presented a procedure to construct a solution from the operator
$[ \, e^{\lambda V(a,b)} \, ]_r$
satisfying the set of assumptions stated in the introduction.
We believe that all of these assumptions are satisfied
for any exactly marginal deformation
and are thus necessary conditions
for exact marginality of the deformation.
We also believe that the set of assumptions
provides a sufficient condition for marginality
to all orders in $\lambda$
because we have succeeded in constructing
solutions of string field theory.
We regard this new characterization
of exact marginality as another important result of the paper,
and we hope that our approach
motivated by string field theory
will provide new perspectives
on the study of marginal deformations.

In section~\ref{singular_explicit} we explicitly constructed
the operator $[ \, e^{\lambda V(a,b)} \, ]_r$
for any marginal operator
satisfying the finiteness condition~(\ref{finiteness}).
We thus believe that the finiteness condition~(\ref{finiteness}) is
a sufficient condition for marginality
to all orders in $\lambda$.
We can actually relax the condition
because we only needed finiteness
of the operator $\dcirc ( V(a,b) )^n \dcirc$ constructed
in~(\ref{dcirc-V^n-dcirc}).
Therefore, we can construct solutions
even if the finiteness condition~(\ref{finiteness})
is violated as long as the operator $\dcirc ( V(a,b) )^n \dcirc$
is well defined for any~$n$.\footnote{
We thank Ashoke Sen for discussions on this point
and for explaining explicit examples.}
It would be an interesting open problem
whether the condition can be further relaxed.
In particular, it is an interesting question
whether the operators
$O_L^{(n)}$ and $O_R^{(n)}$ with $n \ge 3$
can be nonvanishing
by nontrivial collisions of more than two operators.
In~\cite{Recknagel:1998ih},
Recknagel and Schomerus
gave a sufficient condition for exact marginality
which they called {\it self-locality} of the marginal operator.
See \S~2.4 of \cite{Recknagel:1998ih}.
It would be also interesting to investigate the relation
between their characterization of exact marginality
in boundary conformal field theory and ours.

In~\cite{Fuchs:2007yy}, Fuchs, Kroyter and Potting constructed
non-real solutions 
for the marginal deformation
corresponding to turning on the constant mode of the gauge field.
We discuss the relation between their solutions and ours
in appendix~\ref{discussion_FKP} and show that
our solutions $\Psi_L$ and $\Psi_R$ 
for this particular marginal deformation
coincide with theirs.

There are many interesting directions for future work.
It would be interesting to study
the solution corresponding to the deformation
by the cosine potential in detail.
The deformation at the value of $\lambda$
describing lower-dimensional D-branes
is particularly interesting.
In the level-truncation analysis of marginal deformations,
it has been demonstrated that the Siegel gauge condition
is not globally well defined~\cite{Ellwood:2001ne}
and the branch of the marginal deformation
corresponding to turning on the constant mode
of the gauge field truncates
at a finite value of the deformation parameter~\cite{Sen:2000hx}.\footnote{
See \cite{Asano:2006hk, Asano:2006hm} for recent related study.}
It is therefore important to study the convergence property
of the expansion in $\lambda$ for our solutions.

We expect that our work will play a role
in further investigating
background independence in string field theory
by extending previous work~\cite{Sen:1990hh}--\cite{Sen:1993kb}.
We also expect that the generalization of our construction 
to open superstring field theory formulated
by Berkovits~\cite{Berkovits:1995ab}
would be fairly straightforward.
Another important generalization
is the construction of solutions corresponding
to boundary conditions which are not connected
by exactly marginal deformations.
For example, consider the case
where the original CFT flows to a different CFT
by a marginally relevant deformation.
We then expect that the operator $[ \, e^{\lambda V(a,b)} \, ]_r$
satisfying the assumptions~(\ref{1}) and~(\ref{2})
can be constructed at a special value of $\lambda$
and our framework will be useful
in constructing solutions
for such marginally relevant deformations.
Finally,
the approach explored in~\cite{Imamura:2005zm} seems to be
closely related to ours and may be useful in future developments
of our work.

\newpage 

\noindent
{\bf \large Acknowledgments}

\medskip

We would like to thank Ian Ellwood, Leonardo Rastelli,
Volker Schomerus, Ashoke Sen, and Barton Zwiebach
for useful discussions
and for their valuable comments on an earlier version
of the manuscript.
The work of M.K. is supported in part
by the U.S. DOE grant DE-FG02-05ER41360
and by an MIT Presidential Fellowship.

\bigskip

\appendix

\section{Proof of ${}-Q_BA_L=A_L\ast U^{-1}\ast A_R$}
\label{QBAL-appendix}
\setcounter{equation}{0}

In~\S~\ref{singular_general_proof}
we have shown that $U_{l+r}=U_l \ast U^{-1} \ast U_r$ holds
for the general case.
To prove the equation ${}-Q_BA_L=A_L\ast U^{-1}\ast A_R$
in~(\ref{condQAL}),
we have to extend this identity to the case
where $O_L$ and $O_R$
are also inserted.
We first present an explicit proof at $\ord{\lambda^4}$
and then explain how the proof generalizes to all orders.
The equation~(\ref{condQAL}) at $\ord{\lambda^4}$ is
\begin{equation}\label{condQAL-lambda^4}
{}- Q_B A_L^{(4)} = A_L^{(1)} \ast A_R^{(3)}
+ A_L^{(2)} \ast A_R^{(2)} + A_L^{(3)} \ast A_R^{(1)}
- A_L^{(1)} \ast U_0^{(2)} \ast A_R^{(1)} \,.
\end{equation}
We need to prove that
\begin{equation}\label{lambda^4-example}
\begin{split}
& \quad ~ [ \, O_L^{(1)}(1) \, V^{(2)}(1,4) \, O_R^{(1)}(4) \, ]_r
+ [ \, O_L^{(1)}(1) \, V^{(1)}(1,4) \, O_R^{(2)}(4) \, ]_r
+ [ \, O_L^{(2)}(1) \, V^{(1)}(1,4) \, O_R^{(1)}(4) \, ]_r \\
& \quad ~ + [ \, O_L^{(1)}(1) \, O_R^{(3)}(4) \, ]_r
+ [ \, O_L^{(2)}(1) \, O_R^{(2)}(4) \, ]_r
+ [ \, O_L^{(3)}(1) \, O_R^{(1)}(4) \, ]_r \\
& = [ \, W_L^{(1)}(1,1) \, ]_r \,
[ \, W_R^{(3)}(2,4) \, ]_r
+ [ \, W_L^{(2)}(1,2) \, ]_r \,
[ \, W_R^{(2)}(3,4) \, ]_r
+ [ \, W_L^{(3)}(1,3) \, ]_r \,
[ \, W_R^{(1)}(4,4) \, ]_r \\
& \quad ~ {}- [ \, W_L^{(1)}(1,1) \, ]_r \,
[ \, V^{(2)}(2,3) \, ]_r \, [ \, W_R^{(1)}(4,4) \, ]_r \,,
\end{split}
\end{equation}
where we denoted terms of
$[\,O_L(a)\,e^{\lambda V(a,b)}\,]_r$ and
$[\,e^{\lambda V(a,b)}\,O_R(b)\,]_r$ at $\ord{\lambda^n}$ as follows:
\begin{equation}
W_L^{(n)}(a,b) \equiv
\sum_{l=1}^n O_L^{(l)}(a) \, V^{(n-l)}(a,b) \,, \qquad
W_R^{(n)}(a,b) \equiv
\sum_{r=1}^n V^{(n-r)}(a,b) \, O_R^{(r)}(b) \,.
\end{equation}
Recall that $V^{(0)}(a,b) \equiv 1$ even in the limit $b \to a$.
Thus we have $W_L^{(1)}(1,1) = O_L^{(1)}(1)$
and $W_R^{(1)}(4,4) = O_R^{(1)}(4)$.
Here we have used
the locality assumption~(\ref{5})
on $[ \, e^{\lambda V(a,b)} \, ]_r$
and $[ \, O_L (a) \, e^{\lambda V(a,b)} \, ]_r$.
The operator $[ \, e^{\lambda V(a,b)} \, O_R (b) \, ]_r$
defined on ${\cal W}_n$ also takes the same form
when embedded in ${\cal W}_m$ with $m > n$
because $[ \, e^{\lambda V(a,b)} \, O_R (b) \, ]_r
= Q_B \cdot [ \, e^{\lambda V(a,b)} \, ]_r
+ [ \, O_L (a) \, e^{\lambda V(a,b)} \, ]_r$ from
the assumption~(\ref{1}).

We next use the factorization assumption~(\ref{4})
of the following form:
\begin{equation}\label{OLexpexpOR}
[ \, O_L(1) \, e^{\lambda_1 V(1,2)} \,
e^{\lambda_2 V(3,4)} \, O_R(4) \, ]_r
= [ \, O_L(1) \, e^{\lambda_1 V(1,2)} \, ]_r \,
[ \, e^{\lambda_2 V(3,4)} \, O_R(4) \, ]_r \,.
\end{equation}
The operator $O_L(a)$ always appears in the combination
$[ \, O_L(a) \, e^{\lambda V(a,b)} \, \ldots \, ]_r$
with some $b$,
and the value of $\lambda$ for $O_L(a)$ is the same
as the one appearing in the exponential operator.
Similarly, the operator $O_R(b)$ always appears in the combination
$[ \, \ldots \, e^{\lambda V(a,b)} \, O_R(b) \, ]_r$
with some $a$,
and the value of $\lambda$ for $O_R(b)$ is the same
as the one appearing in the exponential operator.
In~(\ref{OLexpexpOR}), for example, the value of $\lambda$
for $O_L(1)$ is $\lambda_1$
and the value of $\lambda$ for $O_R(4)$ is $\lambda_2$.
The relation~(\ref{OLexpexpOR})
at $\ord{\lambda_1^2 \, \lambda_2^2}$ reads
\begin{equation}
[ \, W_L^{(2)}(1,2) \, W_R^{(2)}(3,4) \, ]_r
= [ \, W_L^{(2)}(1,2) \, ]_r \,
[ \, W_R^{(2)}(3,4) \, ]_r \,.
\end{equation}
Since $W_L^{(1)}(a,a) = W_L^{(1)}(a,b)$
and $W_R^{(1)}(b,b) = W_R^{(1)}(a,b)$
for $a < b$,
the operators $[ \, W_L^{(1)}(1,1) \, ]_r$
and $[ \, W_R^{(1)}(4,4) \, ]_r$
can be thought of as the $\ord{\lambda_1}$ term
of $[ \, O_L(1) \, e^{\lambda_1 V(1,1+\alpha)} \, ]_r$
and the $\ord{\lambda_2}$ term
of $[ \, e^{\lambda_2 V(4-\alpha,4)} \, O_R(4) \, ]_r$,
respectively,
with arbitrary $\alpha$ in the range $0 < \alpha < 1$.
Therefore, the right-hand side of~(\ref{lambda^4-example})
can be written using the factorization assumption~(\ref{4}) as follows:
\begin{equation}\label{lambda^4-example-2}
\begin{split}
& \quad ~ [ \, W_L^{(1)}(1,1) \, ]_r \,
[ \, W_R^{(3)}(2,4) \, ]_r
+ [ \, W_L^{(2)}(1,2) \, ]_r \,
[ \, W_R^{(2)}(3,4) \, ]_r
+ [ \, W_L^{(3)}(1,3) \, ]_r \,
[ \, W_R^{(1)}(4,4) \, ]_r \\
& \quad ~ {}- [ \, W_L^{(1)}(1,1) \, ]_r \,
[ \, V^{(2)}(2,3) \, ]_r \, [ \, W_R^{(1)}(4,4) \, ]_r \\
& = [ \, W_L^{(1)}(1,1) \, W_R^{(3)}(2,4) \, ]_r
+ [ \, W_L^{(2)}(1,2) \, W_R^{(2)}(3,4) \, ]_r
+ [ \, W_L^{(3)}(1,3) \, W_R^{(1)}(4,4) \, ]_r \\
& \quad ~ {}- [ \, W_L^{(1)}(1,1) \,
V^{(2)}(2,3) \, W_R^{(1)}(4,4) \, ]_r \,.
\end{split}
\end{equation}
We then apply the replacement assumption~(\ref{3})
of the following forms:
\begin{equation}
\begin{split}
[ \, O_L(1) \, e^{\lambda_1 V(1,1+\alpha)} \,
e^{\lambda_2 V(2,4)} \, O_R(4) \, ]_r
& = [ \, O_L(1) \, e^{\lambda_1 V(1,1+\alpha)} \,
e^{\lambda_2 V(2,3)} \,
e^{\lambda_2 V(3,4)} \, O_R(4) \, ]_r \,, \\
[ \, O_L(1) \, e^{\lambda_1 V(1,3)} \,
e^{\lambda_2 V(4-\alpha,4)} \, O_R(4) \, ]_r
& = [ \, O_L(1) \, e^{\lambda_1 V(1,2)} \,
e^{\lambda_1 V(2,3)} \,
e^{\lambda_2 V(4-\alpha,4)} \, O_R(4) \, ]_r\,,
\end{split}
\end{equation}
where $\alpha$ is again an arbitrary number
in the range $0 < \alpha < 1$.
The first equation at $\ord{\lambda_1\lambda_2^3}$
and the second equation at $\ord{\lambda_1^3\lambda_2}$ give
\begin{equation}
\begin{split}
[ \, W_L^{(1)}(1,1) \, W_R^{(3)}(2,4) \, ]_r
& = [ \, W_L^{(1)}(1,1) \, W_R^{(3)}(3,4) \, ]_r
+ [ \, W_L^{(1)}(1,1) \, V^{(1)}(2,3) \,
W_R^{(2)}(3,4) \, ]_r \\
& \quad ~ + [ \, W_L^{(1)}(1,1) \, V^{(2)}(2,3) \,
W_R^{(1)}(3,4) \, ]_r \,, \\
[ \, W_L^{(3)}(1,3) \, W_R^{(1)}(4,4) \, ]_r
& = [ \, W_L^{(1)}(1,2) \, V^{(2)}(2,3) \,
W_R^{(1)}(4,4) \, ]_r
+ [ \, W_L^{(2)}(1,2) \, V^{(1)}(2,3) \,
W_R^{(1)}(4,4) \, ]_r \\
& \quad ~ + [ \, W_L^{(3)}(1,2) \, W_R^{(1)}(4,4) \, ]_r \,.
\end{split}
\end{equation}
Replacing $W_L^{(1)}(1,1)$ with $W_L^{(1)}(1,2)$
and $W_R^{(1)}(4,4)$ with $W_R^{(1)}(3,4)$,
the right-hand side of~(\ref{lambda^4-example-2})
can be written as follows:
\begin{equation}\label{finalrhs}
\begin{split}
& \quad ~ [ \, W_L^{(1)}(1,1) \, W_R^{(3)}(2,4) \, ]_r
+ [ \, W_L^{(2)}(1,2) \, W_R^{(2)}(3,4) \, ]_r
+ [ \, W_L^{(3)}(1,3) \, W_R^{(1)}(4,4) \, ]_r \\
& \quad ~ {}- [ \, W_L^{(1)}(1,1) \,
V^{(2)}(2,3) \, W_R^{(1)}(4,4) \, ]_r \\
& = [ \, W_L^{(1)}(1,2) \, W_R^{(3)}(3,4) \, ]_r
+ [ \, W_L^{(1)}(1,2) \, V^{(1)}(2,3) \,
W_R^{(2)}(3,4) \, ]_r
+ [ \, W_L^{(1)}(1,2) \, V^{(2)}(2,3) \,
W_R^{(1)}(3,4) \, ]_r \\
& \quad ~
+ [ \, W_L^{(2)}(1,2) \, W_R^{(2)}(3,4) \, ]_r
+ [ \, W_L^{(2)}(1,2) \, V^{(1)}(2,3) \,
W_R^{(1)}(3,4) \, ]_r
+ [ \, W_L^{(3)}(1,2) \, W_R^{(1)}(3,4) \, ]_r \,.
\end{split}
\end{equation}

The terms on the left-hand side of~(\ref{lambda^4-example})
are obtained from the expansion of
$[ \, O_L(1) \, e^{\lambda V(1,4)} \, O_R(4) \, ]_r$
in $\lambda$.
Using the replacement assumption~(\ref{3}), we have
\begin{equation}
[ \, O_L(1) \, e^{\lambda V(1,4)} \, O_R(4) \, ]_r
= [ \, O_L(1) \, e^{\lambda V(1,2)} \, e^{\lambda V(2,3)} \,
e^{\lambda V(3,4)} \, O_R(4) \, ]_r \,.
\end{equation}
By evaluating both sides at $\ord{\lambda^4}$,
the left-hand side of~(\ref{lambda^4-example}) can be written as
\begin{equation}\label{lambda^4-example-3}
\begin{split}
& \quad ~ [ \, O_L^{(1)}(1) \, V^{(2)}(1,4) \, O_R^{(1)}(4) \, ]_r
+ [ \, O_L^{(1)}(1) \, V^{(1)}(1,4) \, O_R^{(2)}(4) \, ]_r
+ [ \, O_L^{(2)}(1) \, V^{(1)}(1,4) \, O_R^{(1)}(4) \, ]_r \\
& \quad ~ + [ \, O_L^{(1)}(1) \, O_R^{(3)}(4) \, ]_r
+ [ \, O_L^{(2)}(1) \, O_R^{(2)}(4) \, ]_r
+ [ \, O_L^{(3)}(1) \, O_R^{(1)}(4) \, ]_r \\
& =
[ \, W_L^{(1)}(1,2) \, W_R^{(3)}(3,4) \, ]_r
+ [ \, W_L^{(1)}(1,2) \, V^{(1)}(2,3) \,
W_R^{(2)}(3,4) \, ]_r
+ [ \, W_L^{(1)}(1,2) \, V^{(2)}(2,3) \,
W_R^{(1)}(3,4) \, ]_r \\
& \quad ~
+ [ \, W_L^{(2)}(1,2) \, W_R^{(2)}(3,4) \, ]_r
+ [ \, W_L^{(2)}(1,2) \, V^{(1)}(2,3) \,
W_R^{(1)}(3,4) \, ]_r
+ [ \, W_L^{(3)}(1,2) \, W_R^{(1)}(3,4) \, ]_r \,.
\end{split}
\end{equation}
We have reproduced~(\ref{finalrhs}) and thus shown
${}- Q_B A_L = A_L \ast U^{-1} \ast A_R$
at $\ord{\lambda^4}$.

We will now show
that this proof can be generalized to
$\ord{\lambda^n}$ for any $n \ge 3$,
while the equation trivially holds for $n=1$ and $n=2$.
Using the replacement assumption~(\ref{3}),
we can rewrite
\begin{equation}
[\,O_L(1)\,e^{\lambda V(1,n)}\,O_R(n)\,]_r
= \bigl[\,O_L(1)\,e^{\lambda V(1,2)}
\prod_{i=2}^{n-2}\,[\,e^{\lambda V(i,i+1)}\,]\,
e^{\lambda V(n-1,n)}\,O_R(n)\,\bigr]_r \,.
\end{equation}
At $\ord{\lambda^n}$, this implies that
the operator insertions for ${}- Q_B A_L^{(n)}$
on ${\cal W}_n$ can be expanded in the basis
\begin{equation}\label{basis}
\Bigl\{ \, \bigl[ \, W_L^{(\ell_1)}(1,2) \,
\prod_{i=2}^{n-2} \, [ \, V^{(\ell_i)}(i,i+1) \, ] \,
W_R^{(\ell_{n-1})}(n-1,n) \, \bigr]_r \, \Bigr\} \,,
\end{equation}
where $\ell_i$'s are non-negative integers
with $\sum_{i=1}^{n-1} \ell_i = n$
and $\ell_1,\, \ell_{n-1}\geq1$.
On the other hand, because of the locality assumption~(\ref{5}),
the terms of $A_L \ast U^{-1} \ast A_R$ at $\ord{\lambda^n}$
can be expressed in terms of products of the form
\begin{equation}\label{WexpW}
[ \, W_L^{(k_1)}(1,b_1) \, ]_r \,
\prod_{j=2}^{m-1} \, [ \, V^{(k_j)}(a_j,b_j) \, ]_r \,
[ \, W_R^{(k_m)}(a_m,n) \, ]_r
\end{equation}
on ${\cal W}_n$,
where positive integers $a_j$, $b_j$  and $k_j$ satisfy
$1\le a_j<b_j\le n$,
$b_j < a_{j+1}$, and $\sum_{j=1}^m k_j = n$.
{}From the factorization assumption~(\ref{4}), we have
\begin{equation}
\begin{split}
[ \, O_L(1) \, e^{\lambda_{1} V(1,b_1)} \,]_r
\prod_{j=2}^{m-1} [\,  e^{\lambda_{j} V(a_j,b_j)}&\,]_r \,
[\,e^{\lambda_{m} V(a_m,n)} \, O_R(n)\,]_r\\
&=
\bigl[ \, O_L(1) \, e^{\lambda_{1} V(1,b_1)} \,
\prod_{j=2}^{m-1} [\,  e^{\lambda_{j} V(a_j,b_j)}\,] \,
e^{\lambda_{m} V(a_m,n)} \,O_R(n)\, \bigr]_r \,.
\end{split}
\end{equation}
At $\ord{\prod_j \lambda^{k_j}}$, this allows us to express~(\ref{WexpW}) as
\begin{equation}
\bigl[ \, W_L^{(k_1)}(1,b_1) \,
\prod_{j=2}^{m-1} \, [ \, V^{(k_j)}(a_j,b_j) \, ] \,
W_R^{(k_m)}(a_m,n) \, \bigr]_r
\end{equation}
on ${\cal W}_n$.
Finally, applying the replacement assumption~(\ref{3})
and using $W_L^{(1)}(1,1)=W_L^{(1)}(1,2)$
and $W_R^{(1)}(n,n)=W_R^{(1)}(n-1,n)$,
the operators can be expanded in the basis~(\ref{basis}).
Now consider the following state for a marginal operator
with regular operator products:
\begin{equation}\label{dummy}
\sum_{l, \, r = 1}^\infty \lambda^{l+r} \,
c_L^{(l)} \, c_R^{(r)} \, U_{l+r} \,,
\end{equation}
where $c_L^{(l)}$ and $c_R^{(r)}$ are parameters.
The operators at $\ord{\lambda^n}$ on ${\cal W}_n$
can be expanded in the basis
\begin{equation}\label{basis-2}
\Bigl\{ \, \omega_L^{(\ell_1)}(1,2) \,
\prod_{i=2}^{n-2} \, [ \, V^{(\ell_i)}(i,i+1) \, ] \,
\omega_R^{(\ell_{n-1})}(n-1,n) \, \Bigr\} \,,
\end{equation}
where
\begin{equation}
\omega_L^{(i)}(1,2) \equiv
\sum_{l=1}^i c_L^{(l)} \, V^{(i-l)}(1,2) \,, \qquad
\omega_R^{(i)}(n-1,n) \equiv
\sum_{r=1}^i c_R^{(i)}\,V^{(i-r)}(n-1,n)  \,,
\end{equation}
and $\ell_i$'s are non-negative integers
with $\sum_{i=1}^{n-1} \ell_i = n$ and $\ell_1,\,\ell_{n-1}\geq1$
as in~(\ref{basis}).
The coefficients when the state~(\ref{dummy})
is expanded in this basis reproduce those of ${}-Q_B A_L$
expanded in the basis~(\ref{basis})
with replacing $W_L^{(i)}$ by $\omega_L^{(i)}$
and $W_R^{(i)}$ by $\omega_R^{(i)}$.
Let us next consider the following state for a marginal operator
with regular operator products:
\begin{equation}\label{dummy-2}
\sum_{l, \, r = 1}^\infty
\Bigl( \, \lambda^l \, c_L^{(l)} \, U_l \, \Bigr)
\ast U^{-1} \ast
\Bigl( \, \lambda^r \, c_R^{(r)} \, U_r \, \Bigr)
\end{equation}
where again $c_L^{(l)}$ and $c_R^{(r)}$ are parameters.
The terms of~(\ref{dummy-2}) at $\ord{\lambda^n}$
can also be expanded in the basis~(\ref{basis-2})
and the coefficients reproduce those of $A_L \ast U^{-1} \ast A_R$
at $\ord{\lambda^n}$ expanded in the basis~(\ref{basis})
with replacing $W_L^{(i)}$ by $\omega_L^{(i)}$
and $W_R^{(i)}$ by $\omega_R^{(i)}$.
The states~(\ref{dummy}) and~(\ref{dummy-2})
are actually equal because of the relation
$U_{l+r} = U_l \ast U^{-1} \ast U_r$:
\begin{equation}
\sum_{l, \, r = 1}^\infty \lambda^{l+r} \,
c_L^{(l)} \, c_R^{(r)} \, U_{l+r}
= \sum_{l, \, r = 1}^\infty
\Bigl( \, \lambda^l \, c_L^{(l)} \, U_l \, \Bigr)
\ast U^{-1} \ast
\Bigl( \, \lambda^r \, c_R^{(r)} \, U_r \, \Bigr) \,.
\end{equation}
We have thus shown that
${}- Q_B A_L = A_L \ast U^{-1} \ast A_R$
to all orders in $\lambda$.

\section{Proof of the assumptions}
\label{proof-appendix}
\setcounter{equation}{0}

In section~\ref{singular_explicit}
we have presented explicit forms
of $[\,e^{\lambda V(a,b)} \,]_r$
and $[\, O_L (a) \, e^{\lambda V(a,b)} \,]_r$,
which are used in constructing $\Psi_L$ and $\Psi$,
for the class of marginal deformations
satisfying the finiteness condition~(\ref{finiteness})
in~\S~\ref{class}.
We have shown that the assumptions
(\ref{1}), (\ref{5}), and (\ref{6})
are satisfied for these operators.
We prove the remaining assumptions
(\ref{2}), (\ref{3}), and (\ref{4})
in this appendix.

\subsection{Assumptions (\ref{3}) and (\ref{4}): replacement and factorization}
\label{explicit_proof}

Let us start by proving the replacement
and factorization assumptions~(\ref{3}) and~(\ref{4}).
To this end, we first need to define
$[ \, \prod_{i=1}^n \, e^{\lambda_i V (a_i, a_{i+1} )} \, ]_r$,
$[ \, V (a_1) \,
\prod_{i=1}^n \, e^{\lambda_i V (a_i, a_{i+1} )} \, ]_r$,
$[ \, \prod_{i=1}^n \, e^{\lambda_i V (a_i, a_{i+1} )} \,
V (a_{n+1}) \, ]_r$,
and $[ \, V (a_1) \,
\prod_{i=1}^n \, e^{\lambda_i V (a_i, a_{i+1} )} \,
V (a_{n+1}) \, ]_r$.
Let us begin with
$[ \, \prod_{i=1}^n \, e^{\lambda_i V (a_i, a_{i+1} )} \, ]_r$.
We define it as follows:
\begin{equation}\label{expliVab}
[ \, \prod_{i=1}^n \, e^{\lambda_i V (a_i, a_{i+1} )} \, ]_r
\equiv \prod_{i=1}^n \, e^{\frac{1}{2} \, \lambda_i^2 \,
\langle \, V (a_i, a_{i+1})^2 \, \rangle_r}
\prod_{i < j} \, e^{\lambda_i \lambda_j \,
\langle \, V (a_i, a_{i+1}) \, V (a_j, a_{j+1}) \, \rangle_r}
\dcirc \prod_{i=1}^n \, e^{\lambda_i V (a_i, a_{i+1} )}
\dcirc \,,
\end{equation}
where
\begin{equation}
\begin{split}
\langle V(a,b)^2\rangle_r\,&\equiv\, 2\lim_{\epsilon \to 0} \,
\biggl[\, \int_a^{b-\epsilon} dt_1\int_{t_1+\epsilon}^{b} dt_2\,
G(t_1,t_2)-\frac{b-a-\epsilon}{\epsilon}-\ln\epsilon \, \biggr] \,,
\\
\langle \, V(a,b) \, V(b,c) \, \rangle_r
& \equiv \lim_{\epsilon \to 0} \, \biggl[ \,
\int_a^{b-\epsilon/2} d t_1 \, \int_{b + \epsilon/2}^c d t_2 \,
G (t_1,t_2) + \ln \, \epsilon \, \biggr] \,,
\\
\langle V(a,b)V(c,d)\rangle_r\,
& \equiv\, \int_a^bdt_1\int_c^ddt_2\,G(t_1,t_2)
\end{split}
\end{equation}
for $a<b<c<d$.
Their explicit expressions on ${\cal W}_n$ are
\begin{equation}
\begin{split}
\langle V(a,b)^2\rangle_r
& =\, \ln G_n (a,b) \,, \\
\langle \, V(a,b) \, V(b,c) \, \rangle_r
& =\, \frac{1}{2} \, \Bigl[ \,
\ln G_n (a,c) - \ln G_n (a,b) - \ln G_n (b,c) \, \Bigr] \,, \\
\langle V(a,b)V(c,d)\rangle_r\,
& =\, \frac{1}{2} \, \Bigl[ \, \ln G_n (a,d) + \ln G_n (b,c)
- \ln G_n (a,c) - \ln G_n (b,d) \, \Bigr] \,,
\end{split}
\end{equation}
where
\begin{equation}
G_n (t_1,t_2)
= \frac{\pi^2}{(n+1)^2\sin^2
\Bigl( \, \dfrac{t_2-t_1}{n+1}\,\pi \Bigr)} \,.
\end{equation}
The operator~(\ref{expliVab}) reduces to $[ \, e^{\lambda \, V(a,b)} \, ]_r$
defined in (\ref{renexpV}) when $n=1$.
It is easy to show that
\begin{equation}
\begin{split}
\langle \, V (a,c)^2 \, \rangle_r
& = \langle \, V (a,b)^2 \, \rangle_r
+2 \,\langle \, V (a,b) \, V (b,c) \, \rangle_r
+ \langle \, V (b,c)^2 \, \rangle_r \,, \\
\langle \, V (a,c) \, V (c,d) \, \rangle_r
& = \langle \, V (a,b) \, V (c,d) \, \rangle_r
+ \langle \, V (b,c) \, V (c,d) \, \rangle_r \,, \\
\langle \, V (a,b) \, V (b,d) \, \rangle_r
& = \langle \, V (a,b) \, V (b,c) \, \rangle_r
+ \langle \, V (a,b) \, V (c,d) \, \rangle_r \,, \\
\langle \, V (a,c) \, V (d,e) \, \rangle_r
& = \langle \, V (a,b) \, V (d,e) \, \rangle_r
+ \langle \, V (b,c) \, V (d,e) \, \rangle_r \,, \\
\langle \, V (a,b) \, V (c,e) \, \rangle_r
& = \langle \, V (a,b) \, V (c,d) \, \rangle_r
+ \langle \, V (a,b) \, V (d,e) \, \rangle_r
\end{split}
\end{equation}
for $a < b < c < d < e$.
The replacement assumption (\ref{3}) is therefore satisfied.
The assumption (\ref{4}) of factorization is also satisfied
because of the definition of
$\langle \, V(a,b) \, V(c,d) \, \rangle_r$
for $a < b < c < d$.

Let us next define the operators
$[ \, V (a_1) \,
\prod_{i=1}^n \, e^{\lambda_i \, V (a_i, a_{i+1} )} \, ]_r$,
$[ \, \prod_{i=1}^n \, e^{\lambda_i \, V (a_i, a_{i+1} )} \,
V (a_{n+1}) \, ]_r$,
and $[ \, V (a_1) \,
\prod_{i=1}^n \, e^{\lambda_i \, V (a_i, a_{i+1} )} \,
V (a_{n+1}) \, ]_r$.
We define them as follows:
\begin{equation}\label{multiple-region-operators}
\begin{split}
& [ \, V (a_1) \,
\prod_{i=1}^n \, e^{\lambda_i \, V (a_i, a_{i+1} )} \, ]_r \\
& \qquad \equiv
\prod_{i=1}^n \, e^{\frac{1}{2} \, \lambda_i^2 \,
\langle \, V (a_i, a_{i+1})^2 \, \rangle_r}
\prod_{i < j} \, e^{\lambda_i \lambda_j \,
\langle \, V (a_i, a_{i+1}) \, V (a_j, a_{j+1}) \, \rangle_r}
\dcirc V (a_1) \, \prod_{i=1}^n \, e^{\lambda_i \, V (a_i, a_{i+1} )}
\dcirc\\
& \qquad \quad ~ {}+ \sum_{i=1}^n \, \lambda_i \,
\langle \, V(a_1) \, V(a_i,a_{i+1}) \, \rangle_r \,
[ \, \prod_{i=1}^n \, e^{\lambda_i \, V (a_i, a_{i+1} )} \, ]_r \,,
\\
& [ \, \prod_{i=1}^n \, e^{\lambda_i \, V (a_i, a_{i+1} )} \,
V (a_{n+1}) \, ]_r \\
& \qquad \equiv
\prod_{i=1}^n \, e^{\frac{1}{2} \, \lambda_i^2 \,
\langle \, V (a_i, a_{i+1})^2 \, \rangle_r}
\prod_{i < j} \, e^{\lambda_i \lambda_j \,
\langle \, V (a_i, a_{i+1}) \, V (a_j, a_{j+1}) \, \rangle_r}
\dcirc \prod_{i=1}^n \, e^{\lambda_i \, V (a_i, a_{i+1} )} \,
V (a_{n+1}) \, \dcirc \\
& \qquad \quad ~ {}+ \sum_{i=1}^n \, \lambda_i \,
\langle \, V(a_i,a_{i+1}) \, V(a_{n+1}) \, \rangle_r \,
[ \, \prod_{i=1}^n \, e^{\lambda_i \, V (a_i, a_{i+1} )} \, ]_r \,,
\\
& [ \, V (a_1) \,
\prod_{i=1}^n \, e^{\lambda_i \, V (a_i, a_{i+1} )} \,
V (a_{n+1}) \, ]_r \\
& \qquad \equiv
\prod_{i=1}^n \, e^{\frac{1}{2} \, \lambda_i^2 \,
\langle \, V (a_i, a_{i+1})^2 \, \rangle_r}
\prod_{i < j} \, e^{\lambda_i \lambda_j \,
\langle \, V (a_i, a_{i+1}) \, V (a_j, a_{j+1}) \, \rangle_r}
\dcirc V(a_1) \,
\prod_{i=1}^n \, e^{\lambda_i \, V (a_i, a_{i+1} )} \,
V (a_{n+1}) \, \dcirc \\
& \qquad \quad ~ {}+ \sum_{i=1}^n \, \lambda_i \,
\langle \, V(a_1) \, V(a_i,a_{i+1}) \, \rangle_r \,
[ \, \prod_{i=1}^n \, e^{\lambda_i \, V (a_i, a_{i+1} )} \,
V(a_{n+1}) \, ]_r \\
& \qquad \quad ~ {}+ \sum_{i=1}^n \, \lambda_i \,
\langle \, V(a_i,a_{i+1}) \, V(a_{n+1}) \, \rangle_r \,
[ \, V(a_1) \,
\prod_{i=1}^n \, e^{\lambda_i \, V (a_i, a_{i+1} )} \, ]_r \\
& \qquad \quad ~
{}- \sum_{i, \, j=1}^n \, \lambda_i \, \lambda_j \,
\langle \, V(a_1) \, V(a_i,a_{i+1}) \, \rangle_r \,
\langle \, V(a_j,a_{j+1}) \, V(a_{n+1}) \, \rangle_r \,
[ \, \prod_{i=1}^n \, e^{\lambda_i \, V (a_i, a_{i+1} )} \, ]_r \\
& \qquad \quad ~ {}+ \langle \, V(a_1) \, V(a_{n+1}) \, \rangle_r \,
[ \, \prod_{i=1}^n \, e^{\lambda_i \, V (a_i, a_{i+1} )} \, ]_r \,,
\end{split}
\end{equation}
where
\begin{equation}
\begin{split}
& \langle \, V(a) \ V(a,b) \, \rangle_r
\equiv \lim_{\epsilon \to 0} \, \biggl[
\, \int_{a+\epsilon}^b dt \, G(a,t) -\frac{1}{\epsilon} \,
\biggr] \,, \qquad
\langle \, V(a,b) \, V(b) \, \rangle_r
\equiv\, \lim_{\epsilon \to 0} \,
\biggl[ \, \int_{a}^{b-\epsilon} dt \,
G(t,b) -\frac{1}{\epsilon} \, \biggr] \,, \\
& \langle \, V(a) \, V(b,c) \, \rangle_r
\equiv \int_b^c dt \, G(a,t) \,, \qquad
\langle \, V(a,b) \, V(c) \, \rangle_r
\equiv \int_a^b dt \, G(t,c) \,,
\qquad
\langle \, V(a) \, V(b) \, \rangle_r
\equiv G(a,b)
\end{split}
\end{equation}
for $a < b < c$.
These definitions are consistent with
$[ \, V(a) \, e^{\lambda \, V(a,b)} \, ]_r$
and $[ \, e^{\lambda \, V(a,b)} \, V(b) \, ]_r$
in~(\ref{renVexpV}).
It is easy to show that
\begin{equation}
\begin{split}
& \langle \, V (a) \, V (a,c) \, \rangle_r
= \langle \, V (a) \, V (a,b) \, \rangle_r
+ \langle \, V (a) \, V (b,c) \, \rangle_r \,, \\
& \langle \, V (a) \, V (b,d) \, \rangle_r
= \langle \, V (a) \, V (b,c) \, \rangle_r
+ \langle \, V (a) \, V (c,d) \, \rangle_r \,, \\
& \langle \, V (a,c) \, V (c) \, \rangle_r
= \langle \, V (a,b) \, V (c) \, \rangle_r
+ \langle \, V (b,c) \, V (c) \, \rangle_r \,, \\
& \langle \, V (a,c) \, V (d) \, \rangle_r
= \langle \, V (a,b) \, V (d) \, \rangle_r
+ \langle \, V (b,c) \, V (d) \, \rangle_r
\end{split}
\end{equation}
for $a < b < c < d$.
The replacement assumption (\ref{3}) is therefore satisfied.
The assumption (\ref{4}) of factorization is also satisfied
because of the definitions of
$\langle \, V(a) \, V(b,c) \, \rangle_r$,
$\langle \, V(a,b) \, V(c) \, \rangle_r$,
and $\langle \, V(a) \, V(b) \, \rangle_r$
for $a < b < c$.

\subsection{Assumption (\ref{2}): calculation of
$Q_B \cdot [ \, O_L(a) \, e^{\lambda \, V(a,b)} \, ]_r$}\label{appQOLexpV}

Let us next prove the assumption (\ref{2}) on the BRST transformation of
$[ \, O_L(a) \, e^{\lambda \, V(a,b)} \, ]_r$:
\begin{equation}
Q_B \cdot [ \, O_L(a) \, e^{\lambda \, V(a,b)} \, ]_r
= {}- [ \, O_L(a) \, e^{\lambda \, V(a,b)} O_R(b)\, ]_r \,,
\end{equation}
where
\begin{equation}
O_L (a) = \lambda \, cV (a)
- \frac{\lambda^2}{2} \, \partial c (a) \,, \qquad
O_R (b) = \lambda \, cV (b)
+ \frac{\lambda^2}{2} \, \partial c (b) \,.
\end{equation}
The operator $[ \, O_L(a) \, e^{\lambda \, V(a,b)} \, ]_r$
can be written as
\begin{equation}\label{O_L-exp}
\begin{split}
[ \, O_L(a) \, e^{\lambda \, V(a,b)} \, ]_r
& = \lambda \, e^{\frac{1}{2}\lambda^2\langle V(a,b)^2\rangle_r}\,
\dcirc cV(a) \, e^{\lambda V(a,b)}\dcirc \\
& \quad ~
+ \lambda^2 \, \langle \, V(a) \, V(a,b) \, \rangle_r \,
[ \, c(a) \, e^{\lambda V(a,b)} \, ]_r
- \frac{\lambda^2}{2} \,
[ \, \partial c(a) \, e^{\lambda V(a,b)} \, ]_r \,.
\end{split}
\end{equation}
The BRST transformation of
$\dcirc cV(a) \, e^{\lambda V(a,b)}\dcirc$
can be calculated in the following way:
\begin{equation}\label{Q_B-dcirc-cV-exp-dcirc}
\begin{split}
& Q_B \cdot \dcirc cV(a) \, e^{\lambda V(a,b)}\dcirc \\
& = Q_B \cdot \lim_{\epsilon \to 0} \, \biggl[ \,
cV(a-\epsilon) \, \dcirc e^{\lambda V(a,b)}\dcirc
- \lambda \, c(a-\epsilon) \int_a^b dt \, G(a-\epsilon,t) \,
\, \dcirc e^{\lambda V(a,b)}\dcirc \, \biggr] \\
& = \lim_{\epsilon \to 0} \, \biggl[ \,
{}- cV(a-\epsilon) \, Q_B \cdot \dcirc e^{\lambda V(a,b)}\dcirc
{}- \lambda \, c \partial c (a-\epsilon)
\int_a^b dt \, G(a-\epsilon,t) \,
\, \dcirc e^{\lambda V(a,b)}\dcirc \\
& \qquad \qquad
{}+ \lambda \, c(a-\epsilon) \int_a^b dt \, G(a-\epsilon,t) \,
\, Q_B \cdot \dcirc e^{\lambda V(a,b)}\dcirc \, \biggr] \,.
\end{split}
\end{equation}
The BRST transformation of $\dcirc e^{\lambda V(a,b)} \dcirc$
appearing in~(\ref{Q_B-dcirc-cV-exp-dcirc})
has been calculated in~(\ref{Q_B-exp}).
The contribution from the first term
$\lambda \, \dcirc e^{\lambda V(a,b)} \, cV(b) \dcirc$
on the right-hand side of~(\ref{Q_B-exp}) is
\begin{equation}
\begin{split}
& \lim_{\epsilon \to 0} \, \biggl[ \, {}- \lambda \,
cV(a-\epsilon) \, \dcirc e^{\lambda V(a,b)} \, cV(b) \dcirc
{}+ \lambda^2 \, c(a-\epsilon) \int_a^b dt \, G(a-\epsilon,t) \,
\, \dcirc e^{\lambda V(a,b)} \, cV(b) \dcirc \, \biggr] \\
& = {}- \lambda \, \dcirc cV(a) \, e^{\lambda V(a,b)} \, cV(b) \dcirc
{}- \lambda \, G(a,b) \,
\dcirc c(a) \, e^{\lambda V(a,b)} \, c(b) \dcirc \,.
\end{split}
\end{equation}
The contribution from the second term
$- \lambda \, \dcirc cV(a) \, e^{\lambda V(a,b)} \dcirc$
on the right-hand side of~(\ref{Q_B-exp})
diverges in the limit $\epsilon \to 0$:
\begin{equation}
\begin{split}
& \lambda \, cV(a-\epsilon) \,
\dcirc cV(a) \, e^{\lambda V(a,b)} \dcirc
{}- \lambda^2 \, c(a-\epsilon) \int_a^b dt \, G(a-\epsilon,t) \,
\, \dcirc cV(a) \, e^{\lambda V(a,b)} \dcirc \\
& = \lambda \,
\dcirc cV(a-\epsilon) \, cV(a) \, e^{\lambda V(a,b)} \dcirc
{}+ \lambda \, G(a-\epsilon,a) \, c(a-\epsilon) \, c(a) \,
\dcirc e^{\lambda V(a,b)} \dcirc \,.
\end{split}
\end{equation}
The first term on the right-hand side vanishes
in the limit $\epsilon \to 0$.
The second term is of $\ord{1/\epsilon}$,
but the sum of this term
and the second term on the right-hand side
of~(\ref{Q_B-dcirc-cV-exp-dcirc}) is finite
in the limit $\epsilon \to 0$:
\begin{equation}
\begin{split}
& \lim_{\epsilon \to 0} \, \biggl[ \,
{}- \lambda \, c \partial c (a-\epsilon)
\int_a^b dt \, G(a-\epsilon,t) \,
\, \dcirc e^{\lambda V(a,b)}\dcirc
{}+ \lambda \, G(a-\epsilon,a) \, c(a-\epsilon) \, c(a) \,
\dcirc e^{\lambda V(a,b)} \dcirc \, \biggr] \\
& = {}- \lambda \, \langle \, V(a) \, V(a,b) \, \rangle_r \,
c \partial c (a) \, \dcirc e^{\lambda V(a,b)}\dcirc
{}+ \frac{\lambda}{2}\, c \partial^2 c (a) \,
\dcirc e^{\lambda V(a,b)}\dcirc \,,
\end{split}
\end{equation}
where we have used
\begin{equation}
\begin{split}
\int_{a}^b dt\, G(a-\epsilon,t)
\,=&\, \frac{1}{\epsilon}
+\langle V(a)\,V(a,b)\rangle_r+\ord \epsilon \,, \\
G(a-\epsilon,a) \, c(a-\epsilon)c(a)
-\frac{1}{\epsilon}c\del c(a-\epsilon)
\,=&\,\frac{1}{2}c\del^2 c(a)+\ord \epsilon \,.
\end{split}
\end{equation}
Contributions from the remaining terms
on the right-hand side of~(\ref{Q_B-exp})
can be easily calculated.
The final result for the BRST transformation of
$\dcirc cV(a) \, e^{\lambda V(a,b)}\dcirc$ is
\begin{equation}
\begin{split}
Q_B \cdot \dcirc cV(a) \, e^{\lambda V(a,b)}\dcirc
& = {}- \lambda \,
\dcirc cV(a) \, e^{\lambda V(a,b)} \, cV(b) \dcirc
{}- \lambda \, G(a,b) \,
\dcirc c(a) \, e^{\lambda V(a,b)} \, c(b) \dcirc \\
& \quad ~ {}- \lambda \, \langle \, V(a) \, V(a,b) \, \rangle_r \,
\dcirc c \partial c (a) \, e^{\lambda V(a,b)}\dcirc
{}+ \frac{\lambda}{2}\,
\dcirc c \partial^2 c (a) \, e^{\lambda V(a,b)}\dcirc \\
& \quad ~ - \lambda^2 \,
\langle \, V(a,b) \, V(b) \, \rangle_r \,
\dcirc cV(a) \, e^{\lambda V(a,b)} \, c(b) \dcirc \\
& \quad ~ - \frac{\lambda^2}{2} \,
\dcirc cV(a) \, e^{\lambda V(a,b)} \, \partial c(b) \dcirc
- \frac{\lambda^2}{2} \,
\dcirc c \partial c V(a) \, e^{\lambda V(a,b)} \dcirc \,.
\end{split}
\end{equation}
Using (\ref{renVexpV}) and (\ref{multiple-region-operators}),
the operator
$Q_B \cdot \dcirc cV(a) \, e^{\lambda V(a,b)}\dcirc$
multiplied by the factor
$\lambda \, e^{\frac{1}{2}\lambda^2\langle V(a,b)^2\rangle_r}$
can be written as follows:
\begin{equation}\label{Q_B-O_L-exp-1}
\begin{split}
& \lambda \, e^{\frac{1}{2}\lambda^2\langle V(a,b)^2\rangle_r} \,
Q_B \cdot \dcirc cV(a) \, e^{\lambda V(a,b)}\dcirc \\
& = {}- \lambda \,
[ \, cV(a) \, e^{\lambda V(a,b)} \, O_R (b) \, ]_r
{}+ \lambda^2 \,
\langle \, V(a) \, V(a,b) \, \rangle_r \,
[ \, c(a) \, e^{\lambda V(a,b)} \, O_R (b) \, ]_r \\
& \quad ~ {}- \lambda^2 \, \langle \, V(a) \, V(a,b) \, \rangle_r \,
[ \, c \partial c (a) \, e^{\lambda V(a,b)} \, ]_r
{}+ \frac{\lambda^2}{2}\,
[ \, c \partial^2 c (a) \, e^{\lambda V(a,b)} \, ]_r \\
& \quad ~ - \frac{\lambda^3}{2} \,
[ \, c \partial c V(a) \, e^{\lambda V(a,b)} \, ]_r
+ \frac{\lambda^4}{2} \, \langle \, V(a) \, V(a,b) \, \rangle_r \,
[ \, c \partial c (a) \, e^{\lambda V(a,b)} \, ]_r \,. \\
\end{split}
\end{equation}
The BRST transformation of
$[ \, c(a) \, e^{\lambda V(a,b)} \, ]_r$ in (\ref{O_L-exp})
can be calculated as follows:
\begin{equation}\label{Q_B-O_L-exp-2}
\begin{split}
& Q_B \cdot [ \, c(a) \, e^{\lambda V(a,b)} \, ]_r
= \lim_{\epsilon \to 0} \,
Q_B \cdot [ \, c(a-\epsilon) \, e^{\lambda V(a,b)} \, ]_r \\
& = \lim_{\epsilon \to 0} \,
[ \, c \partial c (a-\epsilon) \, e^{\lambda V(a,b)} \, ]_r
- \lim_{\epsilon \to 0} \,
[ \, c(a-\epsilon) \, Q_B \cdot e^{\lambda V(a,b)} \, ]_r \\
& = [ \, c \partial c (a) \, e^{\lambda V(a,b)} \, ]_r
- [ \, c(a) \, e^{\lambda V(a,b)} \, O_R (b) \, ]_r
- \frac{\lambda^2}{2} \,
[ \, c \partial c (a) \, e^{\lambda V(a,b)} \, ]_r \,.
\end{split}
\end{equation}
Similarly, the BRST transformation of
$[ \, \partial c(a) \, e^{\lambda V(a,b)} \, ]_r$
in (\ref{O_L-exp}) can be calculated as
\begin{equation}\label{Q_B-O_L-exp-3}
\begin{split}
Q_B \cdot [ \, \partial c(a) \, e^{\lambda V(a,b)} \, ]_r
& = \lim_{\epsilon \to 0} \,
Q_B \cdot [ \, \partial c(a-\epsilon) \, e^{\lambda V(a,b)} \, ]_r \\
& = \lim_{\epsilon \to 0} \,
[ \, c \partial^2 c (a-\epsilon) \, e^{\lambda V(a,b)} \, ]_r
- \lim_{\epsilon \to 0} \,
[ \, \partial c(a-\epsilon) \, Q_B \cdot e^{\lambda V(a,b)} \, ]_r \\
& = [ \, c \partial^2 c (a) \, e^{\lambda V(a,b)} \, ]_r
- [ \, \partial c(a) \, e^{\lambda V(a,b)} \, O_R (b) \, ]_r
- \lambda \, [ \, c \partial c V (a) \, e^{\lambda V(a,b)} \, ]_r \,.
\end{split}
\end{equation}
By combining the results (\ref{Q_B-O_L-exp-1}),
(\ref{Q_B-O_L-exp-2}), and (\ref{Q_B-O_L-exp-3}), we find
\begin{equation}
\begin{split}
Q_B \cdot [ \, O_L(a) \, e^{\lambda \, V(a,b)} \, ]_r
& = {}- \lambda \,
[ \, cV(a) \, e^{\lambda V(a,b)} \, O_R (b) \, ]_r
+ \frac{\lambda^2}{2} \,
[ \, \partial c(a) \, e^{\lambda V(a,b)} \, O_R (b) \, ]_r \\
& = {}- [ \, O_L (a) \, e^{\lambda V(a,b)} \, O_R (b) \, ]_r \,.
\end{split}
\end{equation}
This completes the proof of the assumption~(\ref{2}).

\section{Marginal deformations for the constant mode
of the gauge field}\label{discussion_FKP}
\setcounter{equation}{0}

In~\cite{Fuchs:2007yy}, Fuchs, Kroyter and Potting constructed
solutions for the marginal deformation
corresponding to turning on the constant mode of the gauge field.
We discuss the relation
between their solutions and ours in this appendix.

The marginal operator for this deformation is
\begin{equation}\label{constant-gauge-field}
V(t) = \frac{i}{\sqrt{2 \alpha'}} \, \partial_t X^\mu (t) \,,
\end{equation}
where $X^\mu$ is a space-like direction along the D-brane.\footnote{
It is straightforward to incorporate
the time-like direction into the discussion.}
The solution in~\cite{Fuchs:2007yy} is written formally
as a pure-gauge form using the operator $X^\mu$.
The propagator $\langle \, X^\mu (t_1) \, X^\mu (t_2) \, \rangle$
is logarithmic, and thus the operator $X^\mu$ does not belong
to the complete set of local operators of the boundary CFT.
If we allow to use $X^\mu$, $V(a,b)$ can be written as follows:
\begin{equation}
   V(a,b) \,=\, \frac{i}{\sqrt{2 \alpha'}}\int_a^b dt\,  \partial_t X^\mu (t) \,=\, \frac{i}{\sqrt{2 \alpha'}}\Bigl(X^\mu(b)-X^\mu(a)\Bigr)\,.
\end{equation}
Then the operator $\dcirc e^{\lambda V(a,b)} \dcirc$
can be written as
\begin{equation}
\dcirc e^{\lambda V(a,b)} \dcirc
\,=\, : e^{\lambda V(a,b)} :
\,=\, : e^{ \, -\frac{i \lambda}{\sqrt{2 \alpha'}} X^\mu (a) \,}
e^{ \, \frac{i \lambda}{\sqrt{2 \alpha'}} X^\mu (b) \,} : \,.
\end{equation}
To turn this operator into $[ \, e^{\lambda V(a,b)} \, ]_r$\,,
we have to multiply it by
$e^{\frac{1}{2}\lambda^2\langle V(a,b)^2\rangle_r}$. We notice from
the explicit expression~(\ref{<V(a,b)^2>_r-on-W_n}) that
\begin{equation}
\langle \, V(a,b)^2 \, \rangle_r
= \frac{1}{\alpha'} \, \langle \, X^\mu (a) \,\, X^\mu (b) \, \rangle
\end{equation}
and therefore
\begin{equation}\label{rewriteexpVab}
\begin{split}
[ \, e^{\lambda V(a,b)} \, ]_r
\,&=\,e^{\frac{1}{2}\lambda^2\langle V(a,b)^2\rangle_r}
: e^{ \, -\frac{i \lambda}{\sqrt{2 \alpha'}} X^\mu (a) \,}
e^{ \, \frac{i \lambda}{\sqrt{2 \alpha'}} X^\mu (b) \,} : \\
&=\,e^{\, \frac{\lambda^2}{2 \alpha'}
\langle \, X^\mu (a) \, X^\mu (b) \, \rangle}
: e^{ \, -\frac{i\lambda}{\sqrt{2 \alpha'}} X^\mu (a) \,}
e^{ \, \frac{i \lambda}{\sqrt{2 \alpha'}} X^\mu (b) \,} : \\
&=\, : e^{ \, -\frac{i \lambda}{\sqrt{2 \alpha'}} X^\mu (a) \,} : \,
: e^{ \, \frac{i \lambda}{\sqrt{2 \alpha'}} X^\mu (b) \,} : \,.
\end{split}
\end{equation}
Because of the factor
$e^{\frac{1}{2}\lambda^2\langle V(a,b)^2\rangle_r}$,
the operator
$: e^{ \, -\frac{i \lambda}{\sqrt{2 \alpha'}} X^\mu (a) \,}
e^{ \, \frac{i \lambda}{\sqrt{2 \alpha'}} X^\mu (b) \,} :$
factorized into a product
of two primary fields at $a$ and $b$.
We can interpret the operators
${:e^{-\frac{i \lambda}{\sqrt{2 \alpha'}}  X^\mu(a)} :}$
and
${:e^{\frac{i \lambda}{\sqrt{2 \alpha'}}  X^\mu(b)} :}$
as the boundary-condition changing operators
at $a$ and $b$, respectively.
The conformal properties of the operator
$[ \, e^{\lambda V(a,b)} \, ]_r$ discussed
in~\S~\ref{conformal} are manifest in this expression.
In particular, the conformal dimension of
${:e^{\pm \frac{i \lambda}{\sqrt{2 \alpha'}}  X^\mu(b)} :}$
is $\lambda^2/2$ and thus consistent
with $h(\lambda) = \ord{\lambda^2}$ found in~\S~\ref{conformal}.

Let us see how the operators $O_L$ and $O_R$
arise from this expression.
Using the formula
\begin{equation}
\begin{split}
Q_B \, \cdot
: e^{ \, \pm \frac{i \lambda}{\sqrt{2 \alpha'}} X^\mu \,} :
\,&=\,:\biggl(
\pm\lambda \, \frac{i}{\sqrt{2\alpha'}} \, c\del X^\mu
+\frac{\lambda^2}{2}\,\del c \biggr) \,
e^{ \, \frac{\pm i \lambda}{\sqrt{2 \alpha'}} X^\mu \,} :\\
\,&=\,:\Bigl(\pm\lambda \, cV
+\frac{ \lambda^2}{2}\,\del c \Bigr) \,
e^{ \, \frac{\pm i \lambda}{\sqrt{2 \alpha'}} X^\mu \,} :\,,
\end{split}
\end{equation}
the BRST transformation of $[ \, e^{\lambda V(a,b)} \, ]_r$
can be calculated as follows:
\begin{equation}
\begin{split}
     Q_B \,\cdot\, \bigl[ \, e^{\lambda V(a,b)} \, \bigr]_r
     \,=&\,:e^{ \,- \frac{ i \lambda}{\sqrt{2 \alpha'}} X^\mu(a) \,}:
     \,:\Bigl(\lambda \, cV(b)
     +\frac{\lambda^2}{2}\,\del c(b) \Bigr)
          e^{ \, \frac{ i \lambda}{\sqrt{2 \alpha'}} X^\mu(b) \,}:\\
     \,&\,-:\Bigl(\lambda \, cV(a)
     -\frac{\lambda^2}{2}\,\del c(a) \Bigr)
     e^{ \,- \frac{ i \lambda}{\sqrt{2 \alpha'}} X^\mu(a) \,}:\,:
     e^{ \, \frac{ i \lambda}{\sqrt{2 \alpha'}} X^\mu(b) \,}: \,.
\end{split}
\end{equation}
We have thus reproduced our previous result for $O_L$ and $O_R$:
\begin{equation}
\begin{split}
O^{(1)}_R= O^{(1)}_L=cV\,,\qquad
O^{(2)}_R=-O^{(2)}_L=\frac{\del c}{2}\,,\qquad
O^{(n)}_R= O^{(n)}_L=0\qquad\text{for}\quad n\geq 3\,.
\end{split}
\end{equation}

The operator $[ \, e^{\lambda V(a,b)} \, ]_r$ is written
in~(\ref{rewriteexpVab}) in terms of the exponential operators
in the complete set of local operators
and thus well defined.
When we construct our solution,
we have to expand $[ \, e^{\lambda V(a,b)} \, ]_r$ in $\lambda$
to obtain $[ \, V^{(n)} (a,b) \, ]_r$.
We can write $[ \, V^{(n)} (a,b) \, ]_r$
in terms of local operators in the complete set
as we did in section 4,
but if we allow to use $X^\mu$,
$[ \, e^{\lambda V(a,b)} \, ]_r$
can also be expanded in $\lambda$ as
\begin{equation}
     \bigl[ \, e^{\lambda V(a,b)} \, \bigr]_r=\sum_{n=0}^\infty\lambda^n\biggl(\frac{i}{\sqrt{2\alpha'}}\biggr)^n
     \sum_{k=0}^n\frac{(-1)^k}{k!(n-k)!}:\bigl(X^\mu(a)\bigr)^k:\, :\bigl(X^\mu(b)\bigr)^{n-k}:\,,
\end{equation}
and the state $U^{(n)}$ for $n\geq1$ is
\begin{equation}
\langle \, \phi \,, U^{(n)} \, \rangle
= \sum_{k=0}^n\biggl(\frac{i}{\sqrt{2\alpha'}}\biggr)^n
\frac{(-1)^k}{k!(n-k)!}\,\bigl\langle \, f \circ \phi (0) \,
:\bigl(X^\mu(1)\bigr)^k:\, :\bigl(X^\mu(n)\bigr)^{n-k}:\,
\bigr\rangle_{{\cal W}_n} \,.
\end{equation}
The state $U$ can be formally factorized~\cite{Fuchs:2007yy}
as follows:
\begin{equation}
U = \Lambda_L \ast \Lambda_R \,,
\end{equation}
where
\begin{equation}
\Lambda_L = 1 + \sum_{n=1}^\infty \lambda^n \, \Lambda_L^{(n)} \,,
\qquad
\Lambda_R = 1 + \sum_{n=1}^\infty \lambda^n \, \Lambda_R^{(n)}
\end{equation}
with
\begin{equation}
\begin{split}
\langle \, \phi \,, \Lambda_L^{(n)} \, \rangle
& = \frac{1}{n!} \,
\biggl({}- \frac{i}{\sqrt{2\alpha'}}\biggr)^n \,
\bigl\langle \, f \circ \phi (0) \,
:\bigl(X^\mu(1)\bigr)^n:\, \bigr\rangle_{{\cal W}_n} \,, \\
\langle \, \phi \,, \Lambda_R^{(n)} \, \rangle
& = \frac{1}{n!} \,
\biggl(\frac{i}{\sqrt{2\alpha'}}\biggr)^n \,
\bigl\langle \, f \circ \phi (0) \,
:\bigl(X^\mu(n)\bigr)^n:\, \bigr\rangle_{{\cal W}_n} \,.
\end{split}
\end{equation}
The BRST transformation of $U$ is
\begin{equation}
Q_B U = ( Q_B \Lambda_L ) \ast \Lambda_R
+ \Lambda_L \ast ( Q_B \Lambda_R ) \,,
\end{equation}
and we find
\begin{equation}
A_L = {}-(Q_B \Lambda_L) \ast \Lambda_R \,, \qquad
A_R = \Lambda_L \ast (Q_B \Lambda_R) \,.
\end{equation}
The solutions $\Psi_L$ and $\Psi_R$ can thus be written as
\begin{equation}
\Psi_L = A_L \ast U^{-1}
= {}-(Q_B \Lambda_L) \ast \Lambda_L^{-1} \,, \qquad
\Psi_R = U^{-1} \ast A_R
= \Lambda_R^{-1} \ast (Q_B \Lambda_R) \,.
\end{equation}
These expressions in the pure-gauge form
coincide with the solutions
in~\cite{Fuchs:2007yy}.\footnote{
When the polarization vector $\epsilon_\nu$ of~\cite{Fuchs:2007yy}
is given by $\epsilon_\nu = \eta_{\mu \nu}$,
our $\lambda$ is related to that of \cite{Fuchs:2007yy} as follows:
\begin{equation*}
\lambda_{\, \rm FKP}
= \frac{i}{\sqrt{2}} \, \lambda_{\, \rm ours} \,.
\end{equation*}
Note in particular that their $\lambda$ must be imaginary
for the solution at $\ord{\lambda}$
to satisfy the reality condition.
}
Since the real solution~$\Psi$
constructed in~\S~\ref{singular_general_real} 
is related to $\Psi_L$ and $\Psi_R$ by gauge transformations,
$\Psi$ can also be written in a pure-gauge form:
\begin{equation}
\begin{split}
\Psi & = {}- \Bigl[ \,
Q_B \, \Bigl( \, \frac{1}{\sqrt{U}} \ast \Lambda_L \,
\Bigr) \, \Bigr] \,
\ast \Bigl( \, \Lambda_L^{-1} \ast \sqrt{U} \, \Bigr)\\
& = \Bigl( \, \sqrt{U} \ast \Lambda_R^{-1} \, \Bigr)
\ast \Bigl[\,Q_B \,
\Bigl( \, \Lambda_R \ast \frac{1}{\sqrt{U}}\,\Bigr)\,\Bigr] \,\\
& = \frac{1}{2}\,\Bigl( \, \sqrt{U} \ast \Lambda_R^{-1} \, \Bigr)
\ast \Bigl[\,Q_B \,
\Bigl( \, \Lambda_R \ast \frac{1}{\sqrt{U}}\,\Bigr)\,\Bigr] \,
- \frac{1}{2}\,\Bigl[ \, Q_B \,
\Bigl( \, \frac{1}{\sqrt{U}} \ast \Lambda_L \, \Bigr) \, \Bigr] \,
\ast \Bigl( \, \Lambda_L^{-1} \ast \sqrt{U} \, \Bigr) \,.
\end{split}
\end{equation}
In the last expression, $\Psi$ is manifestly real
because $\Lambda_R^\ddagger=\Lambda_L$.
We have thus solved the problem of finding a real solution
in a pure-gauge form raised in~\cite{Fuchs:2007gw}.

The states $\Lambda_L$ and $\Lambda_R$
cannot be written in terms of local operators
in the complete set,
while the solutions $\Psi_L$ and $\Psi_R$ can be written
without using $X^\mu$,
as we have explicitly demonstrated
in section~\ref{singular_explicit}.
It is, however, highly nontrivial to derive
such an expression of $\Psi_L$ or $\Psi_R$ from
the pure-gauge form in~\cite{Fuchs:2007yy}.
We could attempt, for example, to write $X^\mu (a)$ as 
\begin{equation}
X^\mu (a) = {}- \int_a^\infty dt \, \partial_t X^\mu (t)
\end{equation}
with the prescription that the contribution of
its BRST transformation from the boundary $t=\infty$ vanishes
and with the condition that the ``flux'' to infinity
cancels in the solution.
While this picture could give some useful insight,
it is obviously formal and it seems to be difficult
to make such approaches well defined in general.

We have seen that
the operator $X^\mu$ used in~\cite{Fuchs:2007yy}
as the basic object
in the construction of the solution
is formally the logarithm of the boundary-condition changing operator
corresponding to the marginal deformation.
Thus the solution in~\cite{Fuchs:2007yy} can be generalized
to other marginal deformations
if an expansion of the boundary-condition changing operator
in $\lambda$ is given.
However, the terms in the expansion do not belong
to the complete set of local operators,
and it is not clear how to calculate
correlation functions involving such operators in general.
Let us, for example, consider the deformation by the cosine potential
along a space-like direction $X^\mu$ which is compactified
at the self-dual radius. In this case,
the expansion of the boundary-condition changing
operator can be written in terms of $: (Y^\mu)^n :$\,,
where $Y^\mu$ is the free boson in the different description
we mentioned in~\S~\ref{examples}.
We then need to calculate
correlation functions
involving both $: (Y^\mu)^n :$ and operators
in the $X^\mu$ description, for example, 
when we expand the solution in the component fields.

While the approach in~\cite{Fuchs:2007yy} can be practically
useful for the particular marginal
deformation~(\ref{constant-gauge-field}),
we believe that our approach has an advantage
in the generalization to other marginal deformations.
In particular, we do not need to enlarge
the Hilbert space of the boundary CFT
at any intermediate stage,
which we believe will be a useful feature
when we address the question of background independence
in string field theory.

\end{document}